\documentclass[aps,twocolumn,pra,superscriptaddress,amsmath,amssymb]{revtex4-2}

\usepackage{dcolumn}
\usepackage{amsmath}
\usepackage{mathrsfs}
\usepackage{txfonts}
\usepackage{bm}
\usepackage[T1]{fontenc}
\usepackage{xspace}
\usepackage{ulem}
\usepackage{comment}
\usepackage{braket}
\usepackage{lipsum}
\usepackage{bbold}
\usepackage{mathtools}
\newcommand{\pdiff}[2]{\frac{\partial {#1}}{\partial {#2}}}

\allowdisplaybreaks[4]

\ifx\pdfoutput\undefined
\usepackage[dvipdfmx]{graphicx}
\usepackage[dvipdfmx]{hyperref}
\usepackage[dvipdfmx]{color}
\usepackage[dvipdfmx]{xcolor}
\else
\usepackage{graphicx}
\usepackage{hyperref}
\usepackage{color}
\usepackage{xcolor}
\usepackage[version=3]{mhchem}
\fi
\hypersetup{
        colorlinks=true,
        citecolor=blue,
        urlcolor=blue,
        linkcolor=blue
}

\begin{document}


\title{
Flexocurrent-induced magnetization:
Strain gradient-induced magnetization in time-reversal symmetric systems
}
\author{Shinnosuke Koyama}
\affiliation{%
Department of Physics, Tokyo Metropolitan University,
1-1, Minami-osawa, Hachioji, Tokyo 192-0397, Japan}

\author{Takashi Koretsune}
\affiliation{%
Department of Physics, Tohoku University,
Sendai, Miyagi 980-8578, Japan}

\author{Kazumasa Hattori}
\affiliation{%
Department of Physics, Tokyo Metropolitan University,
1-1, Minami-osawa, Hachioji, Tokyo 192-0397, Japan}

\date{\today}

\begin{abstract}
Symmetry constraints determine which physical responses are allowed in a given system. Magnetization induced by strain fields—such as in piezomagnetic and flexomagnetic effects—has typically been considered in materials that break time-reversal symmetry. Here, we propose that nonuniform strain can induce magnetization even in nonmagnetic metals and semiconductors that preserve time-reversal symmetry. This mechanism differs from the conventional flexomagnetic effect: the strain gradient acts as a driving force on the electrons, generating magnetization in a manner closely analogous to current-induced magnetization. Treating the strain field as an external field, we derive a general expression for the magnetization induced by a strain gradient and demonstrate that this response is symmetry-allowed even in time-reversal symmetric systems. We apply our formulation to nonmagnetic systems that lack spatial inversion symmetry while preserving time-reversal symmetry, using a decorated square lattice, monolayer MoS$_2$, and monolayer Janus MoSSe as representative examples. We find a finite magnetization response to strain gradients, which is consistent with symmetry arguments, supporting the validity of our theoretical framework. These results offer a pathway for controlling magnetization in nonmagnetic materials using strain fields.
\end{abstract}

\pacs{Valid PAY'S appear here}
\maketitle

\section{Introduction}

In condensed matter physics, cross-correlation responses—where the symmetry of an observable differs from that of the applied external field—have attracted significant attention~\cite{spaldin2005}
since the discovery of the magnetoelectric effect in ${\rm Cr_2O_3}$~\cite{folen1961}. 
These responses emerge in systems with symmetry breaking fields
such as the spatial distribution of spins~\cite{cheong2007,tokura2014},
electric dipoles~\cite{eerenstein2006,khomskii2009},
and various multipole moments~\cite{spaldin2008,zimmermann2014}.
They are group-theoretically classified by their macroscopic symmetry under the symmetry broken phases. 
Once the macroscopic symmetry is identified, which cross correlations emerge can be readily obtained~\cite{landau1980}.
To systematically carry out them,
a multipolar expansion of local and cluster degrees of freedom is powerful and provides
a unified description of the couplings between different degrees of freedom~\cite{hayami2018_prb,yatsushiro2021,kuramoto2009,suzuki2018_jpsj}.

The cross-correlation responses are of fundamental importance because they directly reflect symmetry breaking and, at the same time, offer considerable potential for device applications~\cite{bibes2008,narita2018,bowen2014,bukharaev2018,manchon2019,hirohata2020,fert2024}.
In particular, cross-correlated phenomena enable the interconversion among electric, magnetic, and mechanical signals, forming the basis of many spintronic and multiferroic applications. Among these, the control of magnetization is a central issue in spintronics, motivating extensive efforts toward low-power, high-density magnetic manipulation using nonconjugate external fields~\cite{manchon2019,hirohata2020}. Representative examples of magnetization responses driven by nonconjugate fields include the magnetoelectric effect~\cite{curie1894,dzyaloshinski1960,o'dell1962,eerenstein2006,dong2019} and the piezomagnetic effect~\cite{tavger1956,yatsushiro2021}. Since these responses are induced by external fields whose transformation under time reversal differs from that of magnetization, they can occur only in systems with broken time-reversal symmetry~(TRS)~\cite{hayami2018_prb,yatsushiro2021}. As a result, most studies have focused on magnetic materials.

The magnetoelectric and piezomagnetic effects are equilibrium magnetization responses, and whether such responses are allowed is strictly determined by the symmetry of the system. For nonequilibrium responses, however, the conditions imposed by TRS are modified. A representative example is current-induced magnetization (CIM) in metals, also known as the Edelstein effect~\cite{dyakonov1971,ivchenko1978,edelstein1990}. CIM arises in systems that exhibit spin-momentum locking in their electronic band structures, irrespective of whether the underlying symmetry-breaking fields are magnetic or nonmagnetic.
When an external electric field is applied, it acts as a driving force on the electrons and generates a nonequilibrium modification in their distribution function.
In the presence of dissipation—due to, for example, impurity scattering—the system relaxes into a nonequilibrium steady state that carries a finite electric current, which effectively breaks TRS. This nonequilibrium distribution then gives rise to a finite magnetization through spin-momentum locking.
Since CIM is a symmetry-allowed response even in systems that preserve TRS, it provides a powerful route to generate and control spin polarization and spin currents in nonmagnetic materials~\cite{sanchez2013}.

Recently, increasing attention has been directed toward magnetization induced by strain gradients.
This phenomenon, known as the flexomagnetic effect, is defined as the equilibrium linear response of magnetization to a strain gradient~\cite{eliseev2009,lukashev2010,tang2025_mse}. In nanoscale materials, thin films, and nanowires, large strain gradients can naturally arise~\cite{shu2019}, thereby significantly enhancing the response. As a result, small devices can generate substantial strain gradients with minimal deformation, enabling functionalities beyond those achievable with conventional magnetic devices~\cite{chappert2007}. 
Because strain gradients are rank-3 tensors, they can induce nonzero magnetization responses in a broad range of magnetic point groups~\cite{eliseev2011}. Theoretically, strain gradient-induced magnetization has been predicted in various materials, including ${\rm Mn_3GaN}$~\cite{lukashev2010}, ${\rm VSe_2}$~\cite{shi2019}, ${\rm CrI_3}$~\cite{edstrom2022,qiu2023,qiao2024}, ${\rm FeSe}$~\cite{tang2025_prb}, ${\rm CoTe}$~\cite{liu2025}, and Mn-doped ${\rm MoS_2}$~\cite{shen2018}.
Experimentally, the flexomagnetic effect has been observed in alloy films~\cite{belyaev2020,ling2023} and in ${\rm Cr_2O_3}$~\cite{makushko2022}, highlighting its potential for magnetization control via strain gradients.

In equilibrium, magnetization responses induced by strain fields—such as the piezomagnetic and flexomagnetic effects—are generally restricted
to systems that break TRS. 
This restriction arises because both strain and its spatial gradient are even under time reversal, whereas magnetization is odd. However,
when nonequilibrium responses are considered, a finite magnetization can emerge even in systems that preserve TRS. 
Since strain can couple to electric quadrupoles~\cite{callen1963,callen1965,ji2007,rosenberg2019},
a spatially varying strain generates a coupling term in the equations of motion and acts as a driving force on the electrons.
In the presence of dissipation, this driving force leads to a nonequilibrium steady state characterized by an asymmetric electronic distribution. This nonequilibrium state effectively breaks TRS and can induce finite magnetization.
This mechanism serves as a mechanical analog of CIM and suggests a novel route for controlling magnetization using strain fields in nonmagnetic materials.

\begin{figure}[t]
\includegraphics[width=1\columnwidth]{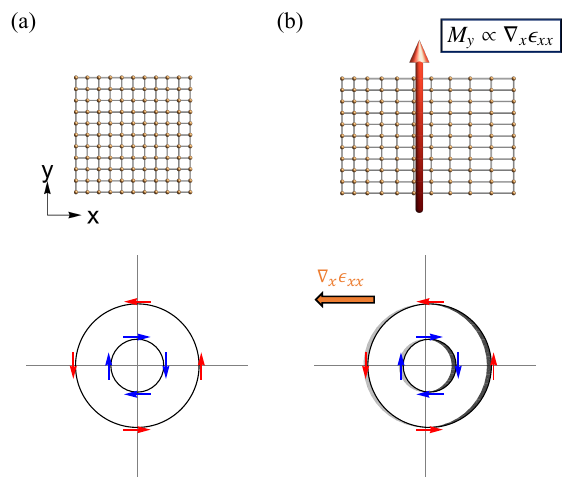}
\caption{
Schematic illustration of the flexocurrent-induced magnetization
under $\nabla_{x}\epsilon_{xx}$.
(a)
A system exhibiting Rashba spin–momentum locking in the absence of external strain.
(b)
In the presence of a strain field with a finite strain gradient $\nabla_x \epsilon_{xx}$,
the strain gradient acts as a driving force.
With momentum relaxation due to scattering mechanisms such as impurity scattering,
a nonequilibrium steady state with an asymmetric distribution is sustained,
resulting in a finite magnetization.
The shaded regions indicate asymmetric distribution of electrons in momentum space
due to the strain gradient.
}
\label{fig:concept}
\end{figure}

In this study, we formulate the magnetization response driven by a strain gradient
in time-reversal symmetric systems without magnetic orders.
Starting from a simple free-fermion Hamiltonian with orbital~(electric quadrupole) degrees of freedom,
we introduce strain fields that couple to the electric quadrupoles.
Employing Luttinger's method~\cite{luttinger1964},
we derive a Kubo formula~\cite{kubo1957} for the
magnetization induced by a strain gradient.
This response is nonequilibrium in nature and
should be distinguished from the flexomagnetic effect.
We refer to this magnetization response as flexocurrent-induced magnetization~(FCIM),
where the term flexocurrent denotes the electric-quadrupole current driven by a strain gradient.
We apply our formalism to three representative systems:
a nonmagnetic two-dimensional square lattice with the C$_{4v}$ point group,
a monolayer ${\rm MoS_2}$ with the D$_{3h}$ point group,
and a monolayer Janus ${\rm MoSSe}$ with the C$_{3v}$ point group.
For all three systems, we evaluate both the atomic orbital and spin components of FCIM.
Our results demonstrate that even in nonmagnetic systems with TRS,
a finite magnetization can be induced by a strain gradient in the presence of spin-momentum locking.
We also perform a simple symmetry analysis to identify the conditions for realizing a finite FCIM,
which supports our numerical findings.

This paper is organized as follows.
In Sec.~\ref{sec:theory},
we review the theoretical framework for FCIM.
Section~\ref{sec:free} introduces a free-fermionic system
that serves as an unperturbed Hamiltonian,
and Sec.~\ref{sec:introduction to strain} introduces 
a perturbative Hamiltonian in which external strains and the electric quadrupole moments are coupled with each other. 
Section~\ref{sec:driving force} demonstrates, using semiclassical theory,
that a strain gradient accelerates electrons and acts as a driving force.
In Sec.~\ref{sec:magnetization response to strain gradient},
we give a derivation of the Kubo formula for FCIM.
We then provide its symmetry analysis in Sec.~\ref{sec:symmetry}.
In Sec.~\ref{sec:result},
we present numerical results
of FCIM
in time-reversal symmetric nonmagnetic systems:
a two-dimensional square lattice with C$_{4v}$ symmetry~(Sec.~\ref{sec:C4v}), 
a monolayer ${\rm MoS}_2$ with D$_{3h}$ symmetry~(Sec.~\ref{sec:MoS2}),
and
a monolayer ${\rm MoSSe}$ with C$_{3v}$ symmetry~(Sec.~\ref{sec:MoSSe}). In Sec.~\ref{sec:discussion},
we discuss the furture perspectives in both experimental and theoretical points of view.
Finally,
Sec.~\ref{sec:summary} summarizes the main conclusions.

\section{Theory}
\label{sec:theory}

\subsection{Free-fermion Hamiltonian}
\label{sec:free}

We first introduce a free-fermion Hamiltonian,
defined as
\begin{align}
    \mathcal{H}
    &=
    \int d\bm{r} \Psi^\dagger (\bm{r}) \hat{H} \Psi (\bm{r}),
    \label{eq:H}
\end{align}
where $\hat{H}$ is an $N\times N$ Hermitian matrix
and
$N$ denotes the number of bands.
The field operators
$\Psi^\dagger(\bm{r})$
and $\Psi(\bm{r})$
are $N$-component vectors
constructed from the fermion creation and annihilation operators
satisfying the following relation:
$\{\Psi^\dagger_{s}(\bm{r}), \Psi_{s'}^\dagger (\bm{r}')\} = \delta(\bm{r}-\bm{r}') \delta_{ss'}$.
We introduce the Fourier transform as
\begin{align}
    \Psi_s (\bm{r}) = \frac{1}{\sqrt{V}} \sum_{\bm{k}} \Psi_{\bm{k},s} e^{i\bm{k}\cdot\bm{r}},
    \label{eq:Fourier}
\end{align}
where $V$ denotes the volume of the system and $\bm{k}$ is the wave vector.
Substituting Eq.~\eqref{eq:Fourier} into
Eq.~\eqref{eq:H}, we obtain
\begin{align}
    \mathcal{H} = \sum_{\bm{k}} \Psi^\dagger_{\bm{k}} \hat{H}_{\bm{k}} \Psi_{\bm{k}},
    \label{eq:H_momentum}
\end{align}
where
$\hat{H}_{\bm{k}} = e^{-i\bm{k}\cdot\bm{r}} \hat{H} e^{i\bm{k}\cdot\bm{r}}$
is the Bloch Hamiltonian for the wave vector $\bm k$.
Introducing an $N\times N$ unitary matrix $U_{\bm{k}}$ that diagonalizes $\hat{H}_{\bm{k}}$,
Eq.~\eqref{eq:H_momentum} can be rewritten as
\begin{align}
    \mathcal{H} = \sum_{\bm{k}} \Gamma_{\bm{k}}^\dagger \mathcal{E}_{\bm{k}} \Gamma_{\bm{k}},
    \label{eq:H_diagonalized}
\end{align}
where $\mathcal{E}_{\bm{k}} = U^\dagger_{\bm{k}} \hat{H}_{\bm{k}} U_{\bm{k}} = {\rm diag} (\varepsilon_{\bm{k},1},\varepsilon_{\bm{k},2}, \cdots, \varepsilon_{\bm{k},N})$, and
$\Gamma_{\bm{k}}^\dagger$ is defined as $\Gamma_{\bm{k}}^\dagger = \Psi_{\bm{k}}^\dagger U_{\bm{k}}^\dagger$.

\subsection{Strain fields}
\label{sec:introduction to strain}

We introduce an external strain field. Throughout this section, we assume that the system has cubic point group symmetry. We omit the discussion for other cases such as hexagonal symmetries, since the derivation is straightforward by similar analyses to those shown below. 

The strain tensor $\epsilon(\bm{r})$ is
a rank-$2$ symmetric tensor
defined as
\begin{align}
    \epsilon(\bm{r}) \equiv
    \begin{pmatrix}
        \epsilon_{xx}(\bm{r}) & \epsilon_{xy}(\bm{r}) & \epsilon_{xz}(\bm{r})\\
        \epsilon_{xy}(\bm{r}) & \epsilon_{yy}(\bm{r}) & \epsilon_{yz}(\bm{r})\\
        \epsilon_{xz}(\bm{r}) & \epsilon_{yz}(\bm{r}) & \epsilon_{zz}(\bm{r})
    \end{pmatrix}.
    \label{eq:strain-def}
\end{align}
Since these tensor components generally do not form the cubic irreducible representations, 
it is useful to rewrite them in the following form:
\begin{align}
    \epsilon_{x^2-y^2}(\bm{r}) &= \epsilon_{xx}(\bm{r}) - \epsilon_{yy}(\bm{r}),\\
    \epsilon_{3z^2-r^2}(\bm{r}) &= \frac{1}{\sqrt{3}} [2\epsilon_{zz}(\bm{r}) - \epsilon_{xx}(\bm{r}) - \epsilon_{yy}(\bm{r}) ],
\end{align}
together with $\epsilon_{xy}(\bm{r})$, $\epsilon_{yz}(\bm{r})$, and $\epsilon_{xz}(\bm{r})$.
Because the strain tensor is even under both spatial inversion and time reversal,
it couples to the electric quadrupole operators $Q_{\lambda}(\bm{r})$ as~\cite{callen1963,callen1965}
\begin{align}
    \mathcal{V}
    =
    \sum_{\lambda}^{{\rm Irrep.}} g_{\lambda} \int d\bm{r} \epsilon_{\lambda}(\bm{r}) Q_{\lambda}(\bm{r}),
    \label{eq:V_continuum}
\end{align}
where $\lambda = \{ xy, yz, xz, x^2-y^2, 3z^2-r^2 \}$, $g_\lambda$ is the coupling constant between $\epsilon_{\lambda}(\bm{r})$ and $Q_{\lambda}(\bm{r})$,
and $Q_{\lambda}(\bm{r})$ denotes the electric quadrupole density
expressed as
\begin{align}
    \label{eq:Q(r)}
    Q_{\lambda}(\bm{r}) = \frac{1}{2}[\Psi^\dagger(\bm{r}) \hat{Q}_{\lambda} \Psi(\bm{r}) + {\rm H.c.}],
\end{align}
with $\hat{Q}_\lambda = \hat{Q}^\dagger_{\lambda}$
\footnote{
The definition of $Q_{\lambda}(\bm{r})$ is fixed by the requirement
that the Hamiltonian be Hermitian.
When modulation of hopping amplitudes is taken into account,
$\hat{Q}_\lambda$ can generally become a nonlocal operator.
Nevertheless,
as long as $Q_\lambda(\bm{r}) = Q^\dagger_\lambda(\bm{r})$~[Eq.~\eqref{eq:Q(r)}] is satisfied,
$\mathcal{V}$ is guaranteed to be Hermitian,
and the formulation in the present study
remains valid even if $\hat{Q}_\lambda$ is a nonlocal operator.
A similar definition of a local density
can also be found in Eq.~(21) of Ref.~\cite{qin2011}.}
.
Each component of $\hat{Q}_{\lambda}$ is defined similarly to that of $\epsilon_{\lambda}$ and acts
on the local orbital subspace of the electron field $\Psi(\bm{r})$.
Here, for simplicity, we assume that the couplings between $\epsilon(\bm{r})$ and $Q(\bm{r})$  are local in this paper,
while it is also possible to derive similar results by taking into account
coupling constants between nonlocal orbital degrees of freedom,
i.e., hopping between neighboring sites and $\epsilon(\bm{r})$~\cite{ogawa2023,uchino2025_arxiv}.
Since the isotropic strain component
$\epsilon_{r^2} = \frac{1}{\sqrt{3}} (\epsilon_{xx} + \epsilon_{yy} + \epsilon_{zz})$ necessarily belongs to the trivial representation
and transforms as a rank-$0$ scalar, 
the coupling in $\lambda = r^2 = x^2+y^2+z^2$ is absent in Eq.~(\ref{eq:V_continuum}). Consequently,
the multipole that couples to $\epsilon_{r^2}$
is not an electric quadrupole but an electric monopole.
From the viewpoint of symmetry,
this coupling merely amounts to expressing
a scalar potential in terms of the strain field.
Accordingly,
the physics generated by the coupling term of $\lambda=r^2$
is essentially identical to that
arising from the coupling between
a scalar potential and an electric monopole.
For this reason,
we exclude this contribution in the present study.

\subsection{Strain gradient as a driving force}
\label{sec:driving force}

Before formulating magnetization responses 
to a strain gradient $\bm{\nabla}\epsilon_\lambda$,
we clarify that $\bm{\nabla}\epsilon_\lambda$ acts as
a driving force on electrons
within the semiclassical theory,
and that
FCIM is determined
by the dissipative part of
the nonequilibrium magnetization response tensor.

The eigenvalue equation of $\hat{H}_{\bm{k}}$
is given by
\begin{align}
    \hat{H}_{\bm{k}} \ket{u_{\bm{k},n}} = \varepsilon_{\bm{k},n} \ket{u_{\bm{k},n}},
\end{align}
where $|u_{\bm{k},n}\rangle$ is the Bloch state corresponding to the $n$th column of $U_{\bm{k}}$ and one should regard
$\hat{H}_{\bm{k}}$ as acting on the Bloch states represented in the bra--ket formalism rather than on the operator $\Psi_{\bm k}$ in Eq.~(\ref{eq:H_momentum}).
Using this notation,
we define a wave packet
well localized around the center position $\bm{r}_c$ and the wave vector $\bm{k}_c$, 
constructed from the $n$th band as~\cite{chang1996,sundaram1999,dong2020}
\begin{align}
    \ket{W_{n}}
    = \int d\bm{q} C_{\bm{k}_c, n}(\bm{q}) e^{i\bm{q}\cdot\hat{\bm{r}}} \ket{u_{\bm{q},n}},
\end{align}
where 
$\ket{W_n}$ satisfies
\begin{align}
    \braket{W_n|\hat{\bm{r}}|W_n} = \bm{r}_c.
\end{align}
Here, the expansion coefficients $C_{\bm{k}_c,n}(\bm{q})$
satisfy
\begin{align}
    &\int d\bm{q} |C_{\bm{k}_c,n}(\bm{q})|^2 = 1,\\
    &\int d\bm{q} |C_{\bm{k}_c,n}(\bm{q})|^2 f(\bm{q}) \simeq f(\bm{k}_c),
\end{align}
for an arbitrary function $f(\bm{q})$.
From Eqs.~\eqref{eq:H} and~\eqref{eq:V_continuum},
the total one-body Hamiltonian is written as
\begin{align}
    \hat{H}_{\rm tot}
    =
    \hat{H} + \sum_\lambda^{{\rm Irrep.}} g_\lambda \hat{Q}_\lambda \epsilon_\lambda (\hat{\bm{r}}).
\end{align}
We assume that the strain field varies slowly on the spatial scale of the wave packet.
Thus, we can expand $\epsilon_\lambda(\hat{\bm{r}})$ around $\bm{r}_c$:
\begin{align}
    \epsilon_\lambda (\hat{\bm{r}})
    =
    \epsilon_\lambda (\bm{r}_c)
    +
    (\hat{\bm{r}} - \bm{r}_c) \cdot \bm{\nabla}\epsilon_\lambda
    +
    \cdots.
    \label{eq:expansion-strain}
\end{align}
Hereafter,
we omit the subscript $c$ for notational simplicity.
The equations of motion of the center position and
the center wave vector read~\cite{dong2020}
\begin{align}
    \label{eq:eom-r}
    \dot{\bm{r}} &= \frac{1}{\hbar} \pdiff{\varepsilon_{\bm{k},n}(\bm{r})}{\bm{k}},\\
    \label{eq:eom-k}
    \hbar\dot{\bm{k}} &= - \bm{\nabla}\varepsilon_{\bm{k},n}(\bm{r}),
\end{align}
where $\varepsilon_{\bm{k},n}(\bm{r})
=\braket{W_n|\hat{H}_{\rm tot}|W_n}$
and $\hbar$ is the Dirac constant.
Here, geometric contributions~(e.g., Berry curvature terms)
are omitted,
as they are not relevant to the discussion below.
Substituting Eq.~\eqref{eq:expansion-strain} into Eq.~\eqref{eq:eom-k}, 
we obtain
\begin{align}
    \hbar\dot{\bm{k}}
    \simeq - \sum_\lambda^{{\rm Irrep.}} g_\lambda (\hat{Q}_{\lambda,\bm{k}})_n \bm{\nabla}\epsilon_\lambda,
    \label{eq:k-motion}
\end{align}
where
$(\hat{Q}_{\lambda,\bm{k}})_n
=\braket{W_{n}|\hat{Q}_\lambda|W_{n}}
\simeq \braket{u_{\bm{k},n}|\hat{Q}_\lambda|u_{\bm{k},n}}$.
Equation~\eqref{eq:k-motion} shows that
a finite strain gradient acts as a driving force and thus accelerates the electron wavepacket, 
leading to a shift of the electronic distribution in momentum space~(an asymmetric distribution).

In the presence of dissipation,
the nonequilibrium steady state with the asymmetric distribution can be sustained.
The term {\it flexocurrent} is defined
as the shorthand for the current induced by $\bm{\nabla} \epsilon_\lambda$.
In systems with spin-momentum splittings
in the electronic band structure~[see Fig.~\ref{fig:concept}(a) for Rashba systems],
this nonequilibrium distribution is
expected to result in a finite magnetization,
as shown in Fig.~\ref{fig:concept}(b).
We refer to this nonequilibrium magnetization response as FCIM.
FCIM is analogous to CIM
and constitutes a disspative part of nonequilibrium response in metals.
Therefore,
this effect can be formulated
by evaluating the dissipative component of the adiabatic
magnetization response tensor
within the Kubo's linear response theory~\cite{kubo1957}.

\subsection{Magnetization response to strain gradient}
\label{sec:magnetization response to strain gradient}

In this section, we formulate the linear response theory when general strain fields $\epsilon(\bm{r})$
are applied to the system, where Hamiltonian is given by Eq.~\eqref{eq:H_momentum}.
We introduce the adiabatic
magnetization response tensor $f^{\lambda}_{\alpha\beta}$,
which relates the macroscopic magnetization $\bar{M}_\alpha$
to the strain gradient
$\nabla_\beta \epsilon_\lambda$ as
\begin{align}
    \label{eq:FCIM-def}
    \bar{M}_{\alpha} = f_{\alpha\beta}^{\lambda} \nabla_{\beta}\epsilon_{\lambda}.
\end{align}
Here, the repeated indices are assumed to be summed over and the 
macroscopic magnetization is defined as
$\bar{M}_\alpha =\frac{1}{V}\int d\bm{r} \braket{M_{\alpha}(\bm{r})}$,
where $\braket{\cdots}$ stands for the thermal average and
\begin{align}
    \label{eq:M(r)}
    M_\alpha (\bm{r}) =
    \Psi^\dagger(\bm{r}) \hat{M}_{\alpha} \Psi(\bm{r})
\end{align}
is the local magnetization density.
The operator $\hat{M}_\alpha$ is a one-body magnetization operator,
such as the spin or orbital magnetic moment.
To evaluate $f_{\alpha\beta}^\lambda$,
we employ Luttinger's method~\cite{luttinger1964}
originally introduced to calculate
the thermal average of observables induced by the spatial gradient
of a scalar potential
within the framework of Kubo linear response theory.
This formalism provides the transport limit of the response~($\bm{q}\to 0$ then
$\omega\to 0$).
This corresponds to focusing on the nonequilibrium response.
We apply this approach to the case
of the strain field.
We note that, by contrast, taking the static limit~($\omega \to 0$ then $\bm{q} \to 0$)
amounts to probing the equilibrium response
and corresponds to the conventional flexomagnetic effect
\cite{lukashev2010,tang2025_mse}.
Introducing the density operator $\rho$,
$\bar{M}_\alpha$ can be expressed as
\begin{align}
\bar{M}_\alpha = \frac{1}{V} \int d\bm{r} {\rm Tr}[\rho M_\alpha(\bm{r})].
\label{eq:Mbar_rho}
\end{align}
Within linear response theory,
$\rho$ is given by
\begin{align}
  \rho = - \rho_{{\rm eq}} \int dt e^{-\delta t} \int_0^{\frac{1}{k_{\rm{B}}T}}ds
  \dot{\mathcal{V}}(-t-i\hbar s),
  \label{eq:rho_luttinger}
\end{align}
where $\rho_{\mathrm{eq}}$ is the equilibrium density operator
in the absence of external fields,
$k_{\rm B}$ is the Boltzmann constant,
$T$ is the temperature,
and $\delta$ is a positive infinitesimal constant.
Here, $\dot{\mathcal{V}}$ denotes the time derivative
of the perturbation~[Eq.~\eqref{eq:V_continuum}], and the time evolution of an operator
is defined in the Heisenberg representation
as $O(t) = e^{i\mathcal{H}t/\hbar} O e^{-i\mathcal{H}t/\hbar}$.
To evaluate $\dot{\mathcal{V}}$,
we calculate the time derivative of $Q_\lambda(\bm{r})$. 
The Heisenberg equation leads to 
\begin{align}
  \dot{Q}_{\lambda}(\bm{r})
  &= \frac{i}{\hbar} [\mathcal{H}, Q_{\lambda}(\bm{r})]\notag\\
  &= -\bm{\nabla} \cdot \bm{J}^{\lambda}(\bm{r}) + \tau^{\lambda}(\bm{r}), \label{eq:Qdot}
\end{align}
where we have introduced the electric quadrupole current $\bm{J}^{\lambda}(\bm{r})$ and torque densities $\tau^{\lambda}(\bm{r})$ as 
\begin{align}
    \label{eq:J(r)}
    \bm{J}^{\lambda}(\bm{r}) = \frac{1}{2}[\Psi^\dagger(\bm{r}) \hat{\bm{J}}^{\lambda} \Psi(\bm{r}) + {\rm H.c.}],\\
    \label{eq:T(r)}
    \tau^{\lambda}(\bm{r}) = \frac{1}{2} [\Psi^\dagger(\bm{r})\hat{\tau}^{\lambda}\Psi(\bm{r}) + {\rm H.c.}],
\end{align}
Here, using the velocity operator defined as $\hat{\bm{v}} = \frac{i}{\hbar}[\hat{H}, \hat{\bm{r}}]$, 
we can express their one-body operators as $\hat{\bm{J}}^{\lambda} = \frac{1}{2} (\hat{\bm{v}} \hat{Q}_{\lambda} + \hat{Q}_{\lambda}\hat{\bm{v}})$
and
$\hat{\tau}^{\lambda} = \frac{i}{\hbar} [\hat{H}, \hat{Q}_{\lambda}]$. 
Note that Eqs.~\eqref{eq:J(r)} and~\eqref{eq:T(r)}
remain valid even when $\hat{Q}_\lambda$ is nonlocal.
Using Eqs.~\eqref{eq:V_continuum}, \eqref{eq:Qdot}, \eqref{eq:J(r)}, and \eqref{eq:T(r)}, 
we obtain 
\begin{align}
  \dot{\mathcal{V}}
  &= \sum_{\lambda}^{{\rm Irrep.}} g_\lambda \int d\bm{r} \epsilon_{\lambda}(\bm{r}) \dot{Q}_\lambda(\bm{r})\notag\\
  &= \sum_{\lambda}^{{\rm Irrep.}} g_\lambda \int 
  d\bm{r} \left[\bm{J}^\lambda(\bm{r}) \cdot \bm{\nabla} \epsilon_\lambda(\bm{r})
  + \tau^\lambda (\bm{r}) \epsilon_\lambda(\bm{r})
  \right],
  \label{eq:Vdot}
\end{align}
where we have carried out integration by parts for the first term.
Hence, $\bar{M}_\alpha$ is given by
\begin{align}
  \bar{M}_\alpha
  = - \frac{1}{V}
  &\sum_{\lambda}^{{\rm Irrep.}} g_\lambda \int dt e^{-\delta t} \int_0^{\frac{1}{k_{\rm{B}}T}} ds\int d\bm{r} \int d\bm{r}'\notag\\
  &\times
  \Bigg\{
    \sum_\beta \braket{J_\beta^\lambda(\bm{r},-i\hbar s) M_\alpha(\bm{r}',t)}_{{\rm eq}} \nabla_\beta \epsilon_\lambda (\bm{r}) \notag\\
  & + \braket{\tau^\lambda(\bm{r},-i\hbar s) M_\alpha(\bm{r}',t)}_{{\rm eq}} \epsilon_\lambda(\bm{r})
  \Bigg\}, \label{eq:M}
\end{align}
where $\braket{\cdots}_{{\rm eq}}$ denotes
the thermal average
in equilibrium before applying the external field $\epsilon(\bm{r})$ at $t=-\infty$.
Thus, $f^\lambda_{\alpha\beta}$ can be expressed as
\begin{align}
  f^{\lambda}_{\alpha\beta} =
  f^{\lambda(1)}_{\alpha\beta} + f^{\lambda(2)}_{\alpha\beta},
\end{align}
with
\begin{align}
  f^{\lambda(1)}_{\alpha\beta}
  &=
  - \frac{g_\lambda}{V} \int dt e^{-\delta t} \int_0^{\frac{1}{k_{\rm{B}}T}} ds 
  \braket{J_{\beta,-\bm{q}=0}^\lambda(-i\hbar s) M_{\alpha,\bm{q}=0}(t)}_{{\rm eq}},
  \label{eq:f1}\\
  f^{\lambda(2)}_{\alpha\beta}
  &=
  - \frac{g_\lambda}{V} \int dt e^{-\delta t} \int_0^{\frac{1}{k_{\rm{B}}T}} ds 
  \pdiff{}{iq_\beta}\left[
  \braket{\tau^\lambda_{-\bm{q}}(-i\hbar s) M_{\alpha,\bm{q}}(t)}_{{\rm eq}} \right]_{\bm{q}= 0}.
  \label{eq:f2}
\end{align}
Here,
$J^\lambda_{\beta,\bm{q}} = \int d\bm{r} e^{-i\bm{q}\cdot\bm{r}} J^\lambda_{\beta}(\bm{r})$,
$\tau^\lambda_{\bm{q}} = \int d\bm{r} e^{-i\bm{q}\cdot\bm{r}} \tau^\lambda(\bm{r})$, and
$M_{\alpha,\bm{q}} = \int d\bm{r} e^{-i\bm{q}\cdot\bm{r}} M_{\alpha}(\bm{r})$.
The derivation of $f^{\lambda(2)}_{\alpha\beta}$ in Eq.~\eqref{eq:f2}
is provided in Appendix~\ref{app:expression of f2}.

The response tensor
$f_{\alpha\beta}^{\lambda}$ contains
both dissipative and nondissipative contributions. 
To evaluate FCIM, we focus on the dissipative contributions.
The dissipative parts of $f^{\lambda(1)}_{\alpha\beta}$ and $f^{\lambda(2)}_{\alpha\beta}$
are denoted by $f^{\lambda(1):{\rm dis}}_{\alpha\beta}$ and $f^{\lambda(2):{\rm dis}}_{\alpha\beta}$, respectively,
and are given by
\begin{align}
  f^{\lambda(1):{\rm dis}}_{\alpha\beta}
  &=
  \frac{\hbar g_\lambda}{V\eta} \sum_{\bm{k}} \sum_{n=1}^{N} (\hat{M}_{\alpha,\bm{k}})_{n} (\hat{J}^{\lambda}_{\beta,\bm{k}})_{n}
  \pdiff{f(\varepsilon_{\bm{k},n})}{\varepsilon_{\bm{k},n}}, \label{eq:f1-dis}\\
  f^{\lambda(2):{\rm dis}}_{\alpha\beta}
  &=
  -\frac{\hbar g_\lambda}{V\eta} \sum_{\bm{k}} \sum_{n=1}^{N} \sum_{m (\neq n)}^{N}
  (\hat{M}_{\alpha,\bm{k}})_{n} {\rm Re} [ (\hat{v}_{\beta,\bm{k}})_{nm} (\hat{Q}_{\lambda,\bm{k}})_{mn} ]\notag\\
  &\qquad\qquad\qquad\qquad\quad
  \times\pdiff{f(\varepsilon_{\bm{k},n})}{\varepsilon_{\bm{k},n}} \label{eq:f2-dis},
\end{align}
where $\eta$ is a phenomenological damping rate originating from nonmagnetic impurity scatterings,
and $f(x) =  [e^{\beta(x-\mu)}+1]^{-1}$
denotes the Fermi distribution function with
the chemical potential $\mu$.
A detailed derivation of Eqs.~\eqref{eq:f1-dis} and~\eqref{eq:f2-dis}
is provided in Appendix~\ref{app:derivation of FCIM}.
We have introduced 
the Bloch representation of a one-body operator
$\hat{o}_{\bm{k}} = e^{-i\bm{k}\cdot\bm{r}} \hat{o} e^{i\bm{k}\cdot\bm{r}}$,
and define its matrix elements on the band basis as
$(\hat{o}_{\bm{k}})_{n} = (U^\dagger_{\bm{k}}\hat{o}_{\bm{k}}U_{\bm{k}})_{nn}$ and
$(\hat{o}_{\bm{k}})_{nm} = (U^\dagger_{\bm{k}}\hat{o}_{\bm{k}}U_{\bm{k}})_{nm}$.
Note that since $\hat{Q}_{\lambda}$ does not contain the momentum operator,
we have $e^{-i\bm{k}\cdot\bm{r}} \hat{Q}_{\lambda}e^{i\bm{k}\cdot\bm{r}} = \hat{Q}_{\lambda}$.
Equation~\eqref{eq:f1-dis} represents the contribution induced by the electric quadrupole current,
which can be regarded as
the counterpart of CIM.
In CIM, the electric quadrupole current in Eq.~\eqref{eq:f1-dis}
is replaced by the electric current~\cite{freimuth2014CIM,zelezny2017CIM}.
Equation~\eqref{eq:f2-dis} corresponds to
the contribution induced by the electric quadrupole torque.
We note that in CIM, no analogous torque term~[Eq.~\eqref{eq:f2-dis}] appears because the scalar potential couples with the charge density,
which always satisfies the continuity equation.
Using the relation
$(\hat{J}^\lambda_{\beta,\bm{k}})_n
= {\rm Re} \sum_{m=1}^{N} (\hat{v}_{\beta,\bm{k}})_{nm} (\hat{Q}_{\lambda,\bm{k}})_{mn}$,
the total response coefficient $f^{\lambda:{\rm dis}}_{\alpha\beta}$ can be written as
\begin{align}
  f^{\lambda:{\rm dis}}_{\alpha\beta}
  &=
  f^{\lambda(1):{\rm dis}}_{\alpha\beta} + f^{\lambda(2):{\rm dis}}_{\alpha\beta}\notag\\
  &=
  \frac{\hbar g_\lambda}{V\eta} \sum_{\bm{k}} \sum_{n=1}^{N}
  (\hat{M}_{\alpha,\bm{k}})_{n} (\hat{v}_{\beta,\bm{k}})_{n} (\hat{Q}_{\lambda,\bm{k}})_{n}
  \pdiff{f(\varepsilon_{\bm{k},n})}{\varepsilon_{\bm{k},n}} \label{eq:f-dis}.
\end{align}
When the torque is absent, i.e., $[\hat{H}_{\bm{k}}, \hat{Q}_{\lambda}] = 0$,
all the off-diagonal interband components $(\hat{Q}_{\lambda,\bm{k}})_{mn}$ with $m\neq n$ vanish,
and Eq.~\eqref{eq:f1-dis} reduces to Eq.~\eqref{eq:f-dis}.

We note that it is straightforward to check that 
Eq.~\eqref{eq:f-dis} agrees exactly 
with the expression obtained from the semiclassical theory.
One can follow the discussion in Ref.~\cite{johansson2024}. 
Using the semiclassical Boltzmann transport equation
under Eqs.~\eqref{eq:eom-r},~\eqref{eq:eom-k}, and~\eqref{eq:k-motion},
one finds that the Fermi surface contribution to the nonequilibrium magnetization
coincides with Eq.~\eqref{eq:f-dis}.

\subsection{Symmetry arguments under spatial inversion and time-reversal operations}
\label{sec:symmetry}

\begin{table}[b]
\centering
\setlength{\tabcolsep}{8pt}
\caption{Symmetry properties under spatial inversion $\mathcal{P}$ and time-reversal $\mathcal{T}$.}
\begin{ruledtabular}
\begin{tabular}{lccc}
Quantities & Symbol & $\mathcal{P}$ & $\mathcal{T}$\\
\hline
Strain gradient & $\bm{\nabla}\epsilon_\lambda$ & $-$ & $+$ \\[4pt]
Electric quadrupole current & $\bm{J}^\lambda$ & $-$ & $-$ \\[4pt]
Electric quadrupole torque & $\tau^\lambda$ & $+$ & $-$ \\[4pt]
Magnetization & $\bm{M}$ & $+$ & $-$
\end{tabular}
\end{ruledtabular}
\label{tab:PandT}
\end{table}

In this section, we discuss how presence or absence of the spatial inversion and time-reversal symmetries gives conditions for a finite FCIM.
In systems with spatial inversion symmetry,
the one-body operators appearing in Eq.~\eqref{eq:f-dis} 
satisfy
$\mathcal{I}\hat{H}_{\bm{k}}\mathcal{I}^{-1} = \hat{H}_{-\bm{k}}$,
$\mathcal{I}\hat{\bm{v}}_{\bm{k}}\mathcal{I}^{-1} = - \hat{\bm{v}}_{-\bm{k}}$, 
$\mathcal{I}\hat{Q}_\lambda\mathcal{I}^{-1} =  \hat{Q}_{\lambda}$, and
$\mathcal{I}\hat{\bm{M}}_{\bm{k}}\mathcal{I}^{-1} = \hat{\bm{M}}_{-\bm{k}}$,
where $\mathcal{I}$ denotes the spatial inversion operator.
Thus, Eq.~\eqref{eq:f-dis} with the dummy variable $\bm{k}\to -\bm{k}$ reads
\begin{align}
    f_{\alpha\beta}^{\lambda:{\rm dis}}
    &=
    \frac{\hbar g_\lambda}{V\eta} \sum_{\bm{k}}\sum_{n=1}^{N}
    \Bigg[
    (+\hat{M}_{\alpha,\bm{k}})_{n} (-\hat{v}_{\beta,\bm{k}})_{n} (+\hat{Q}_{\lambda,\bm{k}})_{n}
    \pdiff{f(\varepsilon_{\bm{k},n})}{\varepsilon_{\bm{k},n}}
    \Bigg]\notag\\
    &=-f_{\alpha\beta}^{\lambda:{\rm dis}}.
\end{align}
Thus, FCIM vanishes 
in systems with spatial inversion symmetry.
Whereas the spatial inversion symmetry forces
FCIM to vanish, the TRS does not impose any
constraint on FCIM.
In systems with TRS,
the operators transform as
$\Theta\hat{H}_{\bm{k}}\Theta^{-1} = \hat{H}_{-\bm{k}}$,
$\Theta\hat{\bm{v}}_{\bm{k}}\Theta^{-1} = - \hat{\bm{v}}_{-\bm{k}}$, 
$\Theta\hat{Q}_\lambda\Theta^{-1} =  \hat{Q}_{\lambda}$, and
$\Theta\hat{\bm{M}}_{\bm{k}}\Theta^{-1} = -\hat{\bm{M}}_{-\bm{k}}$,
where $\Theta$ denotes the time-reversal operator.
Noting that there are two time-reversal odd operators in Eq.~\eqref{eq:f-dis}, one obtains no condition due to time-reversal invariance. 
Therefore, breaking the spatial inversion symmetry
is sufficient to obtain a finite response;
breaking the TRS is not required for this effect.

The same conclusion is reached more directly from symmetry considerations
of the correlation function~[Eq.~\eqref{eq:M}].
From Eq.~\eqref{eq:f1}, $f^{\lambda(1):{\rm dis}}_{\alpha\beta}$ transforms
as $M_\alpha J_\beta^\lambda$,
where $J_\beta^\lambda$ is odd under both spatial inversion
and time-reversal operations,
while $M_\alpha$ is even under spatial inversion
and odd under time-reversal operations.
Therefore, $f_{\alpha\beta}^{\lambda(1):{\rm dis}}$ can be finite in systems
that break the spatial inversion symmetry, 
regardless of whether the TRS is preserved.
Similarly, from Eq.~\eqref{eq:f2}~[or Eq.~\eqref{appeq:f2}],
$f_{\alpha\beta}^{\lambda(2):{\rm dis}}$ transforms
as $M_\alpha r_\beta \tau^\lambda$, leading to the same conclusion.
There exist $21$ crystallographic point groups
without spatial inversion symmetry.
Group-theoretical analysis indicates that
the FCIM response is allowed in all of them.
Moreover, for each of these 21 point groups,
all components of the magnetization can, in principle, be finite.
That is, for each $\alpha \in (x,y,z)$,
there exist at least one index pair $(\beta,\lambda)$
with nonzero $f_{\alpha\beta}^{\lambda:{\rm dis}}$.
We note that these arguments apply only to dissipative contributions,
for which irreversible relaxation processes
effectively break
the TRS and
relax the constraints on $f_{\alpha\beta}^{\lambda}$ owing to the opposite time-reversal parities of
$\nabla_{\beta}\epsilon_{\lambda}$ and $M_\alpha$.
For the nondissipative contribution,
there is no source that breaks the TRS,
and the symmetry constraint must be analyzed in terms of $M_\alpha \nabla_\beta\epsilon_\lambda$.
For this contribution, both spatial inversion and
TRS must be broken
to produce a finite magnetization response.
Table~\ref{tab:PandT} summarizes the symmetry properties of
$\bm{\nabla}\epsilon$, $\bm{J}^\lambda$, $\tau^\lambda$, and $\bm{M}$
under the spatial inversion and time-reversal operations.

It should be emphasized that FCIM is distinct
from the dynamical multiferroic effect~\cite{juraschek2017_prm,juraschek2019_prm}.
The dynamical multiferroic effect refers to the generation
of a magnetization driven by a time-dependent electric polarization,
defined as $\bm{M} \propto \bm{P} \times  \dot{\bm{P}}$.
The electric polarization responsible for this effect
can arise from various microscopic origins,
such as static displacements in ferroelectric materials
and dynamical ionic motions associated with optical phonons.
In contrast, the FCIM discussed in this paper does not require the polarization.
The magnetization observed in nonmagnetic semiconductor silicon
under strain gradients is attributed to the
dynamical mutiferroic effect~\cite{lou2021}.

\section{Application to nonmagnetic systems}
\label{sec:result}

In this section, we apply our theory for FCIM
to time-reversal symmetric systems without spatial inversion symmetry.
As representative examples, we examine three noncentrosymmetric systems
described by tight-binding models and evaluate their magnetization responses to strain gradients.
The first model is based on a decorated square lattice obtained by slightly displacing the bond-centered sites
of the Lieb lattice along the out-of-plane direction.
The second example is a monolayer transition-metal dichalcogenide~(TMDC), ${\rm MoS_2}$, for which we employ the model constructed in Ref.~\cite{fang2015}.
The third model is a monolayer Janus ${\rm MoSSe}$, which has the same
structural polymorph as the second example, ${\rm MoS_2}$.
In the present study, we consider both macroscopic spin and orbital magnetizations,
denoted by $\bar{S}_\alpha$ and $\bar{L}_\alpha$, respectively,
and evaluate the corresponding response coefficients $f^{\lambda:{\rm dis}}_{\alpha x}$.
The definition of the $x$ axis for each model is shown in Figs.~\ref{fig:lattice_C4v}(a) and~\ref{fig:MoS2}(a).
Throughout Sec.~\ref{sec:result}, we set $\hbar = 1$ and
define electric quadrupole operators to satisfy ${\rm Tr} [\hat{Q}_\lambda \hat{Q}_{\lambda'}] = 2\delta_{\lambda,\lambda'}$.

\subsection{Decorated square lattice}
\label{sec:C4v}

\begin{figure}[t]
\includegraphics[width=0.9\columnwidth]{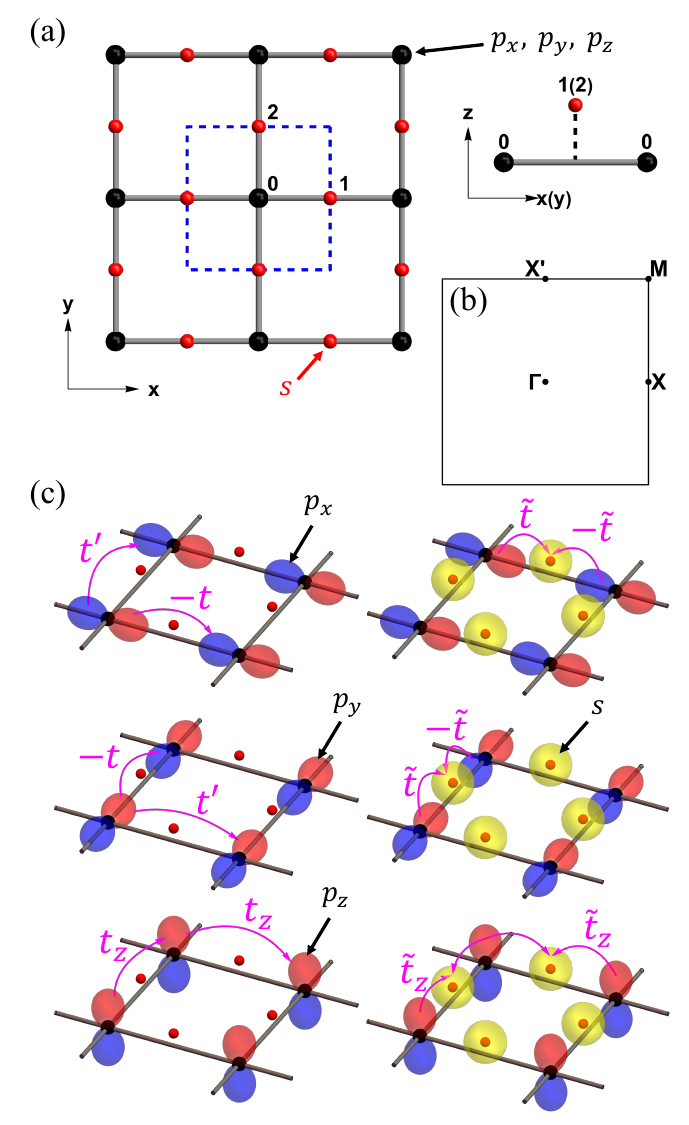}
\caption{
(a) Schematic picture of the decorated square lattice. Black and red circles represent sites hosting $p$- and $s$-orbital degrees of freedom, respectively.
The red sites are slightly elevated along the $z$ direction.
The blue dashed square indicates the unit cell. 
The numbers represent the sublattice indices.
(b) Two-dimensional first Brillouin zone of the decorated square lattice.
(c) Nearest-neighbor hoppings for $p$-$p$ and $p$-$s$ orbitals allowed in the present C$_{4v}$ system.
}
\label{fig:lattice_C4v}
\end{figure}

\begin{figure*}[t]
\includegraphics[width=2\columnwidth]{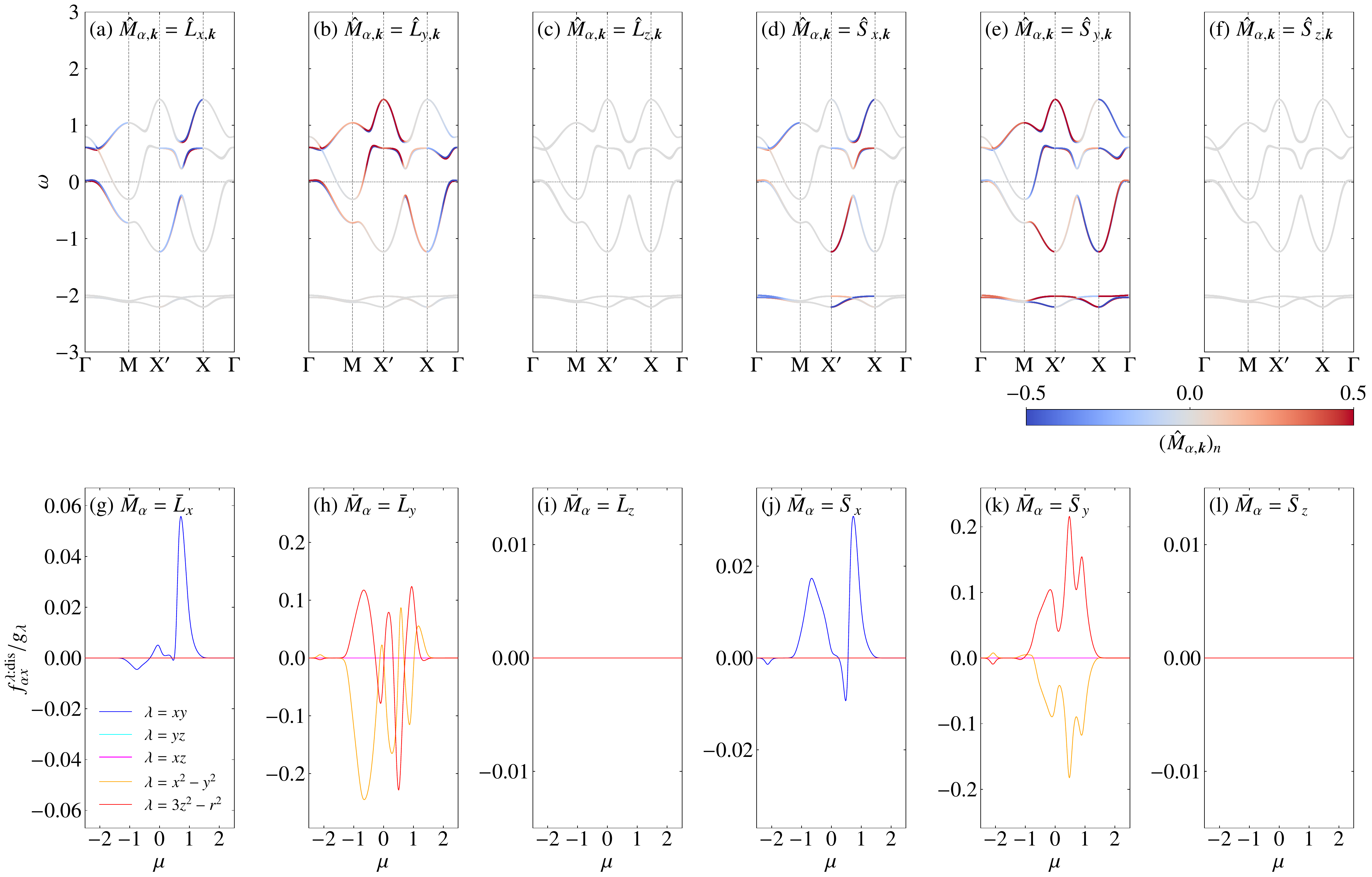}
\caption{
[(a)--(f)]
Band dispersions and color maps of
(a) $(\hat{L}_{x,\bm{k}})_n$, (b) $(\hat{L}_{y,\bm{k}})_n$,
(c) $(\hat{L}_{z,\bm{k}})_n$, (d) $(\hat{S}_{x,\bm{k}})_n$,
(e) $(\hat{S}_{y,\bm{k}})_n$, and
(f) $(\hat{S}_{z,\bm{k}})_n$
at
$\varepsilon_s = -2$, $\varepsilon_{p_x} = \varepsilon_{p_y} = 0$, $\varepsilon_{p_z} = 0.6$,
$t = 0.5$, $t'=0.2$, $t_z = 0.1$, $\tilde{t}_z = 0.1$, $\tilde{t}=0.2$, and $\lambda_{{\rm SO}}=0.4$.
[(g)--(l)]
Chemical potential dependence of $f_{\alpha x}^{\lambda:{\rm dis}}/ g_{\lambda}$ for
(g) $\bar{L}_x$, (h) $\bar{L}_y$, (i) $\bar{L}_z$,
(j) $\bar{S}_x$, (k) $\bar{S}_y$, and (l) $\bar{S}_z$,
with $k_{{\rm B}} T =0.05$ and $\eta = 0.05$.
}
\label{fig:flexo_C4v}
\end{figure*}

First, let us discuss the decorated square system.
We consider a lattice structure composed of three sublattices; $l=0$ (black) on the $xy$ plane,
while $l=1$ (red) and 2 (red) slightly above the $xy$ plane at the center of each bond between the $l=0$ sites, as illustrated in Fig.~\ref{fig:lattice_C4v}(a),
which is called the buckled Lieb lattice~\cite{feng2020}.
In the present study, 
we set the length of the primitive translation vectors to unity
and assume that the sublattices $l=0$ host $p$-orbital, while those for $l=1$ and 2 do $s$-orbitals.
Since the sites for the $l=1$ and 2 sublattices are slightly elevated along the $z$ direction, the system belongs to the C$_{4v}$ point group, where a macroscopic electric polarization along the $z$ direction is allowed. 
The creation and annihilation operators for the $p$-orbital on the sublattice $l=0$ in the $i$th unit cell
are denoted as $p^\dagger_{i,\gamma\sigma}$ and $p_{i,\gamma\sigma}$,
where $\sigma=\uparrow,\downarrow$ and $\gamma = x, y, z$  represent the spin and the orbital indices, respectively.
The corresponding operators for the $s$-orbitals on the sublattices $l=1$ and $2$ are represented as
$s^\dagger_{i,l\sigma}$ and $s_{i,l\sigma}$, respectively.
Thus, the Hamiltonian $\mathcal{H}$ is written as
\begin{align}
	\mathcal{H} &= \mathcal{H}_0 + \mathcal{H}_{\rm SO} + \mathcal{H}_t, \label{eq:Hamil} \\
    \mathcal{H}_0 &= \sum_{i,\sigma} \sum_{\gamma=x,y,z} \varepsilon_{\gamma} p^\dag_{i,\gamma\sigma} p_{i,\gamma\sigma}
	+ \sum_{i,\sigma} \sum_{l=1,2} \varepsilon_{s} s^\dag_{i,l\sigma} s_{i,l\sigma},
	\\
	\mathcal{H}_{\rm SO} &= \lambda_{{\rm SO}}\sum_{i,\sigma\sigma'}\sum_{\gamma\gamma'=x,y,z}
    p^\dag_{i,\gamma\sigma} (\hat{\bm{L}}\cdot \hat{\bm{S}})_{\gamma\sigma,\gamma'\sigma'} p_{i,\gamma'\sigma'},\\
	\mathcal{H}_t &= \sum_{i,\sigma}
    \Big\{t (p_{i,x\sigma}^\dag p_{i+\hat{a}_1, x\sigma} + p_{i,y\sigma}^\dag p_{i+\hat{a}_2, y\sigma} + {\rm H.c.})\notag\\
	&-    t' (p_{i,x\sigma}^\dag p_{i+\hat{a}_2,x\sigma} + p_{i,y\sigma}^\dag p_{i+\hat{a}_1,y\sigma} + {\rm H.c.})\notag\\
	&-    t_z (p_{i,z\sigma}^\dag p_{i+\hat{a}_1,z\sigma} + p_{i,z\sigma}^\dag p_{i+\hat{a}_2,z\sigma} + {\rm H.c.})\notag\\
	&-    \tilde{t}_z [p_{i,z\sigma}^\dag( s_{i,1\sigma} + s_{i,2\sigma} + s_{i-\hat{a}_1,1\sigma} + s_{i-\hat{a}_2,2\sigma})
	+{\rm H.c.}]\notag\\
	&-    \tilde{t} [p_{i,x\sigma}^\dag (s_{i,1\sigma} - s_{i-\hat{a}_1,1\sigma})
	+ p_{i,y\sigma}^\dag (s_{i,2\sigma} - s_{i-\hat{a}_2,2\sigma})
	+ {\rm H.c.}]\Big\}.\label{eq:Hamil-TB}
\end{align}
Here, $i$ labels the unit cell and the $i$th unit cell has four nearest-neighbor
sites $i \pm \hat{a}_{1(2)}$ along the $\pm x~(\pm y)$ direction. 
$\varepsilon_{x}=\varepsilon_{y}$, $\varepsilon_{z}$, and $\varepsilon_s$
are the local energy levels for the $p_{x,y}$-, $p_z$-, and $s$-electrons, respectively.
The hopping parameters $t$ and the spin-orbit coupling~(SOC) $\lambda_{\rm SO}$ are supposed to be positive.
Equation~\eqref{eq:Hamil-TB} includes all nearest-neighbor hoppings between the orbitals $p$-$p$ and $p$-$s$  that are allowed by the C$_{4v}$ symmetry.
Each hopping process in Eq.~\eqref{eq:Hamil-TB} is illustrated in Fig.~\ref{fig:lattice_C4v}(b).
For numerical calculations, the parameters are fixed as
$\varepsilon_s = -2$, $\varepsilon_{p_x} = \varepsilon_{p_y} = 0$, $\varepsilon_{p_z} = 0.6$,
$t = 0.5$, $t'=0.2$, $t_z = 0.1$, $\tilde{t}_z = 0.1$, $\tilde{t}=0.2$, and $\lambda_{{\rm SO}}=0.4$ as representative ones.

Within the $p$-orbital sector, local electric quadrupole moments are  active,
leading to a coupling with the strain~[see Eq.~\eqref{eq:V_continuum}].
We introduce the electric quadrupole operators $Q_{i,\lambda}$ for the $p$-electron at the $i$th unit cell, which are written by
\begin{align}
    Q_{i,x^2-y^2} &= \sum_{\sigma} (p_{i,x\sigma}^\dag p_{i,x\sigma}-p_{i,y\sigma}^\dag p_{i,y\sigma}),\\
    Q_{i,3z^2-r^2} &= \sum_{\sigma}\frac{1}{\sqrt{3}}(2p_{i,z\sigma}^\dag p_{i,z\sigma} -p_{i,x\sigma}^\dag p_{i,x\sigma} - p_{i,y\sigma}^\dag p_{i,y\sigma}),\\
    Q_{i,xy} &= \sum_{\sigma} (p_{i,x\sigma}^\dag p_{i,y\sigma}+{\rm H.c.}),\\
    Q_{i,yz} &= \sum_{\sigma} (p_{i,y\sigma}^\dag p_{i,z\sigma}+{\rm H.c.}),\\
    Q_{i,xz} &= \sum_{\sigma} (p_{i,z\sigma}^\dag p_{i,x\sigma}+{\rm H.c.}).
\end{align}
$Q_{i,\lambda}$ corresponds to $Q_{\lambda}(\bm{r})$ in the continuum systems.
In the absence of strain, the system preserves the TRS, while lacking inversion symmetry.
Therefore, $f^{\lambda:{\rm dis}}_{\alpha\beta}$ can take a finite value.

Before presenting the numerical result for $f_{\alpha x}^{\lambda:{\rm dis}}$,
we first estimate in advance
the components of $f_{\alpha x}^{\lambda:{\rm dis}}$
that are allowed by symmetry under the C$_{4v}$ point group.
As discussed in Sec.~\ref{sec:symmetry},
TRS is irrelevant to the response of $f_{\alpha\beta}^{\lambda:{\rm dis}}$,
and therefore it does not need to be considered.
From the definition in Eq.~\eqref{eq:FCIM-def},
the scalar quantity constructed from
$f_{\alpha\beta}^{\lambda:{\rm dis}}$ and
$\bar{M}_\alpha \nabla_\beta \epsilon_\lambda$
necessarily belongs to the trivial representation
i.e., $A_1$ representation which is invariant in the C$_{4v}$ group.
According to Neumann's principles,
in the absence of any symmetry-breaking order parameters,
$f_{\alpha\beta}^{\lambda:{\rm dis}}$
belongs to the trivial representation.
Thus, $\bar{M}_\alpha \nabla_\beta \epsilon_\lambda$
must also belong to the trivial representation.
From Table~\ref{tab:IR_C4v},
the combinations of $\bar{M}_\alpha \nabla_\beta\epsilon_\lambda$
that transform
according to the $A_1$ representation are
\begin{align}
    &\bar{M}_y \nabla_y \epsilon_{xy} - \bar{M}_x \nabla_x \epsilon_{xy},\notag\\
    &\bar{M}_y \nabla_x \epsilon_{x^2-y^2} + \bar{M}_x \nabla_y \epsilon_{x^2-y^2},\notag\\
    &\bar{M}_y \nabla_x \epsilon_{3z^2-r^2} - \bar{M}_x \nabla_y \epsilon_{3z^2-r^2},\notag\\
    &\bar{M}_z (\nabla_x \epsilon_{yz} - \nabla_y \epsilon_{xz})\notag.
\end{align}
Thus, in the C$_{4v}$ point group,
the nonzero components of $f_{\alpha x}^{\lambda:{\rm dis}}$
are
$f_{xx}^{xy:{\rm dis}}$, $f_{yx}^{x^2-y^2:{\rm dis}}$,
$f_{yx}^{3z^2-r^2:{\rm dis}}$, and $f_{zx}^{yz:{\rm dis}}$.
This conclusion can also be verified
by explicitly applying the symmetry operations of the C$_{4v}$ point group
to $f_{\alpha\beta}^{\lambda:{\rm dis}}$.

\begin{table}[b]
\centering
\setlength{\tabcolsep}{8pt}
\caption{Classification of $\bm{\nabla}\epsilon_\lambda$ and $\bar{\bm{M}}$
according to the irreducible representations of the C$_{4v}$ point group.
}
\begin{ruledtabular}
\begin{tabular}{cl}
\text{Basis} & \text{Irrep.} \\[2pt]
\hline
$\nabla_{x} \epsilon_{xz} + \nabla_y \epsilon_{yz}$ & $A_1$ \\[2pt]
$\nabla_x \epsilon_{yz} - \nabla_y \epsilon_{xz}$ & $A_2$ \\[2pt]
$\nabla_x \epsilon_{xz} - \nabla_y \epsilon_{yz}$ & $B_1$ \\[2pt]
$\nabla_x \epsilon_{yz} + \nabla_y \epsilon_{xz}$ & $B_2$ \\[2pt]
$(\nabla_{y} \epsilon_{xy} , \nabla_x \epsilon_{xy})$ & $E$ \\[2pt]
$(\nabla_{x} \epsilon_{x^2-y^2}, -\nabla_{y} \epsilon_{x^2-y^2})$ & $E$ \\[2pt]
$(\nabla_{x} \epsilon_{3z^2-r^2}, \nabla_{y} \epsilon_{3z^2-r^2})$ & $E$ \\[2pt]
\hline
$\bar{M}_z$ & $A_2$ \\[2pt]
$(\bar{M}_y, -\bar{M}_x)$ & $E$
\end{tabular}
\end{ruledtabular}
\label{tab:IR_C4v}
\end{table}

Figures~\ref{fig:flexo_C4v}(a)–\ref{fig:flexo_C4v}(f) show the electronic band dispersions
along with color maps of the expectation values of the angular momenta
$(\hat{M}_{\alpha,\bm{k}})_n=(\hat{L}_{\alpha,\bm{k}})_n, (\hat{S}_{\alpha,\bm{k}})_n ~(\alpha=x,y,z)$
for each band.
These plots clearly indicate finite expectation values of $\hat{L}_{x,\bm{k}}$, $\hat{L}_{y,\bm{k}}$, $\hat{S}_{x,\bm{k}}$, and $\hat{S}_{y,\bm{k}}$, as shown in
Figs.~\ref{fig:flexo_C4v}(a),~\ref{fig:flexo_C4v}(b),~\ref{fig:flexo_C4v}(d), and~\ref{fig:flexo_C4v}(e).
This behavior can be understood from the symmetry properties of the C$_{4v}$ point group, which allow Rashba-type momentum dependent couplings.
Both $(k_x, k_y)$ and $(\hat{M}_y,-\hat{M}_x)$ form the basis of the $E$ representation, and thus momentum-dependent terms of the form $k_x \hat{M}_y - k_y \hat{M}_x$ are symmetry-allowed. 
Accordingly, the momentum-space Hamiltonian can contain terms such as
$\alpha_{L} (k_x\hat{L}_y - k_y \hat{L}_x) + \alpha_S (k_x\hat{S}_y - k_y \hat{S}_x)$, 
where $\alpha_L$ and $\alpha_S$ denote coupling constants.
The orbital contribution arises from indirect hopping processes 
in which the $p_x$- and $p_y$-orbitals hybridize with the $p_z$-orbital
through the $s$-orbitals at the bond center.
This mechanism does not rely on the SOC and appears when both $\tilde{t}_z$ and $\tilde{t}$ are finite.
An effective $p$-orbital Hamiltonian obtained by the Schriefer-Wolff transformation yields $\alpha_L \sim \tilde{t}_z \tilde{t}$.
The second term corresponds to the Rashba spin-orbit interaction with $\alpha_S \sim  \lambda_{\rm SO}\alpha_{L}$.
Therefore, in systems with nonzero $\tilde{t}_z$, $\tilde{t}$, and $\lambda_{\rm SO}$,
finite expectation values of $\hat{L}_{x,\bm{k}}$, $\hat{L}_{y,\bm{k}}$, $\hat{S}_{x,\bm{k}}$, and $\hat{S}_{y,\bm{k}}$ are naturally expected.
In contrast, the expectation values of $\hat{L}_{z,\bm{k}}$ and $\hat{S}_{z,\bm{k}}$ vanish
for all bands and momenta, as shown in Figs.~\ref{fig:flexo_C4v}(c) and \ref{fig:flexo_C4v}(f).
Since $\hat{M}_z$ belongs to the $A_2$ representation,
the orbital terms such as $k_x k_y (k_x^2-k_y^2) \hat{L}_z$ are symmetry-allowed in principle.
However, in the absence of hopping between $p_x$- and $p_y$-orbitals,
such terms do not appear, and hence $(\hat{L}_{z,\bm{k}})_n$ are zero everywhere in the momentum space.
The same argument applies to $\hat{S}_z$.

Figures~\ref{fig:flexo_C4v}(g)--\ref{fig:flexo_C4v}(l) show the chemical potential dependence of
$f_{\alpha x}^{\lambda:{\rm dis}} / g_{\lambda}$ for $\bar{L}_\alpha$ and $\bar{S}_\alpha$. See Eq.~\eqref{eq:Mbar_rho}.
As shown in
Figs.~\ref{fig:flexo_C4v}(g),~\ref{fig:flexo_C4v}(h),~\ref{fig:flexo_C4v}(j), and~\ref{fig:flexo_C4v}(k),
the responses of $\bar{L}_x$, $\bar{L}_y$, $\bar{S}_x$, and $\bar{S}_y$ 
are finite.
This behavior can be understood from the fact that 
$(\hat{L}_{x,\bm{k}})_n$, $(\hat{L}_{y,\bm{k}})_n$, $(\hat{S}_{x,\bm{k}})_n$, and $(\hat{S}_{y,\bm{k}})_n$ are finite~[see Figs.~\ref{fig:flexo_C4v}(a), \ref{fig:flexo_C4v}(b), \ref{fig:flexo_C4v}(d), and \ref{fig:flexo_C4v}(e)].
Furthermore,  $\bar{L}_x$ and $\bar{S}_x$ are induced by $\nabla_{x} \epsilon_{xy}$,
whereas $\bar{L}_y$ and $\bar{S}_y$ are induced by $\nabla_{x} \epsilon_{x^2-y^2}$ and $\nabla_{x} \epsilon_{3z^2-r^2}$, 
in complete agreement with the symmetry arguments.
In contrast, as shown in Figs.~\ref{fig:flexo_C4v}(i) and \ref{fig:flexo_C4v}(l), the responses of $\bar{L}_z$ and $\bar{S}_z$ are zero for all chemical potentials.
This is consistent with the fact that $(\hat{L}_{z,\bm{k}})_n$ and $(\hat{S}_{z,\bm{k}})_n$ are zero for all bands and momenta~[see Figs.~\ref{fig:flexo_C4v}(c) and~\ref{fig:flexo_C4v}(f)].

\subsection{Monolayer ${\rm MoS_2}$}
\label{sec:MoS2}

\begin{figure}[t]
\includegraphics[width=1\columnwidth]{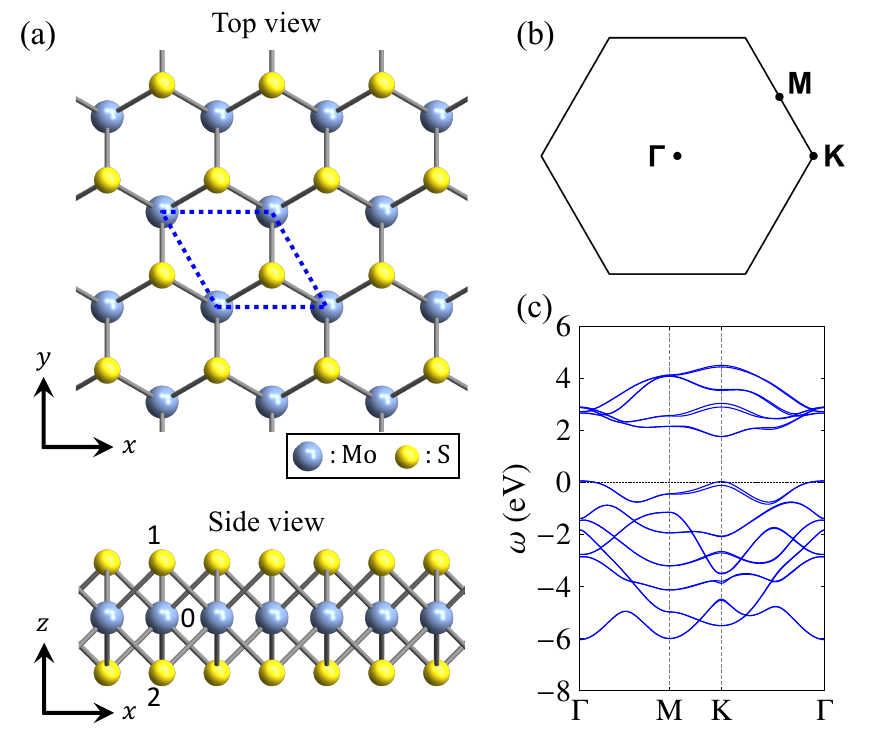}
\caption{
(a) Top and side views of monolayer ${\rm MoS_2}$.
The blue dashed rhombus denotes the unit cell.
The numbers in the side view represent the sublattice indices.
(b) First Brillouin zone of monolayer ${\rm MoS_2}$.
(c) Band structure of monolayer ${\rm MoS_2}$ calculated using the tight-binding model in Ref.~\cite{fang2015}.
}
\label{fig:MoS2}
\end{figure}

\begin{table}[b]
\centering
\setlength{\tabcolsep}{8pt}
\caption{Classification of $\bm{\nabla}\epsilon_\lambda$ and $\bar{\bm{M}}$
according to the irreducible representations of the D$_{3h}$ point group.
Note that the inner product in $E' \otimes E''$ gives the $A_2''$ representation
in the present arrangement in this table.
}
\begin{ruledtabular}
\begin{tabular}{cl}
\text{Basis} & \text{Irrep.} \\[2pt]
\hline
$2\nabla_x \epsilon_{xy} + \nabla_y \epsilon_{x^2-y^2}$ & $A_1'$\\[2pt]
$\nabla_x \epsilon_{x^2-y^2} - 2 \nabla_{y}\epsilon_{xy}$ & $A_{2}'$\\[2pt]
$( \nabla_x \epsilon_{x^2-y^2}+2\nabla_y\epsilon_{xy}, 2\nabla_{x}\epsilon_{xy} - \nabla_y\epsilon_{x^2-y^2} )$ & $E'$\\[2pt]
$( \nabla_x \epsilon_{3z^2-r^2}, \nabla_{y}\epsilon_{3z^2-r^2} )$ & $E'$\\[2pt]
$\nabla_x\epsilon_{yz} - \nabla_y \epsilon_{xz}$ & $A_1''$\\[2pt]
$\nabla_x \epsilon_{xz}+\nabla_y\epsilon_{yz}$ & $A_2''$\\[2pt]
$(\nabla_x\epsilon_{yz}+\nabla_y\epsilon_{xz}, \nabla_x\epsilon_{xz}-\nabla_y\epsilon_{yz})$ & $E''$\\[2pt]
\hline
$\bar{M}_z$ & $A_2'$\\[2pt]
$(\bar{M}_y, -\bar{M}_x)$ & $E''$
\end{tabular}
\end{ruledtabular}
\label{tab:IR_D3h}
\end{table}

\begin{figure*}[t]
\includegraphics[width=2\columnwidth]{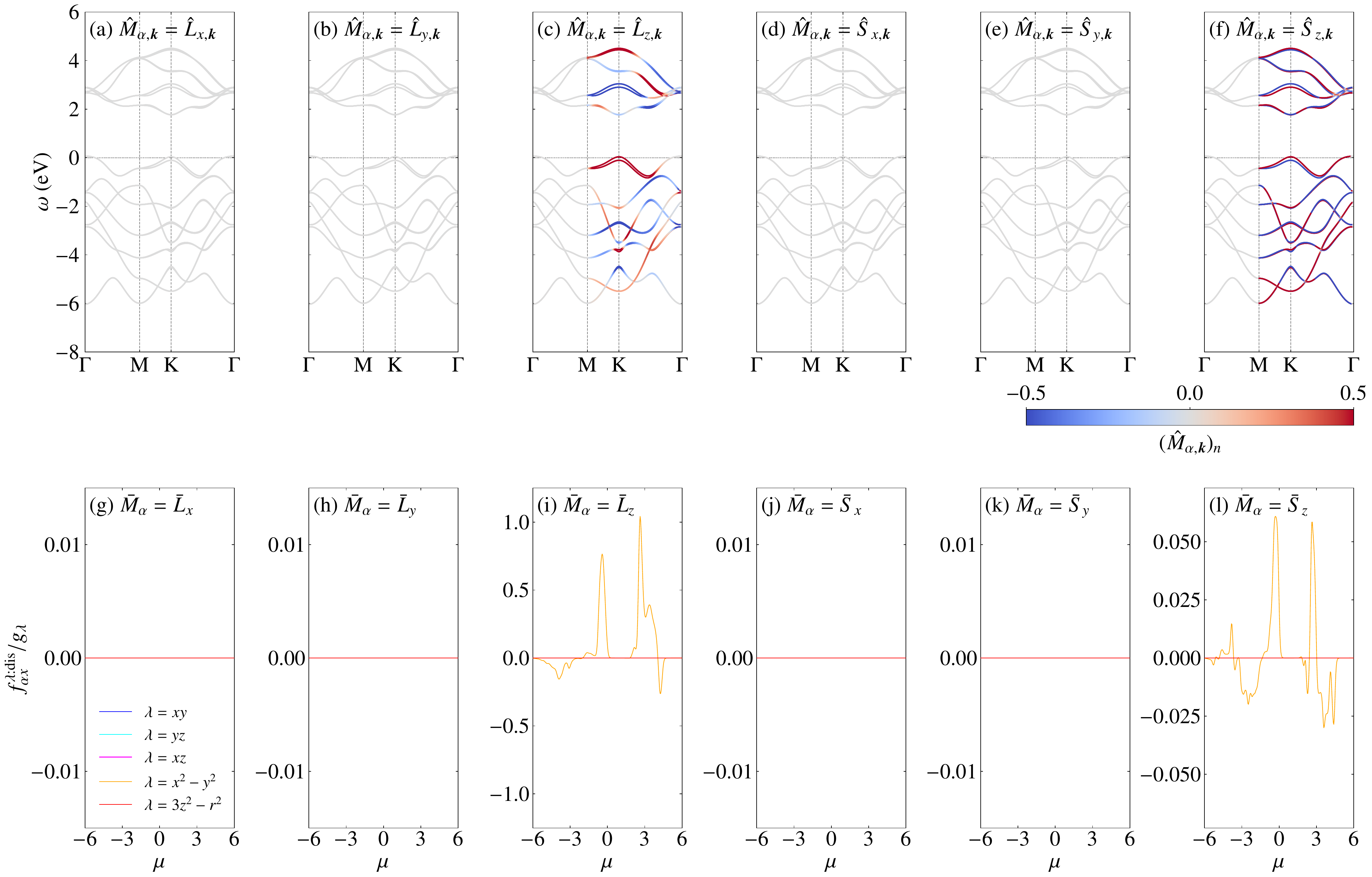}
\caption{
[(a)--(f)]
Band dispersions and color maps of
(a) $(\hat{L}_{x,\bm{k}})_n$, (b) $(\hat{L}_{y,\bm{k}})_n$,
(c) $(\hat{L}_{z,\bm{k}})_n$, (d) $(\hat{S}_{x,\bm{k}})_n$,
(e) $(\hat{S}_{y,\bm{k}})_n$, and
(f) $(\hat{S}_{z,\bm{k}})_n$
for monolayer ${\rm MoS_2}$.
[(g)--(l)]
Chemical potential dependence of $f_{\alpha x}^{\lambda:{\rm dis}}/ g_{\lambda}$ for
(g) $\bar{L}_x$, (h) $\bar{L}_y$, (i) $\bar{L}_z$,
(j) $\bar{S}_x$, (k) $\bar{S}_y$, and (l) $\bar{S}_z$,
with $k_{{\rm B}} T =0.05$ and $\eta = 0.05$.
The length of primitive translation vectors
is set to $3.18~\AA$ in (g)--(l)~\cite{fang2015}.}
\label{fig:flexo_D3h}
\end{figure*}

Next, we evaluate the magnetization response to strain gradients
using a tight-binding model for a monolayer ${\rm MoS_2}$.
The monolayer ${\rm MoS_2}$ exhibits several structural polymorphs,
such as $1H$, $1T$, and $1T'$~\cite{samy2021MoS2,zhang2024MoS2}.
In this study, we focus on the $1H$ structure with the D$_{3h}$ point group as illustrated in Fig.~\ref{fig:MoS2}(a).
This structure consists of ABA stacking, where the top and bottom chalcogen atoms, labeled as sublattices $1$ and $2$,
are projected onto the same in-plane position in the top view.
The $1H$-${\rm MoS_2}$ is a nonmagnetic direct-gap semiconductor with the band gap located
at the ${\rm K}$ and ${\rm K}'$ points in the Brillouin zone.
Following Ref.~\cite{fang2015},
we set the length of the primitive translation vectors to $3.18 \ \AA$ 
and construct the tight-binding model
including $d$-orbitals on the ${\rm Mo}$ sites
and the $p$-orbitals on the ${\rm S}$ sites.
This setup effectively captures all orbitals relevant to the Fermi level
in monolayer ${\rm MoS_2}$.
We have confirmed that the calculated band structure
and the magnitude of spin splitting are consistent with Ref.~\cite{fang2015}.
The band dispersion of ${\rm MoS_2}$ with SOC is shown in Fig.~\ref{fig:MoS2}(c).
The electric quadrupole degrees of freedom are active in
both the $p$- and $d$-orbitals
and are coupled with an external strain field.
For simplicity, we consider only the electric quadrupoles constructed from
the $d$-orbitals at the ${\rm Mo}$ sites, since the local nature of the $d$ electrons is more important in the discussion of the strain effects than that of the $p$ electrons at the ${\rm S}$ sites. The explicit forms of the electric quadrupole moments are given in Appendix~\ref{app:quadrupole}.
Since the D$_{3h}$ point group lacks spatial inversion symmetry, nonzero values of $f_{\alpha x}^{\lambda:{\rm dis}}$
are expected.
We note that CIM is symmetry-forbidden in D$_{3h}$ point group.

Before presenting the numerical results,
we determine, on symmetry grounds, which 
 components of $f_{\alpha x}^{\lambda:{\rm dis}}$
are finite in the D$_{3h}$ point group.
As discussed in Sec.~\ref{sec:C4v},
if TRS is not included
as a symmetry operation,
the scalar quantity composed of $f_{\alpha\beta}^{\lambda:{\rm dis}} \bar{M}_\alpha \nabla_\beta \epsilon_{\lambda}$
belongs to
the trivial representation.
Since monolayer ${\rm MoS_2}$ does not exhibit
any spontaneous order,
$f_{\alpha\beta}^{\lambda:{\rm dis}}$
belongs to the trivial representation.
Consequently, the combination of
$\bar{M}_\alpha \nabla_\beta\epsilon_\lambda$
must also form a basis of the trivial representation.
From Table~\ref{tab:IR_D3h},
the combinations of $\bar{M}_\alpha \nabla_\beta\epsilon_\lambda$
that form the basis of the $A_{1}'$ representation are
\begin{align}
    &\bar{M}_y (\nabla_x\epsilon_{yz} + \nabla_y \epsilon_{xz})
    -\bar{M}_x (\nabla_x\epsilon_{xz}-\nabla_y\epsilon_{yz}),\notag\\
    &\bar{M}_z (\nabla_x \epsilon_{x^2-y^2} - 2\nabla_y \epsilon_{xy}).
\end{align}
Therefore,
$f_{xx}^{xz:{\rm dis}} = -f_{yx}^{yz:{\rm dis}}$
and
$f_{zx}^{x^2-y^2:{\rm dis}}$ are finite, while the others vanish.

Figures~\ref{fig:flexo_D3h}(a)--\ref{fig:flexo_D3h}(f) show the color maps of $(\hat{M}_{\alpha,\bm{k}})_n$ for each band.
In Figs.~\ref{fig:flexo_D3h}(a), \ref{fig:flexo_D3h}(b), \ref{fig:flexo_D3h}(d) and \ref{fig:flexo_D3h}(e),
one finds that $(\hat{L}_{x,\bm{k}})_n$, $(\hat{L}_{y,\bm{k}})_n$, $(\hat{S}_{x,\bm{k}})_n$, and $(\hat{S}_{y,\bm{k}})_n$
are zero for all bands and momenta.
This behavior can be understood from the symmetry-allowed couplings
between the wave vector and the magnetization under the D$_{3h}$ point group.
The lowest-order coupling term involving $\hat{M}_x$($\hat{M}_y$) takes the following form:
$k_z[2k_xk_y \hat{M}_y - (k_x^2-k_y^2) \hat{M}_x]$.
Since there is no $k_z$ component in two-dimensional systems, the expectation values of
$\hat{L}_x$, $\hat{L}_y$, $\hat{S}_x$, and $\hat{S}_y$ are zero.
In contrast, Figs.~\ref{fig:flexo_D3h}(c) and \ref{fig:flexo_D3h}(f) show
that $(\hat{L}_{z,\bm{k}})_n$ and $(\hat{S}_{z,\bm{k}})_n$ are nonzero.
This can be understood from the fact that $\bm{k}$ and $\hat{M}_z$ can couple
under the D$_{3h}$ symmetry in the following form:
$k_x (k_x^2 - 3 k_y^2) \hat{M}_z$.
It is also confirmed that both $(\hat{L}_{z,\bm{k}})_n$ and $(\hat{S}_{z,\bm{k}})_n$
vanish along the $\mathrm{\Gamma}$-$\mathrm{M}$ path,
which is consistent with the relation $k_x^2-3k_y^2 = 0$ along this line
and with the results of Ref.~\cite{liu2013}.
Figures~\ref{fig:flexo_D3h}(g)--\ref{fig:flexo_D3h}(l) show the chemical potential dependence of
$f_{\alpha x}^{\lambda:{\rm dis}}/g_\lambda$ for $\bar{L}_{\alpha}$ and $\bar{S}_{\alpha}$.
From Figs.~\ref{fig:flexo_D3h}(g), \ref{fig:flexo_D3h}(h), \ref{fig:flexo_D3h}(j), and \ref{fig:flexo_D3h}(k),
we find that the responses of $\bar{M}_\alpha = \bar{L}_x$, $\bar{L}_y$, $\bar{S}_x$, and $\bar{S}_y$ vanish,
which can be easily understood from the fact that $(\hat{M}_{\alpha,\bm{k}})_n$
is identically zero for all bands and momenta.
For $\bar{M}_\alpha = \bar{L}_z$ and $\bar{S}_z$,
the responses induced by $\nabla_x\epsilon_{x^2-y^2}$ are finite,
as shown in Figs.~\ref{fig:flexo_D3h}(h) and \ref{fig:flexo_D3h}(l),
in full agreement with the symmetry arguments. 
In addition, the magnitude of $f_{\alpha x}^{\lambda:{\rm dis}} / g_\lambda$
is strongly enhanced near $\mu \sim 0$.
This enhancement originates from the large value of $(\hat{M}_{z,\bm{k}})_n$
around the $\mathrm{M}$ and $\mathrm{K}$ points near
the valence band edge~[see Figs.~\ref{fig:flexo_D3h}(c) and \ref{fig:flexo_D3h}(d)].
Therefore,
it is possible to observe a finite FCIM for
$\nabla_x \epsilon_{xx}$ or $\nabla_{x}\epsilon_{yy}$ via hole doping
in monolayer ${\rm MoS_2}$
or metalic TMDCs such as ${\rm NbS_2}$~\cite{yan2019,chen2025}.

\subsection{Monolayer ${\rm MoSSe}$}
\label{sec:MoSSe}

\begin{figure}[t]
\includegraphics[width=1\columnwidth]{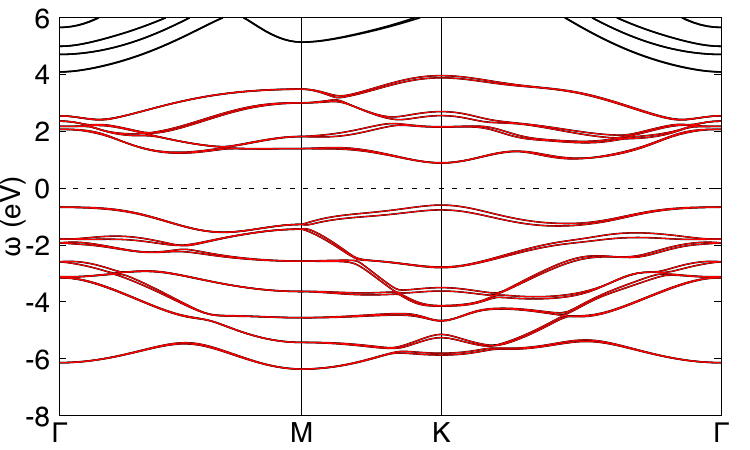}
\caption{
Band structure of monolayer Janus ${\rm MoSSe}$ obtained from DFT calculation.
The red curves indicate the bands for the effective tight-binding
model.
}
\label{fig:MoSSe}
\end{figure}

\begin{figure*}[t]
\includegraphics[width=2\columnwidth]{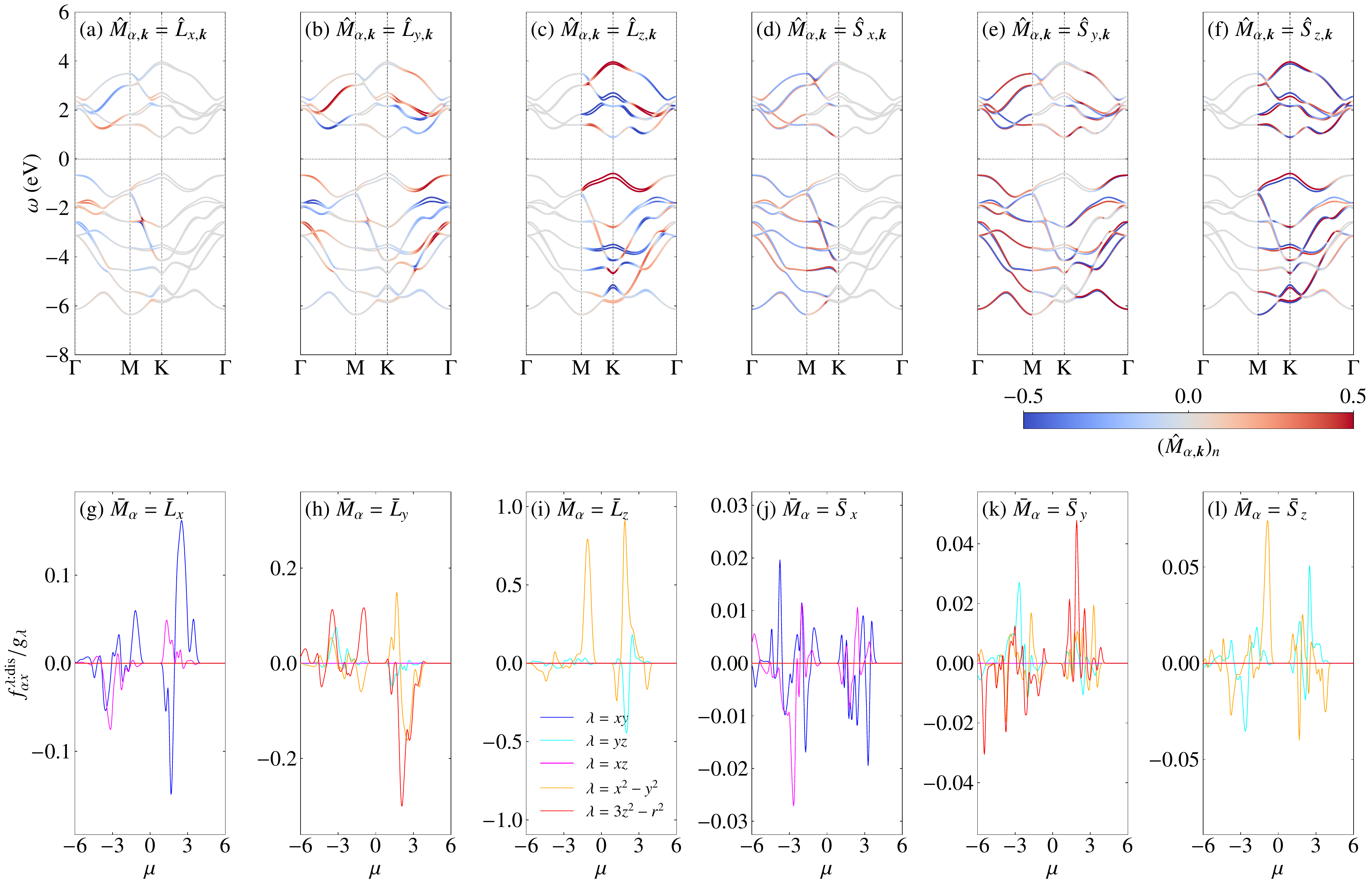}
\caption{
[(a)--(f)]
Band dispersions and color maps of
(a) $(\hat{L}_{x,\bm{k}})_n$, (b) $(\hat{L}_{y,\bm{k}})_n$,
(c) $(\hat{L}_{z,\bm{k}})_n$, (d) $(\hat{S}_{x,\bm{k}})_n$,
(e) $(\hat{S}_{y,\bm{k}})_n$, and
(f) $(\hat{S}_{z,\bm{k}})_n$
for monolayer Janus ${\rm MoSSe}$.
[(g)--(l)]
Chemical potential dependence of $f_{\alpha x}^{\lambda:{\rm dis}}/ g_{\lambda}$ for
(g) $\bar{L}_x$, (h) $\bar{L}_y$, (i) $\bar{L}_z$,
(j) $\bar{S}_x$, (k) $\bar{S}_y$, and (l) $\bar{S}_z$,
with $k_{{\rm B}} T =0.05$ and $\eta = 0.05$.
The length of primitive translation vectors
is set to $3.251~\AA$ in (g)--(l).
}
\label{fig:flexo_C3v}
\end{figure*}

Finally, we present the FCIM results
for a monolayer Janus ${\rm MoSSe}$~\cite{yin2021janus,li2018janus}.
The monolayer Janus ${\rm MoSSe}$ is
a member of the TMDC family and
can be obtained from ${\rm MoS_2}$ by
substituting the ${\rm S}$ atoms on
the top or bottom chalcogen layer with ${\rm Se}$~\cite{lu2017}.
In this paper, we focus on the $1H$-${\rm MoSSe}$ structure
obtained by replacing the top-layer ${\rm S}$ with ${\rm Se}$
in the monolayer ${\rm MoS_2}$ shown in Fig.~\ref{fig:MoS2}.
The $1H$-${\rm MoSSe}$ is, similarly to $1H$-${\rm MoS_2}$,
a nonmagnetic semiconductor with a direct band gap
at the ${\rm K}$ and ${\rm K}'$ points.
In contrast, the substitution of the top-layer
${\rm S}$ by ${\rm Se}$ breaks the mirror symmetry with respect to
the horizontal plane, lowering the point group
from D$_{3h}$ to C$_{3v}$.
To determine the band structure of ${\rm MoSSe}$,
we perform density functional theory~(DFT) calculations
using Quantum ESPRESSO~\cite{giannozzi_2017}.
We first relax the crystal structure while keeping the interlayer Mo-Mo distance at 20 \AA,
and obtain an in-plane lattice constant of $3.251~\AA$
for the length of the primitive translation vector.
The resulting band structure is shown in Fig.~\ref{fig:MoSSe}.
Focusing on the ${\rm K}$ point,
we find a band gap of $1.48$ eV, and estimate
that the spin splittings of the uppermost valence and
the lowermost conduction band are $0.17$ eV and $0.013$ eV, respectively,
in good agreement with the previous study~\cite{cheng2013}.
As in Sec.~\ref{sec:MoS2},
to describe the bands near the Fermi level
we construct an effective tight-binding model
in a basis consisting of the $d$-orbitals on the ${\rm Mo}$ sites
and the $p$-orbitals on the ${\rm S}$ and ${\rm Se}$ sites
using Wannier$90$~\cite{wannier90,pizzi_2020}. 
The resulting effective Hamiltonian is found to
accurately reproduce the DFT bands highlighted by the
red curves in Fig.~\ref{fig:MoSSe}.
Although electric quadrupole degrees of freedom are active
in the $d$-orbitals on the ${\rm Mo}$ sites
and the $p$-orbitals on the ${\rm S}$ and ${\rm Se}$ sites,
we consider only the electric quadrupole degrees of freedom
originating from the $d$-orbitals, as in monolayer ${\rm MoS_2}$~(Sec.~\ref{sec:MoS2}),
and calculate $f_{\alpha x}^{\lambda:{\rm dis}}$.

Before showing the numerical results,
we determine in advance the components of $f_{\alpha x}^{\lambda:{\rm dis}}$
that can be nonzero under the C$_{3v}$ symmetry.
Since C$_{3v}$ is a subgroup of D$_{3h}$,
all responses allowed under D$_{3h}$ are also allowed 
under C$_{3v}$.
The important difference is that a finite electric dipole moment is
allowed along the $z$ direction in the C$_{3v}$ point group symmetry.
Consequently,
the responses $f_{\alpha\beta}^{\lambda:{\rm dis}}$ that belong to the $A_{2}''$
representation under D$_{3h}$ are also allowed.
This can be understood from the fact that
an electric dipole along the $z$ direction belongs to the $A_{2}''$
representation in D$_{3h}$.
Thus, from Table~\ref{tab:IR_D3h}, such $A_{2}''$ components in $\bar{M}_\alpha \nabla_\beta \epsilon_\lambda$ are  
\begin{align}
    &\bar{M}_y (\nabla_x \epsilon_{x^2-y^2} + 2 \nabla_y \epsilon_{xy})
    -
    \bar{M}_x (2\nabla_x\epsilon_{xy} - \nabla_y \epsilon_{x^2-y^2}),\notag\\
    &\bar{M}_y \nabla_x \epsilon_{3z^2-r^2}
    -
    \bar{M}_x \nabla_y \epsilon_{3z^2-r^2},\notag\\
    &\bar{M}_z (\nabla_x \epsilon_{yz} - \nabla_y \epsilon_{xz}).
\end{align}
As a result,
under the C$_{3v}$ symmetry,
the nonvanishing components of $f_{\alpha x}^{\lambda:{\rm dis}}$
include, in addition to those already allowed under the D$_{3h}$ symmetry,
the additional components 
$f_{xx}^{xy:{\rm dis}} = - 2f_{y x}^{x^2-y^2:{\rm dis}}$,
$f_{yx}^{3z^2-r^2:{\rm dis}}$, and
$f_{zx}^{yz:{\rm dis}}$.

Figures~\ref{fig:flexo_C3v}(a)--\ref{fig:flexo_C3v}(f)
show the color maps of $(\hat{M}_{\alpha,\bm{k}})_n$
for each band.
Focusing first on $(\hat{L}_{z,\bm{k}})_n$ and $(\hat{S}_{z,\bm{k}})_n$,
we find that the color maps are similar to those of the monolayer ${\rm MoS_2}$
[see Figs.~\ref{fig:flexo_D3h}(c) and~\ref{fig:flexo_D3h}(f)].
This can be attributed to the fact that
the coupling between the in-plane momentum
and $\hat{M}_z$ allowed under the C$_{3v}$
is identical to that under the D$_{3h}$.
In the C$_{3v}$ symmetry,
coupling terms between $\hat{M}_\alpha$ and $\bm{k}$ composed of
odd powers of the in-plane momentum $k_{x,y}$
and $\hat{M}_z$ are, in principle,
symmetry-allowed.
These terms transform as the $A_{2}''$ representation of D$_{3h}$. 
Since $\hat{M}_z$ belongs to $A_{2}'$,
such terms can emerge only if
the odd-momentum polynomial in $k_x$ and $k_y$
transforms as $A_{1}''$.
However, because the $A_{1}''$ representation
changes sign under mirror operation
with respect to the horizontal plane,
it cannot be constructed solely
from $k_x$ and $k_y$.
Consequently,
the symmetry-allowed couplings between $\hat{M}_z$ and the in-plane momentum
in C$_{3v}$ are exactly the same as those in D$_{3h}$.
Therefore, the color map structures of $(\hat{L}_{z,\bm{k}})_n$ and $(\hat{S}_{z,\bm{k}})_n$
in monolayer ${\rm MoSSe}$ closely resemble those in monolayer ${\rm MoS_2}$.
In contrast, as can be seen 
in Figs.~\ref{fig:flexo_C3v}(a),~\ref{fig:flexo_C3v}(b),
\ref{fig:flexo_C3v}(d), and~\ref{fig:flexo_C3v}(e),
the in-plane components of the magnetization expectations
are nonzero in monolayer ${\rm MoSSe}$ unlike ${\rm MoS_2}$.
Since $(\hat{M}_y, -\hat{M}_x)$ belongs to the $E''$ representation,
it can be coupled to a basis constructed from odd powers
of $(k_x, k_y)$ that belong to $E'$.
At the lowest order,
$(k_x,k_y)$ belongs to $E'$,
and thus a Rashba-type term $k_x \hat{M}_y - k_y \hat{M}_x$
is allowed under the C$_{3v}$ symmetry.
As a result,
the in-plane magnetization expectation values
can be finite even in monolayer TMDCs.
Indeed, Figs.~\ref{fig:flexo_C3v}(a) and~\ref{fig:flexo_C3v}(d)
show that along the K–$\mathrm{\Gamma}$ path~($k_y=0$)
the expectation values of
$(\hat{L}_{x,\bm{k}})_n$ and $(\hat{S}_{x,\bm{k}})_n$ disappear,
consistently capturing the characteristic of a Rashba system.

Figures~\ref{fig:flexo_C3v}(g)--\ref{fig:flexo_C3v}(l)
present the chemical potential dependence of
$f_{\alpha x}^{\lambda:{\rm dis}}/g_\lambda$.
As shown in Figs.~\ref{fig:flexo_C3v}(i) and~\ref{fig:flexo_C3v}(l),
the magnetization response to $\nabla_x\epsilon_{x^2-y^2}$
is nonzero, and its behavior is similar to that of ${\rm MoS_2}$.
This can be understood from the fact
that the color-map patterns of the out-of-plane magnetization expectation values 
are identical in
Figs.~\ref{fig:flexo_D3h}(f) and~\ref{fig:flexo_C3v}(f).
In contrast to ${\rm MoS_2}$, the response to $\nabla_x\epsilon_{yz}$ is also finite,
which is in agreement with the $C_{3v}$ symmetry analysis.
For the in-plane magnetization,
Figs.~\ref{fig:flexo_C3v}(g),~\ref{fig:flexo_C3v}(h),~\ref{fig:flexo_C3v}(j), and~\ref{fig:flexo_C3v}(k)
show that $\bar{M}_x$ becomes finite in response to $\nabla_x\epsilon_{xy}$
and $\nabla_x\epsilon_{xz}$,
and
$\bar{M}_y$ becomes finite in response to
$\nabla_x\epsilon_{x^2-y^2}$, $\nabla_x\epsilon_{yz}$, and $\nabla_x\epsilon_{3z^2-r^2}$.
These results are in full agreement with
the symmetry-allowed responses derived above.
Moreover,
the symmetry argument predicts the relations
$f_{xx}^{xz:{\rm dis}}=-f_{yx}^{yz:{\rm dis}}$ and
$f_{xx}^{xy:{\rm dis}}=-2f_{yx}^{x^2-y^2:{\rm dis}}$.
We have confirmed that
these relations are satisfied separately for $\bar{\bm{L}}$ and $\bar{\bm{S}}$.

\section{Discussion}
\label{sec:discussion}

In this section,
we discuss the experimental detectability of FCIM.
We also remark on the limitations of our formulation
and possible directions for future extensions.

\subsection{Experimental detectability of FCIM}

In time-reversal symmetric metals and semiconductors,
FCIM can be experimentally detectable
because this effect is the only magnetization response induced by a strain field.
When a strain field is applied,
the observed magnetization contains contributions arising from
the strain and the strain gradient.
Among these, the equilibrium magnetization vanishes
in systems with TRS.
This can be understood from the fact that
strain~(and its gradient) and magnetization transform differently
under time-reversal operation~(see Table~\ref{tab:PandT}).
On the other hand,
nonequilibrium magnetization can be induced,
which is given by
\begin{align}
    \label{eq:mag-total}
    \bar{M}_\alpha = P^\lambda_\alpha \epsilon_\lambda
    + f^{\lambda:{\rm dis}}_{\alpha\beta} \nabla_\beta \epsilon_{\lambda},
\end{align}
where $P^\lambda_\alpha = \partial\bar{M}_\alpha/\partial\epsilon_\lambda$
originates from the first term in Eq.~\eqref{appeq:M2-2}.
As discussed in Appendix~\ref{app:expression of f2},
the dissipative part of $P_\alpha^\lambda$ vanishes irrespective of symmetry,
and moreover, its nondissipative part also disappears in time-reversal symmetric systems.
Therefore, the only magnetization response allowed in systems with TRS
is the FCIM for slowly varying strain fields. We expect that 
nuclear magnetic resonance or magneto-optical Kerr effect 
can detect the magnetization induced by strain fields.

\subsection{Additional contributions to FCIM}

In this study, we have incorporated only the local strain–electric quadrupole coupling.
This choice enables a transparent formulation and
allows us to elucidate the essential structure of the response in a controlled manner.
In realistic crystal systems, however,
strain additionally modifies internal microscopic parameters
such as hopping amplitudes~\cite{ogawa2023,uchino2025_arxiv}
and the $g$-tensor~\cite{wilson1961}.
These effects generate effective strain-induced terms
in the Hamiltonian.
Through these strain--induced modulations of microscopic parameters,
the electric quadrupole operators can acquire additional contributions
and become nonlocal.
These additional terms are expected to generate additional
contributions to the FCIM, and taking them into account
would provide important quantitative corrections to the present results.
Since our formulation remains valid even
when $Q_\lambda(\bm{r})$ is nonlocal,
one can evaluate such additional contributions by directly using Eq.~\eqref{eq:f-dis}.

Furthermore, in our numerical calculations,
we have considered only the atomic orbital and spin contributions to the magnetization.
However,
the orbital magnetic moment of the Bloch electron
$U^\dagger_{\bm{k}}\hat{\bm{L}}^{{\rm orb}}U_{\bm{k}}
= -\frac{e}{4}U_{\bm{k}}^\dagger(\hat{\bm{r}}\times \hat{\bm{p}} - \hat{\bm{p}}\times\hat{\bm{r}})U_{\bm{k}}$
also constitutes an additional source
of magnetization~\cite{thonhauser2005,xiao2005,shi2007,xiao2010,fuchs2010,osumi2021,johansson2024,pezo2022}.
Following Refs.~\cite{bhowal2021,pezo2022},
its expectation value is given by
\begin{align}
    (\hat{L}^{{\rm orb}}_{\alpha,\bm{k}})_n
    &=
    (U^\dagger_{\bm{k}}\hat{L}^{{\rm orb}}_{\alpha}U_{\bm{k}})_{nn}\notag\\
    &=-\epsilon_{\alpha\beta\gamma}\frac{ie}{2\hbar}
    \left[ \pdiff{U_{\bm{k}}^\dagger}{k_\beta} (\hat{H}_{\bm{k}} - \varepsilon_{\bm{k},n}) \pdiff{U_{\bm{k}}}{k_\gamma}\right]_{nn},
    \label{eq:Lorb-momentum}
\end{align}
where $\epsilon_{\alpha\beta\gamma}$ is the Levi--Civita symbol.
This is a geometric quantity that shares the same transformation properties
as the Berry curvature~\cite{xiao2010,fuchs2010}.
Consequently, this magnetization vanishes identically in systems
that possess spatial inversion and time-reversal symmetries.
In the systems considered in this study,
spatial inversion symmetry is absent
and therefore Eq.~\eqref{eq:Lorb-momentum} can generally take a finite value.
A rigorous evaluation of FCIM
should incorporate not only the local orbital and spin magnetization,
but also the contribution described by Eq.~\eqref{eq:Lorb-momentum}.

\subsection{Extension to insulating systems}

We have formulated FCIM in electronic systems
and confirmed that a finite response emerges in metals.
However,
since the electric quadrupole can also be expressed
in terms of localized spins~\cite{Lauchli2006,tsunetsugu2006} or
lattice displacements~\cite{martin1972,baroni2001},
FCIM can also arise in insulating systems
through bosonic quasiparticles as elementary excitations~(e.g., chiral phonons, magnons, and triplons).
In particular, some insulating systems that host bosonic quasiparticles
are nonmagnetic~\cite{ueda2023,ohe2024,kageyama1999}, for which
piezomagnetic and flexomagnetic effects are 
forbidden by TRS.
Therefore, in these systems,
FCIM is expected to
be the only mechanism responsible for
magnetization responses induced by strain.
Extending the present formalism to insulating systems
and quantitatively evaluating the resulting FCIM
remain important subjects for future study.

\section{Summary}
\label{sec:summary}

In summary, we have developed a general formalism
for the magnetization induced by strain gradients.
Starting from free-fermionic systems,
we introduced the coupling between
strain and electric quadrupoles as an
external perturbation and
formulated the magnetization response within Kubo's linear response theory
under a spatially varying strain.
The resulting response tensor is determined solely by the diagonal components
of the electric quadrupole, and this expression 
is consistent with the one derived from semiclassical theory.
Importantly, the FCIM is symmetry-allowed
in nonmagnetic systems that preserve TRS,
in close analogy with the CIM.
A group-theoretical analysis further indicates that
FCIM is finite in all $21$ noncentrosymmetric point groups
in clear contrast to CIM.
We have applied the present FCIM formalism
to three nonmagnetic tight-binding models:
a buckled Lieb lattice,
a monolayer ${\rm MoS_2}$,
and a monolayer Janus ${\rm MoSSe}$.
These systems lack spatial inversion symmetry,
and finite magnetization responses emerge as a consequence
of spin-momentum locking,
in full agreement with the symmetry-based predictions.
In particular, for ${\rm MoS_2}$ and ${\rm MoSSe}$,
the response is strongly enhanced near the band edge,
indicating that this effect should be experimentally
observable.

\begin{acknowledgments}
  The authors thank S. Hayami and Y. Miyata for fruitful discussions.
  This work was supported by Grant-in-Aid for Scientific Research from
  JSPS, KAKENHI Grant Nos.~JP23H04866 and JP23H04869.
\end{acknowledgments}


\appendix

\section{Expression of $f^{\lambda(2)}_{\alpha\beta}$}
\label{app:expression of f2}

In this section, we show the derivation of Eq~\eqref{eq:f2}  in detail.
We denote the second term of Eq.~\eqref{eq:M} as $\bar{M}_\alpha^{(2)}$, which is given by
\begin{widetext}
\begin{align}
    \bar{M}_\alpha^{(2)} = 
    -\frac{1}{V} \sum_{\lambda}^{{\rm Irrep.}} g_\lambda
    \int dt e^{-\delta t} \int_0^{\frac{1}{k_{\rm B} T}} d\beta' \int d\bm{r} \int d\bm{r}'
    \braket{\tau^\lambda (\bm{r},-i\hbar\beta') M_\alpha(\bm{r}',t)}_{{\rm eq}} \epsilon_\lambda(\bm{r}).
    \label{appeq:M2}
\end{align}
We assume that the response at $\bm{r}'$ is primarily influenced by the strain gradient in its immediate vicinity and
expand $\epsilon_\lambda(\bm{r})$ around $\bm{r}'$, corresponding to a long-wavelength expansion.
Then, $\epsilon_\lambda(\bm{r})$ can be written as
\begin{align}
    \epsilon_\lambda (\bm{r}) = \epsilon_\lambda(\bm{r}') + \bm{\nabla} \epsilon_\lambda(\bm{r})\Big|_{\bm{r}=\bm{r}'} \cdot (\bm{r} - \bm{r}') + \cdots.
\end{align}
Substituting this expansion into Eq.~\eqref{appeq:M2} yields
\begin{align}
    \bar{M}_\alpha^{(2)}
    =
    &-\frac{1}{V} \sum_{\lambda}^{{\rm Irrep.}} g_\lambda
    \int dt e^{-\delta t} \int_0^{\frac{1}{k_{\rm B} T}} d\beta' \int d\bm{r} \int d\bm{r}'
    \braket{\tau^\lambda (\bm{r},-i\hbar\beta') M_\alpha(\bm{r}',t)}_{{\rm eq}} \epsilon_\lambda(\bm{r}')\notag\\
    &-\frac{1}{V} \sum_{\lambda}^{{\rm Irrep.}} g_\lambda
    \int dt e^{-\delta t} \int_0^{\frac{1}{k_{\rm B} T}} d\beta'
    \left[\pdiff{}{i\bm{q}}\braket{\tau^\lambda_{-\bm{q}}(-i\hbar\beta') M_{\alpha,\bm{q}}(t)}_{{\rm eq}}\Big|_{\bm{q} = 0}\right]
    \cdot \bm{\nabla} \epsilon_\lambda,
    \label{appeq:M2-2}
\end{align}
where we have used
\begin{align}
    \int d\bm{r}\int d\bm{r}'
    \braket{\tau^\lambda (\bm{r},-i\hbar\beta') M_\alpha(\bm{r}',t)}_{{\rm eq}} (\bm{r}-\bm{r}')
    =
    \pdiff{}{i\bm{q}}\braket{\tau^\lambda_{-\bm{q}}(-i\hbar\beta') M_{\alpha,\bm{q}}(t)}_{{\rm eq}}\Big|_{\bm{q} = 0}.
\end{align}
By defining
$f^{\lambda(2)}_{\alpha\beta} = \partial \bar{M}_\alpha^{(2)} / \partial(\nabla_\beta \epsilon_\lambda)$,
we obtain Eq.~\eqref{eq:f2}.

We comment on the first term in Eq.~\eqref{appeq:M2-2}.
This term represents
the magnetization induced by the strain,
indicating that the strain itself can generate a finite magnetization.
$P_{\alpha}^\lambda=\partial \bar{M}_\alpha / \partial \epsilon_\lambda$
in Eq.~\eqref{eq:mag-total}
corresponds to the first term in Eq.~\eqref{appeq:M2-2}.
$P_\alpha^\lambda$ contains both dissipative and nondissipative contributions.
Evaluating the dissipative part following Appendix~\ref{app:derivation of FCIM},
we find that it depends on $(\hat{\tau}^\lambda_{\bm{k}})_n$.
Since $(\hat{\tau}^\lambda_{\bm{k}})_n = 0$ for all bands,
the dissipative contribution vanishes.
In constrast, the nondissipative contribution depends on
$(\hat{\tau}^\lambda_{\bm{k}})_{nm}$ with $n\neq m$
and is in general nonzero.
However, in systems with TRS,
this contribution vanishes.
This can be understood from the fact
that the nondissipative part transform as ${M}_\alpha \epsilon_\lambda$.
According to Table~\ref{tab:PandT},
$M_\alpha \epsilon_\lambda$ is
odd under time-reversal operation.
Therefore, $P_\alpha^\lambda$ vanishes in time-reversal symmetric systems.

\section{Derivation of $f^{\lambda:{\rm dis}}_{\alpha\beta}$}
\label{app:derivation of FCIM}

In this section, we derive the explicit expressions for
$f^{\lambda:{\rm dis}}_{\alpha\beta} = f^{\lambda(1):{\rm dis}}_{\alpha\beta} + f^{\lambda(2):{\rm dis}}_{\alpha\beta}$.

First, evaluating $f^{\lambda(1)}_{\alpha\beta}$ in Eq.~\eqref{eq:f1}, we find
\begin{align}
  f^{\lambda(1)}_{\alpha\beta}
  &=
  \frac{i \hbar g_\lambda}{V}
  \sum_{\bm{k}} \sum_{nm=1}^{N}
  \frac{f(\varepsilon_{\bm{k},n}) - f(\varepsilon_{\bm{k},m})}
  {\varepsilon_{\bm{k},n} - \varepsilon_{\bm{k},m}}
  \frac{(\hat{M}_{\alpha,\bm{k}})_{nm} (\hat{J}^\lambda_{\beta,\bm{k}})_{mn}}
  {\varepsilon_{\bm{k},n} - \varepsilon_{\bm{k},m}+i\delta}.
\end{align}
To take into account nonmagnetic impurity scattering at low temperatures
phenomenologically,
we introduce a small but finite positive $\eta$ and replace $i\delta$ with $i\eta$.
Using the relation ${\rm Im}\frac{1}{x+i\eta} = - \frac{\eta}{x^2 + \eta^2}$,
the dissipative part of $f^{\lambda(1)}_{\alpha\beta}$ is expressed as
\begin{align}
  f^{\lambda(1):{\rm dis}}_{\alpha\beta}
  &=
  \frac{\hbar g_\lambda}{V} \sum_{\bm{k}} \sum_{nm=1}^{N}
  \frac{f(\varepsilon_{\bm{k},n}) - f(\varepsilon_{\bm{k},m})}{\varepsilon_{\bm{k},n} - \varepsilon_{\bm{k},m}}
  (\hat{M}_{\alpha,\bm{k}})_{nm} (\hat{J}^\lambda_{\beta,\bm{k}})_{mn}
  \frac{\eta}{(\varepsilon_{\bm{k},n} - \varepsilon_{\bm{k},m})^2 + \eta^2}\notag\\
  &=
  \frac{\hbar g_\lambda}{V\eta}
  \sum_{\bm{k}} \sum_{n=1}^{N}
  (\hat{M}_{\alpha,\bm{k}})_{n} (\hat{J}^\lambda_{\beta,\bm{k}})_{n}
  \pdiff{f(\varepsilon_{\bm{k},n})}{\varepsilon_{\bm{k},n}},
  \label{appeq:f1-dis}
\end{align}
where we use
$\frac{f(\varepsilon_{\bm{k},n}) - f(\varepsilon_{\bm{k},m})}{\varepsilon_{\bm{k},n} - \varepsilon_{\bm{k},m}} \to \pdiff{f(\varepsilon_{\bm{k},n})}{\varepsilon_{\bm{k},n}}$ for $\varepsilon_{\bm{k},m} \sim \varepsilon_{\bm{k},n}$.
Strictly speaking, Eq.~\eqref{appeq:f1-dis} also contains
a finite interband contribution for small but finite $\eta$.
We have kept only the intraband contribution
that diverges as $1/\eta$ in the limit $\eta \to 0$
and neglected the interband contribution,
which vanishes as $O(\eta)$.
This expression coincides with Eq.~\eqref{eq:f1-dis}.

Next, we evaluate $f^{\lambda(2)}_{\alpha\beta}$ given in Eq.~\eqref{eq:f2}, which can be written as follows: 
\begin{align}
  f^{\lambda(2)}_{\alpha\beta}
  &=
  - \frac{g_\lambda}{V} \int dt e^{-\delta t} \int_0^{\frac{1}{k_{\rm B} T}} ds
  \left[\pdiff{}{iq_\beta}\braket{\tau^\lambda_{-\bm{q}}(-i\hbar s)
  M_{\alpha,\bm{q}}(t)}_{{\rm eq}}\Big|_{\bm{q} = 0}\right]\notag\\
  &=
  -\frac{g_\lambda}{V}
    \int dt e^{-\delta t} \int_0^{\frac{1}{k_{\rm B} T}} ds \int d\bm{r} \int d\bm{r}'
    \left[
    r_\beta\braket{\tau^\lambda(\bm{r},-i\hbar s) M_{\alpha}(\bm{r}',t)}_{{\rm eq}}
    -
    r'_\beta\braket{\tau^\lambda(\bm{r},-i\hbar s)M_{\alpha}(\bm{r}',t)}_{{\rm eq}}
    \right]\notag\\
  &=
  -\frac{g_\lambda}{V}
    \int dt e^{-\delta t} \int_0^{\frac{1}{k_{\rm B} T}} ds
    \left[
    \braket{D^{\tau^{\lambda}}_{\beta}(-i\hbar s) M_{\alpha,\bm{q}=0}(t)}_{{\rm eq}}
    -
    \braket{\tau^{\lambda}_{-\bm{q}=0}(-i\hbar s)D^{M_\alpha}_{\beta}(t)}_{{\rm eq}}
    \right] \label{appeq:f2},
\end{align}
where we introduce
$\bm{D}^{\tau^{\lambda}} =\int  \bm{r} \tau^\lambda(\bm{r}) d\bm{r}$
and
$\bm{D}^{M_{\alpha}} =\int \bm{r}' M_\alpha(\bm{r}') d\bm{r}'$.
To facilitate the calculation in momentum space,
it is useful to rewrite these operators in the following forms.
Using Eqs.~\eqref{eq:M(r)} and~\eqref{eq:T(r)}, we obtain
\begin{align}
    \bm{D}^{\tau^{\lambda}}
    &=
    \frac{1}{2}\int 
    \left\{
    [\hat{\bm{r}}\Psi(\bm{r})]^\dagger \hat{\tau}^{\lambda} \Psi(\bm{r})
    + [\hat{\tau}^{\lambda}\Psi(\bm{r})]^\dagger \hat{\bm{r}}\Psi(\bm{r})
    \right\}d\bm{r}\notag\\
    &=
    \frac{1}{2}\int \Psi^\dagger(\bm{r})(\hat{\bm{r}}\hat{\tau}^{\lambda}
    + \hat{\tau}^{\lambda}\hat{\bm{r}}) \Psi(\bm{r})d\bm{r},\\
    \bm{D}^{M_{\alpha}} &=
    \frac{1}{2}\int 
    \left\{
    [\hat{\bm{r}}\Psi(\bm{r}')]^\dagger \hat{M}_{\alpha} \Psi(\bm{r}')
    + [\hat{M}_{\alpha}\Psi(\bm{r}')]^\dagger \hat{\bm{r}}\Psi(\bm{r}')
    \right\}d\bm{r}'\notag\\
    &=
    \frac{1}{2}\int \Psi^\dagger(\bm{r}') (\hat{\bm{r}}\hat{M}_{\alpha}
    + \hat{M}_{\alpha}\hat{\bm{r}}) \Psi(\bm{r}')d\bm{r}',
\end{align}
where we have used $\hat{\tau}^\lambda = (\hat{\tau}^{\lambda})^\dagger$, $\hat{M}_\alpha = \hat{M}_\alpha^\dagger$,
and
$\Psi^\dagger(\bm{r}') \hat{M}_\alpha \Psi(\bm{r}') = [\hat{M}_\alpha \Psi(\bm{r}')]^\dagger \Psi(\bm{r}')$.
Applying Eq.~\eqref{eq:Fourier} to Eq.~\eqref{appeq:f2} and employing the Bloch basis, we find
\begin{align}
  f^{\lambda(2)}_{\alpha\beta}
  &=
  \frac{i \hbar g_\lambda}{2V} \sum_{\bm{k}} \sum_{nm=1}^{N}
  \frac{f(\varepsilon_{\bm{k},n}) - f(\varepsilon_{\bm{k},m})}{\varepsilon_{\bm{k},n} - \varepsilon_{\bm{k},m}} \frac{1}{\varepsilon_{\bm{k},n} - \varepsilon_{\bm{k},m} + i\delta}
  \left[ (\hat{M}_{\alpha,\bm{k}})_{nm} (\hat{r}_{\beta} \hat{\tau}^{\lambda}_{\bm{k}} + \hat{\tau}^\lambda_{\bm{k}}\hat{r}_{\beta})_{mn}
  - (\hat{r}_{\beta} \hat{M}_{\alpha,\bm{k}} + \hat{M}_{\alpha,\bm{k}}\hat{r}_{\beta})_{nm} (\hat{\tau}^{\lambda}_{\bm{k}})_{mn} \right]\notag\\
  &=
  \frac{i \hbar g_\lambda}{2V} \sum_{\bm{k}} \sum_{nm=1}^{N}
  \frac{f(\varepsilon_{\bm{k},n}) - f(\varepsilon_{\bm{k},m})}{\varepsilon_{\bm{k},n} - \varepsilon_{\bm{k},m}} \frac{1}{\varepsilon_{\bm{k},n} - \varepsilon_{\bm{k},m} + i\delta}
  \Bigg\{
    (\hat{M}_{\alpha,\bm{k}})_{nm} \left[\sum_{l (\neq m)}^{N}(\hat{r}_{\beta})_{ml} (\hat{\tau}^{\lambda}_{\bm{k}})_{ln} + \sum_{l(\neq n)}^{N} (\hat{\tau}^\lambda_{\bm{k}})_{ml} (\hat{r}_{\beta})_{ln} \right]
    \notag\\
  &\qquad\qquad\qquad -
    \left[\sum_{l(\neq n)}^{N} (\hat{r}_{\beta})_{nl} (\hat{M}_{\alpha,\bm{k}})_{lm} + \sum_{l(\neq m)}^{N} (\hat{M}_{\alpha,\bm{k}})_{nl}(\hat{r}_{\beta})_{lm} \right] (\hat{\tau}^{\lambda}_{\bm{k}})_{mn}
  \Bigg\},
\end{align}
where the contribution from the diagonal elements of $\hat{r}_{\beta}$ cancels out.
Using $(\hat{r}_{\beta})_{nm} = (A^\beta_{\bm{k}})_{nm}$ for $n \neq m$~\cite{sipe2000,parker2019}, where $(A^\beta_{\bm{k}})_{nm}$
is the interband Berry connection defined as
\begin{align}
  (A^\beta_{\bm{k}})_{nm} = i \left(U^\dagger_{\bm{k}} \pdiff{U_{\bm{k}}}{k_{\beta}}\right)_{nm},
\end{align}
$f^{\lambda(2)}_{\alpha\beta}$ can be expressed as
\begin{align}
  f^{\lambda(2)}_{\alpha\beta}
  &=
  \frac{i \hbar g_\lambda}{2V} \sum_{\bm{k}} \sum_{nm=1}^{N}
  \frac{f(\varepsilon_{\bm{k},n}) - f(\varepsilon_{\bm{k},m})}{\varepsilon_{\bm{k},n} - \varepsilon_{\bm{k},m}} \frac{1}{\varepsilon_{\bm{k},n} - \varepsilon_{\bm{k},m} + i\delta}
  \Bigg\{
    (\hat{M}_{\alpha,\bm{k}})_{nm} \left[\sum_{l (\neq m)}^{N}(A^{\beta}_{\bm{k}})_{ml} (\hat{\tau}^{\lambda}_{\bm{k}})_{ln} + \sum_{l(\neq n)}^{N} (\hat{\tau}^\lambda_{\bm{k}})_{ml} (A^{\beta}_{\bm{k}})_{ln} \right]
    \notag\\
  &\qquad\qquad\qquad -
    \left[\sum_{l(\neq n)}^{N} (A^{\beta}_{\bm{k}})_{nl} (\hat{M}_{\alpha,\bm{k}})_{lm} + \sum_{l(\neq m)}^{N} (\hat{M}_{\alpha,\bm{k}})_{nl}(A^{\beta}_{\bm{k}})_{lm} \right] (\hat{\tau}^{\lambda}_{\bm{k}})_{mn}
  \Bigg\}\notag\\
  &=
  \frac{i \hbar g_\lambda}{2V} \sum_{\bm{k}} \sum_{nm=1}^{N}
  \frac{f(\varepsilon_{\bm{k},n}) - f(\varepsilon_{\bm{k},m})}{\varepsilon_{\bm{k},n} - \varepsilon_{\bm{k},m}} \frac{1}{\varepsilon_{\bm{k},n} - \varepsilon_{\bm{k},m} + i\delta}
  \Bigg\{
    (\hat{M}_{\alpha,\bm{k}})_{nm} \left[\sum_{l (\neq m)}^{N} \frac{\varepsilon_{\bm{k},l} - \varepsilon_{\bm{k},n}}{\varepsilon_{\bm{k},m}-\varepsilon_{\bm{k},l}}(\hat{v}_{\beta,\bm{k}})_{ml} (\hat{Q}_{\lambda,\bm{k}})_{ln}
    + \sum_{l(\neq n)}^{N} \frac{\varepsilon_{\bm{k},m} - \varepsilon_{\bm{k},l}}{\varepsilon_{\bm{k},l} - \varepsilon_{\bm{k},n}} (\hat{Q}_{\lambda,\bm{k}})_{ml} (\hat{v}_{\beta,\bm{k}})_{ln} \right]
    \notag\\
  &\qquad\qquad\qquad -
    \left[\sum_{l(\neq n)}^{N} \frac{\varepsilon_{\bm{k},m} - \varepsilon_{\bm{k},n}}{\varepsilon_{\bm{k},n} - \varepsilon_{\bm{k},l}} (\hat{v}_{\beta,\bm{k}})_{nl} (\hat{M}_{\alpha,\bm{k}})_{lm}
    + \sum_{l(\neq m)}^{N} \frac{\varepsilon_{\bm{k},m} - \varepsilon_{\bm{k},n}}{\varepsilon_{\bm{k},l} - \varepsilon_{\bm{k},m}} (\hat{M}_{\alpha,\bm{k}})_{nl}(\hat{v}_{\beta,\bm{k}})_{lm} \right] (\hat{Q}_{\lambda,\bm{k}})_{mn}
  \Bigg\}.
\end{align}
Here, we have used the following relations:
\begin{align}
  (A^\beta_{\bm{k}})_{nm} &= - i\hbar 
  \frac{(\hat{v}_{\beta,\bm{k}})_{nm}}
  {\varepsilon_{\bm{k},n} - \varepsilon_{\bm{k},m}},  \quad (n \neq m),\\
  (\hat{\tau}_{\bm{k}})_{nm} &= \frac{i}{\hbar}
  (\varepsilon_{\bm{k},n} - \varepsilon_{\bm{k},m}) (\hat{Q}_{\lambda,\bm{k}})_{nm}.
\end{align}
Similarly to the derivation of $f^{\lambda(1):{\rm dis}}_{\alpha\beta}$,
the dissipative part of $f^{\lambda(2)}_{\alpha\beta}$ is obtained as
\begin{align}
  f^{\lambda(2):{\rm dis}}_{\alpha\beta}
  &=
  \frac{\hbar g_\lambda}{2V\eta} \sum_{\bm{k}} \sum_{n=1}^{N}\sum_{l(\neq n)}^{N}
  (\hat{M}_{\alpha,\bm{k}})_{n} \left[ - (\hat{v}_{\beta,\bm{k}})_{nl} (\hat{Q}_{\lambda,\bm{k}})_{ln} - (\hat{Q}_{\lambda,\bm{k}})_{nl} (\hat{v}_{\beta,\bm{k}})_{ln} \right]
  \pdiff{f(\varepsilon_{\bm{k},n})}{\varepsilon_{\bm{k},n}}
  \notag\\
  &=
  -\frac{\hbar g_\lambda}{V\eta} \sum_{\bm{k}} \sum_{n=1}^{N}\sum_{l(\neq n)}^{N}
  (\hat{M}_{\alpha,\bm{k}})_{n} {\rm Re} \left[(\hat{v}_{\beta,\bm{k}})_{nl} (\hat{Q}_{\lambda,\bm{k}})_{ln} \right]
  \pdiff{f(\varepsilon_{\bm{k},n})}{\varepsilon_{\bm{k},n}}.
\end{align}
This expression corresponds to Eq.~\eqref{eq:f2-dis}.

\section{Electric quadrupole operators in monolayer ${\rm MoS_2}$}
\label{app:quadrupole}

We list the expressions of electric quadrupole operators for the $d$ electrons in ${\rm MoS_2}$.
We denote by $d_{i,\lambda\sigma}^\dagger$ and $d_{i,\lambda\sigma}$ the creation and annihilation operators, respectively,
for the $d$-orbitals on the ${\rm Mo}$ atom in unit cell $i$.
The corresponding quadrupole operators in unit cell $i$ are given by
\begin{align}
    Q_{i,x^2-y^2} &= \frac{1}{\sqrt{7}}\sum_{\sigma}
    \left[
    -\big(2d_{i, 3z^2-r^2\sigma}^\dagger d_{i, x^2-y^2\sigma} + \mathrm{H.c.}\big)
    - \sqrt{3}d_{i,yz\sigma}^\dagger d_{i,yz\sigma}
    + \sqrt{3}d_{i,xz\sigma}^\dagger d_{i,xz\sigma}
    \right],
    \\[4pt]
    Q_{i,3z^2-r^2} &= \frac{1}{\sqrt{7}} \sum_{\sigma}
    \left(
      2 d_{i,3z^2-r^2\sigma}^\dagger d_{i,3z^2-r^2\sigma}
      - 2 d_{i,x^2-y^2\sigma}^\dagger d_{u,x^2-y^2\sigma}
      + d_{i,yz\sigma}^\dagger d_{i,yz\sigma} 
      + d_{i,xz\sigma}^\dagger d_{i,xz\sigma} - 2 d_{i,xy\sigma}^\dagger d_{i,xy\sigma} 
    \right),\\[4pt]
    Q_{i,xy} &= \frac{1}{\sqrt{7}}\sum_{\sigma}
    \left(
      -2 d_{i,3z^2-r^2\sigma}^\dagger d_{i,xy\sigma} + \sqrt{3} d_{i,yz\sigma}^\dagger d_{i,xz\sigma}
      + \mathrm{H.c.}
    \right),\\[4pt]
    Q_{i,yz} &= \frac{1}{\sqrt{7}} \sum_{\sigma}
    \left(
      d_{i,3z^2-r^2\sigma}^\dagger d_{i,yz\sigma} - \sqrt{3} d_{i,x^2-y^2\sigma}^\dagger d_{i,yz\sigma}
      + \sqrt{3} d_{i,xz\sigma}^\dagger d_{i,xy\sigma}
      + \mathrm{H.c.}
    \right),\\[4pt]
    Q_{i,xz} &= \frac{1}{\sqrt{7}} \sum_{\sigma}
    \left(
      d_{i,3z^2-r^2\sigma}^\dagger d_{i,xz\sigma} + \sqrt{3} d_{i,x^2-y^2\sigma}^\dagger d_{i,xz\sigma}
      + \sqrt{3} d_{i,yz\sigma}^\dagger d_{i,xy\sigma}
      + \mathrm{H.c.}
    \right).
\end{align}

\end{widetext}

\bibliography{refs.bib}

\begin{thebibliography}{96}%
\makeatletter
\providecommand \@ifxundefined [1]{%
 \@ifx{#1\undefined}
}%
\providecommand \@ifnum [1]{%
 \ifnum #1\expandafter \@firstoftwo
 \else \expandafter \@secondoftwo
 \fi
}%
\providecommand \@ifx [1]{%
 \ifx #1\expandafter \@firstoftwo
 \else \expandafter \@secondoftwo
 \fi
}%
\providecommand \natexlab [1]{#1}%
\providecommand \enquote  [1]{``#1''}%
\providecommand \bibnamefont  [1]{#1}%
\providecommand \bibfnamefont [1]{#1}%
\providecommand \citenamefont [1]{#1}%
\providecommand \href@noop [0]{\@secondoftwo}%
\providecommand \href [0]{\begingroup \@sanitize@url \@href}%
\providecommand \@href[1]{\@@startlink{#1}\@@href}%
\providecommand \@@href[1]{\endgroup#1\@@endlink}%
\providecommand \@sanitize@url [0]{\catcode `\\12\catcode `\$12\catcode `\&12\catcode `\#12\catcode `\^12\catcode `\_12\catcode `\%12\relax}%
\providecommand \@@startlink[1]{}%
\providecommand \@@endlink[0]{}%
\providecommand \url  [0]{\begingroup\@sanitize@url \@url }%
\providecommand \@url [1]{\endgroup\@href {#1}{\urlprefix }}%
\providecommand \urlprefix  [0]{URL }%
\providecommand \Eprint [0]{\href }%
\providecommand \doibase [0]{https://doi.org/}%
\providecommand \selectlanguage [0]{\@gobble}%
\providecommand \bibinfo  [0]{\@secondoftwo}%
\providecommand \bibfield  [0]{\@secondoftwo}%
\providecommand \translation [1]{[#1]}%
\providecommand \BibitemOpen [0]{}%
\providecommand \bibitemStop [0]{}%
\providecommand \bibitemNoStop [0]{.\EOS\space}%
\providecommand \EOS [0]{\spacefactor3000\relax}%
\providecommand \BibitemShut  [1]{\csname bibitem#1\endcsname}%
\let\auto@bib@innerbib\@empty
\bibitem [{\citenamefont {Spaldin}\ and\ \citenamefont {Fiebig}(2005)}]{spaldin2005}%
  \BibitemOpen
  \bibfield  {author} {\bibinfo {author} {\bibfnamefont {N.~A.}\ \bibnamefont {Spaldin}}\ and\ \bibinfo {author} {\bibfnamefont {M.}~\bibnamefont {Fiebig}},\ }\bibfield  {title} {\bibinfo {title} {{The Renaissance of Magnetoelectric Multiferroics}},\ }\href {https://doi.org/10.1126/science.1113357} {\bibfield  {journal} {\bibinfo  {journal} {Science}\ }\textbf {\bibinfo {volume} {309}},\ \bibinfo {pages} {391} (\bibinfo {year} {2005})}\BibitemShut {NoStop}%
\bibitem [{\citenamefont {Folen}\ \emph {et~al.}(1961)\citenamefont {Folen}, \citenamefont {Rado},\ and\ \citenamefont {Stalder}}]{folen1961}%
  \BibitemOpen
  \bibfield  {author} {\bibinfo {author} {\bibfnamefont {V.~J.}\ \bibnamefont {Folen}}, \bibinfo {author} {\bibfnamefont {G.~T.}\ \bibnamefont {Rado}},\ and\ \bibinfo {author} {\bibfnamefont {E.~W.}\ \bibnamefont {Stalder}},\ }\bibfield  {title} {\bibinfo {title} {{Anisotropy of the Magnetoelectric Effect in ${\mathrm{Cr}}_{2}$${\mathrm{O}}_{3}$}},\ }\href {https://doi.org/10.1103/PhysRevLett.6.607} {\bibfield  {journal} {\bibinfo  {journal} {Phys. Rev. Lett.}\ }\textbf {\bibinfo {volume} {6}},\ \bibinfo {pages} {607} (\bibinfo {year} {1961})}\BibitemShut {NoStop}%
\bibitem [{\citenamefont {Cheong}\ and\ \citenamefont {Mostovoy}(2007)}]{cheong2007}%
  \BibitemOpen
  \bibfield  {author} {\bibinfo {author} {\bibfnamefont {S.-W.}\ \bibnamefont {Cheong}}\ and\ \bibinfo {author} {\bibfnamefont {M.}~\bibnamefont {Mostovoy}},\ }\bibfield  {title} {\bibinfo {title} {{Multiferroics: a magnetic twist for ferroelectricity}},\ }\href {https://doi.org/10.1038/nmat1804} {\bibfield  {journal} {\bibinfo  {journal} {Nat. Mater.}\ }\textbf {\bibinfo {volume} {6}},\ \bibinfo {pages} {13} (\bibinfo {year} {2007})}\BibitemShut {NoStop}%
\bibitem [{\citenamefont {Tokura}\ \emph {et~al.}(2014)\citenamefont {Tokura}, \citenamefont {Seki},\ and\ \citenamefont {Nagaosa}}]{tokura2014}%
  \BibitemOpen
  \bibfield  {author} {\bibinfo {author} {\bibfnamefont {Y.}~\bibnamefont {Tokura}}, \bibinfo {author} {\bibfnamefont {S.}~\bibnamefont {Seki}},\ and\ \bibinfo {author} {\bibfnamefont {N.}~\bibnamefont {Nagaosa}},\ }\bibfield  {title} {\bibinfo {title} {{Multiferroics of spin origin}},\ }\href {https://doi.org/10.1088/0034-4885/77/7/076501} {\bibfield  {journal} {\bibinfo  {journal} {Rep. Prog. Phys.}\ }\textbf {\bibinfo {volume} {77}},\ \bibinfo {pages} {076501} (\bibinfo {year} {2014})}\BibitemShut {NoStop}%
\bibitem [{\citenamefont {Eerenstein}\ \emph {et~al.}(2006)\citenamefont {Eerenstein}, \citenamefont {Mathur},\ and\ \citenamefont {Scott}}]{eerenstein2006}%
  \BibitemOpen
  \bibfield  {author} {\bibinfo {author} {\bibfnamefont {W.}~\bibnamefont {Eerenstein}}, \bibinfo {author} {\bibfnamefont {N.}~\bibnamefont {Mathur}},\ and\ \bibinfo {author} {\bibfnamefont {J.}~\bibnamefont {Scott}},\ }\bibfield  {title} {\bibinfo {title} {{Multiferroic and magnetoelectric materials}},\ }\href {https://doi.org/10.1038/nature05023} {\bibfield  {journal} {\bibinfo  {journal} {Nature}\ }\textbf {\bibinfo {volume} {442}},\ \bibinfo {pages} {759} (\bibinfo {year} {2006})}\BibitemShut {NoStop}%
\bibitem [{\citenamefont {Khomskii}(2009)}]{khomskii2009}%
  \BibitemOpen
  \bibfield  {author} {\bibinfo {author} {\bibfnamefont {D.}~\bibnamefont {Khomskii}},\ }\bibfield  {title} {\bibinfo {title} {{Classifying multiferroics: Mechanisms and effects}},\ }\href {https://physics.aps.org/articles/v2/20} {\bibfield  {journal} {\bibinfo  {journal} {Physics}\ }\textbf {\bibinfo {volume} {2}},\ \bibinfo {pages} {20} (\bibinfo {year} {2009})}\BibitemShut {NoStop}%
\bibitem [{\citenamefont {Spaldin}\ \emph {et~al.}(2008)\citenamefont {Spaldin}, \citenamefont {Fiebig},\ and\ \citenamefont {Mostovoy}}]{spaldin2008}%
  \BibitemOpen
  \bibfield  {author} {\bibinfo {author} {\bibfnamefont {N.~A.}\ \bibnamefont {Spaldin}}, \bibinfo {author} {\bibfnamefont {M.}~\bibnamefont {Fiebig}},\ and\ \bibinfo {author} {\bibfnamefont {M.}~\bibnamefont {Mostovoy}},\ }\bibfield  {title} {\bibinfo {title} {{The toroidal moment in condensed-matter physics and its relation to the magnetoelectric effect*}},\ }\href {https://doi.org/10.1088/0953-8984/20/43/434203} {\bibfield  {journal} {\bibinfo  {journal} {J. Phys.: Condens. Matter}\ }\textbf {\bibinfo {volume} {20}},\ \bibinfo {pages} {434203} (\bibinfo {year} {2008})}\BibitemShut {NoStop}%
\bibitem [{\citenamefont {Zimmermann}\ \emph {et~al.}(2014)\citenamefont {Zimmermann}, \citenamefont {Meier},\ and\ \citenamefont {Fiebig}}]{zimmermann2014}%
  \BibitemOpen
  \bibfield  {author} {\bibinfo {author} {\bibfnamefont {A.}~\bibnamefont {Zimmermann}}, \bibinfo {author} {\bibfnamefont {D.}~\bibnamefont {Meier}},\ and\ \bibinfo {author} {\bibfnamefont {M.}~\bibnamefont {Fiebig}},\ }\bibfield  {title} {\bibinfo {title} {{Ferroic nature of magnetic toroidal order}},\ }\href {https://doi.org/10.1038/ncomms5796} {\bibfield  {journal} {\bibinfo  {journal} {Nat. Commun.}\ }\textbf {\bibinfo {volume} {5}},\ \bibinfo {pages} {4796} (\bibinfo {year} {2014})}\BibitemShut {NoStop}%
\bibitem [{\citenamefont {Landau}\ and\ \citenamefont {Lifshitz}(1980)}]{landau1980}%
  \BibitemOpen
  \bibfield  {author} {\bibinfo {author} {\bibfnamefont {L.~D.}\ \bibnamefont {Landau}}\ and\ \bibinfo {author} {\bibfnamefont {E.~M.}\ \bibnamefont {Lifshitz}},\ }\href@noop {} {\bibfield  {journal} {\bibinfo  {journal} {{{\it Statistical Physics}}}\ } (\bibinfo {year} {Pergamon Press, Oxford, 1980})}\BibitemShut {NoStop}%
\bibitem [{\citenamefont {Hayami}\ \emph {et~al.}(2018)\citenamefont {Hayami}, \citenamefont {Yatsushiro}, \citenamefont {Yanagi},\ and\ \citenamefont {Kusunose}}]{hayami2018_prb}%
  \BibitemOpen
  \bibfield  {author} {\bibinfo {author} {\bibfnamefont {S.}~\bibnamefont {Hayami}}, \bibinfo {author} {\bibfnamefont {M.}~\bibnamefont {Yatsushiro}}, \bibinfo {author} {\bibfnamefont {Y.}~\bibnamefont {Yanagi}},\ and\ \bibinfo {author} {\bibfnamefont {H.}~\bibnamefont {Kusunose}},\ }\bibfield  {title} {\bibinfo {title} {{Classification of atomic-scale multipoles under crystallographic point groups and application to linear response tensors}},\ }\href {https://doi.org/10.1103/PhysRevB.98.165110} {\bibfield  {journal} {\bibinfo  {journal} {Phys. Rev. B}\ }\textbf {\bibinfo {volume} {98}},\ \bibinfo {pages} {165110} (\bibinfo {year} {2018})}\BibitemShut {NoStop}%
\bibitem [{\citenamefont {Yatsushiro}\ \emph {et~al.}(2021)\citenamefont {Yatsushiro}, \citenamefont {Kusunose},\ and\ \citenamefont {Hayami}}]{yatsushiro2021}%
  \BibitemOpen
  \bibfield  {author} {\bibinfo {author} {\bibfnamefont {M.}~\bibnamefont {Yatsushiro}}, \bibinfo {author} {\bibfnamefont {H.}~\bibnamefont {Kusunose}},\ and\ \bibinfo {author} {\bibfnamefont {S.}~\bibnamefont {Hayami}},\ }\bibfield  {title} {\bibinfo {title} {{Multipole classification in 122 magnetic point groups for unified understanding of multiferroic responses and transport phenomena}},\ }\href {https://doi.org/10.1103/PhysRevB.104.054412} {\bibfield  {journal} {\bibinfo  {journal} {Phys. Rev. B}\ }\textbf {\bibinfo {volume} {104}},\ \bibinfo {pages} {054412} (\bibinfo {year} {2021})}\BibitemShut {NoStop}%
\bibitem [{\citenamefont {Kuramoto}\ \emph {et~al.}(2009)\citenamefont {Kuramoto}, \citenamefont {Kusunose},\ and\ \citenamefont {Kiss}}]{kuramoto2009}%
  \BibitemOpen
  \bibfield  {author} {\bibinfo {author} {\bibfnamefont {Y.}~\bibnamefont {Kuramoto}}, \bibinfo {author} {\bibfnamefont {H.}~\bibnamefont {Kusunose}},\ and\ \bibinfo {author} {\bibfnamefont {A.}~\bibnamefont {Kiss}},\ }\bibfield  {title} {\bibinfo {title} {{Multipole Orders and Fluctuations in Strongly Correlated Electron Systems}},\ }\href {https://doi.org/10.1143/JPSJ.78.072001} {\bibfield  {journal} {\bibinfo  {journal} {J. Phys. Soc. Jpn.}\ }\textbf {\bibinfo {volume} {78}},\ \bibinfo {pages} {072001} (\bibinfo {year} {2009})}\BibitemShut {NoStop}%
\bibitem [{\citenamefont {Suzuki}\ \emph {et~al.}(2018)\citenamefont {Suzuki}, \citenamefont {Ikeda},\ and\ \citenamefont {Oppeneer}}]{suzuki2018_jpsj}%
  \BibitemOpen
  \bibfield  {author} {\bibinfo {author} {\bibfnamefont {M.-T.}\ \bibnamefont {Suzuki}}, \bibinfo {author} {\bibfnamefont {H.}~\bibnamefont {Ikeda}},\ and\ \bibinfo {author} {\bibfnamefont {P.~M.}\ \bibnamefont {Oppeneer}},\ }\bibfield  {title} {\bibinfo {title} {{First-principles Theory of Magnetic Multipoles in Condensed Matter Systems}},\ }\href {https://doi.org/10.7566/JPSJ.87.041008} {\bibfield  {journal} {\bibinfo  {journal} {J. Phys. Soc. Jpn.}\ }\textbf {\bibinfo {volume} {87}},\ \bibinfo {pages} {041008} (\bibinfo {year} {2018})}\BibitemShut {NoStop}%
\bibitem [{\citenamefont {Bibes}\ and\ \citenamefont {Barth^^c3^^a9l^^c3^^a9my}(2008)}]{bibes2008}%
  \BibitemOpen
  \bibfield  {author} {\bibinfo {author} {\bibfnamefont {M.}~\bibnamefont {Bibes}}\ and\ \bibinfo {author} {\bibfnamefont {A.}~\bibnamefont {Barth^^c3^^a9l^^c3^^a9my}},\ }\bibfield  {title} {\bibinfo {title} {{Towards a magnetoelectric memory}},\ }\href {https://doi.org/10.1038/nmat2189} {\bibfield  {journal} {\bibinfo  {journal} {Nat. Mater.}\ }\textbf {\bibinfo {volume} {7}},\ \bibinfo {pages} {425} (\bibinfo {year} {2008})}\BibitemShut {NoStop}%
\bibitem [{\citenamefont {Narita}\ and\ \citenamefont {Fox}(2018)}]{narita2018}%
  \BibitemOpen
  \bibfield  {author} {\bibinfo {author} {\bibfnamefont {F.}~\bibnamefont {Narita}}\ and\ \bibinfo {author} {\bibfnamefont {M.}~\bibnamefont {Fox}},\ }\bibfield  {title} {\bibinfo {title} {{A Review on Piezoelectric, Magnetostrictive, and Magnetoelectric Materials and Device Technologies for Energy Harvesting Applications}},\ }\href {https://doi.org/https://doi.org/10.1002/adem.201700743} {\bibfield  {journal} {\bibinfo  {journal} {Adv. Eng. Mater.}\ }\textbf {\bibinfo {volume} {20}},\ \bibinfo {pages} {1700743} (\bibinfo {year} {2018})}\BibitemShut {NoStop}%
\bibitem [{\citenamefont {Bowen}\ \emph {et~al.}(2014)\citenamefont {Bowen}, \citenamefont {Kim}, \citenamefont {Weaver},\ and\ \citenamefont {Dunn}}]{bowen2014}%
  \BibitemOpen
  \bibfield  {author} {\bibinfo {author} {\bibfnamefont {C.~R.}\ \bibnamefont {Bowen}}, \bibinfo {author} {\bibfnamefont {H.~A.}\ \bibnamefont {Kim}}, \bibinfo {author} {\bibfnamefont {P.~M.}\ \bibnamefont {Weaver}},\ and\ \bibinfo {author} {\bibfnamefont {S.}~\bibnamefont {Dunn}},\ }\bibfield  {title} {\bibinfo {title} {{Piezoelectric and ferroelectric materials and structures for energy harvesting applications}},\ }\href {https://doi.org/10.1039/C3EE42454E} {\bibfield  {journal} {\bibinfo  {journal} {Energy Environ. Sci.}\ }\textbf {\bibinfo {volume} {7}},\ \bibinfo {pages} {25} (\bibinfo {year} {2014})}\BibitemShut {NoStop}%
\bibitem [{\citenamefont {Bukharaev}\ \emph {et~al.}(2018)\citenamefont {Bukharaev}, \citenamefont {Zvezdin}, \citenamefont {Pyatakov},\ and\ \citenamefont {Fetisov}}]{bukharaev2018}%
  \BibitemOpen
  \bibfield  {author} {\bibinfo {author} {\bibfnamefont {A.~A.}\ \bibnamefont {Bukharaev}}, \bibinfo {author} {\bibfnamefont {A.~K.}\ \bibnamefont {Zvezdin}}, \bibinfo {author} {\bibfnamefont {A.~P.}\ \bibnamefont {Pyatakov}},\ and\ \bibinfo {author} {\bibfnamefont {Y.~K.}\ \bibnamefont {Fetisov}},\ }\bibfield  {title} {\bibinfo {title} {{Straintronics: a new trend in micro- and nanoelectronics and materials science}},\ }\href {https://doi.org/10.3367/UFNe.2018.01.038279} {\bibfield  {journal} {\bibinfo  {journal} {Physics-Uspekhi}\ }\textbf {\bibinfo {volume} {61}},\ \bibinfo {pages} {1175} (\bibinfo {year} {2018})}\BibitemShut {NoStop}%
\bibitem [{\citenamefont {Manchon}\ \emph {et~al.}(2019)\citenamefont {Manchon}, \citenamefont {\ifmmode~\check{Z}\else \v{Z}\fi{}elezn\'y}, \citenamefont {Miron}, \citenamefont {Jungwirth}, \citenamefont {Sinova}, \citenamefont {Thiaville}, \citenamefont {Garello},\ and\ \citenamefont {Gambardella}}]{manchon2019}%
  \BibitemOpen
  \bibfield  {author} {\bibinfo {author} {\bibfnamefont {A.}~\bibnamefont {Manchon}}, \bibinfo {author} {\bibfnamefont {J.}~\bibnamefont {\ifmmode~\check{Z}\else \v{Z}\fi{}elezn\'y}}, \bibinfo {author} {\bibfnamefont {I.~M.}\ \bibnamefont {Miron}}, \bibinfo {author} {\bibfnamefont {T.}~\bibnamefont {Jungwirth}}, \bibinfo {author} {\bibfnamefont {J.}~\bibnamefont {Sinova}}, \bibinfo {author} {\bibfnamefont {A.}~\bibnamefont {Thiaville}}, \bibinfo {author} {\bibfnamefont {K.}~\bibnamefont {Garello}},\ and\ \bibinfo {author} {\bibfnamefont {P.}~\bibnamefont {Gambardella}},\ }\bibfield  {title} {\bibinfo {title} {{Current-induced spin-orbit torques in ferromagnetic and antiferromagnetic systems}},\ }\href {https://doi.org/10.1103/RevModPhys.91.035004} {\bibfield  {journal} {\bibinfo  {journal} {Rev. Mod. Phys.}\ }\textbf {\bibinfo {volume} {91}},\ \bibinfo {pages} {035004} (\bibinfo {year} {2019})}\BibitemShut {NoStop}%
\bibitem [{\citenamefont {Hirohata}\ \emph {et~al.}(2020)\citenamefont {Hirohata}, \citenamefont {Yamada}, \citenamefont {Nakatani}, \citenamefont {Prejbeanu}, \citenamefont {Di^^c3^^a9ny}, \citenamefont {Pirro},\ and\ \citenamefont {Hillebrands}}]{hirohata2020}%
  \BibitemOpen
  \bibfield  {author} {\bibinfo {author} {\bibfnamefont {A.}~\bibnamefont {Hirohata}}, \bibinfo {author} {\bibfnamefont {K.}~\bibnamefont {Yamada}}, \bibinfo {author} {\bibfnamefont {Y.}~\bibnamefont {Nakatani}}, \bibinfo {author} {\bibfnamefont {I.-L.}\ \bibnamefont {Prejbeanu}}, \bibinfo {author} {\bibfnamefont {B.}~\bibnamefont {Di^^c3^^a9ny}}, \bibinfo {author} {\bibfnamefont {P.}~\bibnamefont {Pirro}},\ and\ \bibinfo {author} {\bibfnamefont {B.}~\bibnamefont {Hillebrands}},\ }\bibfield  {title} {\bibinfo {title} {{Review on spintronics: Principles and device applications}},\ }\href {https://doi.org/https://doi.org/10.1016/j.jmmm.2020.166711} {\bibfield  {journal} {\bibinfo  {journal} {J. Magn. Magn. Mater.}\ }\textbf {\bibinfo {volume} {509}},\ \bibinfo {pages} {166711} (\bibinfo {year} {2020})}\BibitemShut {NoStop}%
\bibitem [{\citenamefont {Fert}\ \emph {et~al.}(2024)\citenamefont {Fert}, \citenamefont {Ramesh}, \citenamefont {Garcia}, \citenamefont {Casanova},\ and\ \citenamefont {Bibes}}]{fert2024}%
  \BibitemOpen
  \bibfield  {author} {\bibinfo {author} {\bibfnamefont {A.}~\bibnamefont {Fert}}, \bibinfo {author} {\bibfnamefont {R.}~\bibnamefont {Ramesh}}, \bibinfo {author} {\bibfnamefont {V.}~\bibnamefont {Garcia}}, \bibinfo {author} {\bibfnamefont {F.}~\bibnamefont {Casanova}},\ and\ \bibinfo {author} {\bibfnamefont {M.}~\bibnamefont {Bibes}},\ }\bibfield  {title} {\bibinfo {title} {{Electrical control of magnetism by electric field and current-induced torques}},\ }\href {https://doi.org/10.1103/RevModPhys.96.015005} {\bibfield  {journal} {\bibinfo  {journal} {Rev. Mod. Phys.}\ }\textbf {\bibinfo {volume} {96}},\ \bibinfo {pages} {015005} (\bibinfo {year} {2024})}\BibitemShut {NoStop}%
\bibitem [{\citenamefont {Curie}(1894)}]{curie1894}%
  \BibitemOpen
  \bibfield  {author} {\bibinfo {author} {\bibfnamefont {P.}~\bibnamefont {Curie}},\ }\bibfield  {title} {\bibinfo {title} {{Sur la sym{\'e}trie dans les ph{\'e}nom{\`e}nes physiques, sym{\'e}trie d'un champ {\'e}lectrique et d'un champ magn{\'e}tique}},\ }\href {https://doi.org/10.1051/jphystap:018940030039300} {\bibfield  {journal} {\bibinfo  {journal} {{J. Phys. Theor. Appl.}}\ }\textbf {\bibinfo {volume} {3}},\ \bibinfo {pages} {393} (\bibinfo {year} {1894})}\BibitemShut {NoStop}%
\bibitem [{\citenamefont {Dzyaloshinski^^c7^^90}(1960)}]{dzyaloshinski1960}%
  \BibitemOpen
  \bibfield  {author} {\bibinfo {author} {\bibfnamefont {I.~E.}\ \bibnamefont {Dzyaloshinski^^c7^^90}},\ }\bibfield  {title} {\bibinfo {title} {{On the magneto-electrical effects in antiferromagnets}},\ }\href@noop {} {\bibfield  {journal} {\bibinfo  {journal} {Sov. Phys. JETP}\ }\textbf {\bibinfo {volume} {10}},\ \bibinfo {pages} {628} (\bibinfo {year} {1960})}\BibitemShut {NoStop}%
\bibitem [{\citenamefont {O'Dell}(1962)}]{o'dell1962}%
  \BibitemOpen
  \bibfield  {author} {\bibinfo {author} {\bibfnamefont {T.~H.}\ \bibnamefont {O'Dell}},\ }\bibfield  {title} {\bibinfo {title} {{The electrodynamics of magneto-electric media}},\ }\href {https://doi.org/10.1080/14786436208213701} {\bibfield  {journal} {\bibinfo  {journal} {Philos. Mag.-J. Theor. Exp. Appl. Phys.}\ }\textbf {\bibinfo {volume} {7}},\ \bibinfo {pages} {1653} (\bibinfo {year} {1962})}\BibitemShut {NoStop}%
\bibitem [{\citenamefont {Dong}\ \emph {et~al.}(2019)\citenamefont {Dong}, \citenamefont {Xiang},\ and\ \citenamefont {Dagotto}}]{dong2019}%
  \BibitemOpen
  \bibfield  {author} {\bibinfo {author} {\bibfnamefont {S.}~\bibnamefont {Dong}}, \bibinfo {author} {\bibfnamefont {H.}~\bibnamefont {Xiang}},\ and\ \bibinfo {author} {\bibfnamefont {E.}~\bibnamefont {Dagotto}},\ }\bibfield  {title} {\bibinfo {title} {{Magnetoelectricity in multiferroics: a theoretical perspective}},\ }\href {https://doi.org/10.1093/nsr/nwz023} {\bibfield  {journal} {\bibinfo  {journal} {Natl. Sci. Rev.}\ }\textbf {\bibinfo {volume} {6}},\ \bibinfo {pages} {629} (\bibinfo {year} {2019})}\BibitemShut {NoStop}%
\bibitem [{\citenamefont {Tavger}\ and\ \citenamefont {Zaitsev}(1956)}]{tavger1956}%
  \BibitemOpen
  \bibfield  {author} {\bibinfo {author} {\bibfnamefont {B.~A.}\ \bibnamefont {Tavger}}\ and\ \bibinfo {author} {\bibfnamefont {V.~M.}\ \bibnamefont {Zaitsev}},\ }\bibfield  {title} {\bibinfo {title} {{Magnetic symmetry of crystals}},\ }\href@noop {} {\bibfield  {journal} {\bibinfo  {journal} {Sov. Phys. JETP}\ }\textbf {\bibinfo {volume} {3}},\ \bibinfo {pages} {430} (\bibinfo {year} {1956})}\BibitemShut {NoStop}%
\bibitem [{\citenamefont {Dyakonov}\ and\ \citenamefont {Perel}(1971)}]{dyakonov1971}%
  \BibitemOpen
  \bibfield  {author} {\bibinfo {author} {\bibfnamefont {M.}~\bibnamefont {Dyakonov}}\ and\ \bibinfo {author} {\bibfnamefont {V.}~\bibnamefont {Perel}},\ }\bibfield  {title} {\bibinfo {title} {{Current-induced spin orientation of electrons in semiconductors}},\ }\href {https://doi.org/https://doi.org/10.1016/0375-9601(71)90196-4} {\bibfield  {journal} {\bibinfo  {journal} {Physics Letters A}\ }\textbf {\bibinfo {volume} {35}},\ \bibinfo {pages} {459} (\bibinfo {year} {1971})}\BibitemShut {NoStop}%
\bibitem [{\citenamefont {{Ivchenko}}\ and\ \citenamefont {{Pikus}}(1978)}]{ivchenko1978}%
  \BibitemOpen
  \bibfield  {author} {\bibinfo {author} {\bibfnamefont {E.~L.}\ \bibnamefont {{Ivchenko}}}\ and\ \bibinfo {author} {\bibfnamefont {G.~E.}\ \bibnamefont {{Pikus}}},\ }\bibfield  {title} {\bibinfo {title} {{New photogalvanic effect in gyrotropic crystals}},\ }\href@noop {} {\bibfield  {journal} {\bibinfo  {journal} {JETP Lett.}\ }\textbf {\bibinfo {volume} {27}},\ \bibinfo {pages} {604} (\bibinfo {year} {1978})}\BibitemShut {NoStop}%
\bibitem [{\citenamefont {Edelstein}(1990)}]{edelstein1990}%
  \BibitemOpen
  \bibfield  {author} {\bibinfo {author} {\bibfnamefont {V.}~\bibnamefont {Edelstein}},\ }\bibfield  {title} {\bibinfo {title} {{Spin polarization of conduction electrons induced by electric current in two-dimensional asymmetric electron systems}},\ }\href {https://doi.org/https://doi.org/10.1016/0038-1098(90)90963-C} {\bibfield  {journal} {\bibinfo  {journal} {Solid State Commun.}\ }\textbf {\bibinfo {volume} {73}},\ \bibinfo {pages} {233} (\bibinfo {year} {1990})}\BibitemShut {NoStop}%
\bibitem [{\citenamefont {S^^c3^^a1nchez}\ \emph {et~al.}(2013)\citenamefont {S^^c3^^a1nchez}, \citenamefont {Vila}, \citenamefont {Desfonds}, \citenamefont {Gambarelli}, \citenamefont {Attan^^c3^^a9}, \citenamefont {Teresa}, \citenamefont {Mag^^c3^^a9n},\ and\ \citenamefont {Fert}}]{sanchez2013}%
  \BibitemOpen
  \bibfield  {author} {\bibinfo {author} {\bibfnamefont {J.}~\bibnamefont {S^^c3^^a1nchez}}, \bibinfo {author} {\bibfnamefont {L.}~\bibnamefont {Vila}}, \bibinfo {author} {\bibfnamefont {G.}~\bibnamefont {Desfonds}}, \bibinfo {author} {\bibfnamefont {S.}~\bibnamefont {Gambarelli}}, \bibinfo {author} {\bibfnamefont {J.}~\bibnamefont {Attan^^c3^^a9}}, \bibinfo {author} {\bibfnamefont {J.~D.}\ \bibnamefont {Teresa}}, \bibinfo {author} {\bibfnamefont {C.}~\bibnamefont {Mag^^c3^^a9n}},\ and\ \bibinfo {author} {\bibfnamefont {A.}~\bibnamefont {Fert}},\ }\bibfield  {title} {\bibinfo {title} {{Spin-to-charge conversion using Rashba coupling at the interface between non-magnetic materials}},\ }\href {https://doi.org/10.1038/ncomms3944} {\bibfield  {journal} {\bibinfo  {journal} {Nat. Commun.}\ }\textbf {\bibinfo {volume} {4}},\ \bibinfo {pages} {2944} (\bibinfo {year} {2013})}\BibitemShut {NoStop}%
\bibitem [{\citenamefont {Eliseev}\ \emph {et~al.}(2009)\citenamefont {Eliseev}, \citenamefont {Morozovska}, \citenamefont {Glinchuk},\ and\ \citenamefont {Blinc}}]{eliseev2009}%
  \BibitemOpen
  \bibfield  {author} {\bibinfo {author} {\bibfnamefont {E.~A.}\ \bibnamefont {Eliseev}}, \bibinfo {author} {\bibfnamefont {A.~N.}\ \bibnamefont {Morozovska}}, \bibinfo {author} {\bibfnamefont {M.~D.}\ \bibnamefont {Glinchuk}},\ and\ \bibinfo {author} {\bibfnamefont {R.}~\bibnamefont {Blinc}},\ }\bibfield  {title} {\bibinfo {title} {{Spontaneous flexoelectric/flexomagnetic effect in nanoferroics}},\ }\href {https://doi.org/10.1103/PhysRevB.79.165433} {\bibfield  {journal} {\bibinfo  {journal} {Phys. Rev. B}\ }\textbf {\bibinfo {volume} {79}},\ \bibinfo {pages} {165433} (\bibinfo {year} {2009})}\BibitemShut {NoStop}%
\bibitem [{\citenamefont {Lukashev}\ and\ \citenamefont {Sabirianov}(2010)}]{lukashev2010}%
  \BibitemOpen
  \bibfield  {author} {\bibinfo {author} {\bibfnamefont {P.}~\bibnamefont {Lukashev}}\ and\ \bibinfo {author} {\bibfnamefont {R.~F.}\ \bibnamefont {Sabirianov}},\ }\bibfield  {title} {\bibinfo {title} {{Flexomagnetic effect in frustrated triangular magnetic structures}},\ }\href {https://doi.org/10.1103/PhysRevB.82.094417} {\bibfield  {journal} {\bibinfo  {journal} {Phys. Rev. B}\ }\textbf {\bibinfo {volume} {82}},\ \bibinfo {pages} {094417} (\bibinfo {year} {2010})}\BibitemShut {NoStop}%
\bibitem [{\citenamefont {Tang}\ \emph {et~al.}(2025{\natexlab{a}})\citenamefont {Tang}, \citenamefont {Gong},\ and\ \citenamefont {Yi}}]{tang2025_mse}%
  \BibitemOpen
  \bibfield  {author} {\bibinfo {author} {\bibfnamefont {Z.}~\bibnamefont {Tang}}, \bibinfo {author} {\bibfnamefont {Q.}~\bibnamefont {Gong}},\ and\ \bibinfo {author} {\bibfnamefont {M.}~\bibnamefont {Yi}},\ }\bibfield  {title} {\bibinfo {title} {{Flexomagnetism: Progress, challenges, and opportunities}},\ }\href {https://doi.org/https://doi.org/10.1016/j.mser.2024.100878} {\bibfield  {journal} {\bibinfo  {journal} {Mater. Sci. Eng. R}\ }\textbf {\bibinfo {volume} {162}},\ \bibinfo {pages} {100878} (\bibinfo {year} {2025}{\natexlab{a}})}\BibitemShut {NoStop}%
\bibitem [{\citenamefont {Shu}\ \emph {et~al.}(2019)\citenamefont {Shu}, \citenamefont {Liang}, \citenamefont {Rao}, \citenamefont {Fei}, \citenamefont {Ke},\ and\ \citenamefont {Wang}}]{shu2019}%
  \BibitemOpen
  \bibfield  {author} {\bibinfo {author} {\bibfnamefont {L.}~\bibnamefont {Shu}}, \bibinfo {author} {\bibfnamefont {R.}~\bibnamefont {Liang}}, \bibinfo {author} {\bibfnamefont {Z.}~\bibnamefont {Rao}}, \bibinfo {author} {\bibfnamefont {L.}~\bibnamefont {Fei}}, \bibinfo {author} {\bibfnamefont {S.}~\bibnamefont {Ke}},\ and\ \bibinfo {author} {\bibfnamefont {Y.}~\bibnamefont {Wang}},\ }\bibfield  {title} {\bibinfo {title} {{Flexoelectric materials and their related applications: A focused review}},\ }\href {https://doi.org/10.1007/s40145-018-0311-3} {\bibfield  {journal} {\bibinfo  {journal} {J. Adv. Ceram.}\ }\textbf {\bibinfo {volume} {8}},\ \bibinfo {pages} {153} (\bibinfo {year} {2019})}\BibitemShut {NoStop}%
\bibitem [{\citenamefont {Chappert}\ \emph {et~al.}(2007)\citenamefont {Chappert}, \citenamefont {Fert},\ and\ \citenamefont {Dau}}]{chappert2007}%
  \BibitemOpen
  \bibfield  {author} {\bibinfo {author} {\bibfnamefont {C.}~\bibnamefont {Chappert}}, \bibinfo {author} {\bibfnamefont {A.}~\bibnamefont {Fert}},\ and\ \bibinfo {author} {\bibfnamefont {F.~V.}\ \bibnamefont {Dau}},\ }\bibfield  {title} {\bibinfo {title} {{The emergence of spin electronics in data storage}},\ }\href {https://doi.org/10.1038/nmat2024} {\bibfield  {journal} {\bibinfo  {journal} {Nat. Mater.}\ }\textbf {\bibinfo {volume} {6}},\ \bibinfo {pages} {813} (\bibinfo {year} {2007})}\BibitemShut {NoStop}%
\bibitem [{\citenamefont {Eliseev}\ \emph {et~al.}(2011)\citenamefont {Eliseev}, \citenamefont {Glinchuk}, \citenamefont {Khist}, \citenamefont {Skorokhod}, \citenamefont {Blinc},\ and\ \citenamefont {Morozovska}}]{eliseev2011}%
  \BibitemOpen
  \bibfield  {author} {\bibinfo {author} {\bibfnamefont {E.~A.}\ \bibnamefont {Eliseev}}, \bibinfo {author} {\bibfnamefont {M.~D.}\ \bibnamefont {Glinchuk}}, \bibinfo {author} {\bibfnamefont {V.}~\bibnamefont {Khist}}, \bibinfo {author} {\bibfnamefont {V.~V.}\ \bibnamefont {Skorokhod}}, \bibinfo {author} {\bibfnamefont {R.}~\bibnamefont {Blinc}},\ and\ \bibinfo {author} {\bibfnamefont {A.~N.}\ \bibnamefont {Morozovska}},\ }\bibfield  {title} {\bibinfo {title} {{Linear magnetoelectric coupling and ferroelectricity induced by the flexomagnetic effect in ferroics}},\ }\href {https://doi.org/10.1103/PhysRevB.84.174112} {\bibfield  {journal} {\bibinfo  {journal} {Phys. Rev. B}\ }\textbf {\bibinfo {volume} {84}},\ \bibinfo {pages} {174112} (\bibinfo {year} {2011})}\BibitemShut {NoStop}%
\bibitem [{\citenamefont {Shi}\ \emph {et~al.}(2019)\citenamefont {Shi}, \citenamefont {Guo}, \citenamefont {Zhang},\ and\ \citenamefont {Guo}}]{shi2019}%
  \BibitemOpen
  \bibfield  {author} {\bibinfo {author} {\bibfnamefont {W.}~\bibnamefont {Shi}}, \bibinfo {author} {\bibfnamefont {Y.}~\bibnamefont {Guo}}, \bibinfo {author} {\bibfnamefont {Z.}~\bibnamefont {Zhang}},\ and\ \bibinfo {author} {\bibfnamefont {W.}~\bibnamefont {Guo}},\ }\bibfield  {title} {\bibinfo {title} {{Strain Gradient Mediated Magnetism and Polarization in Monolayer ${\rm VSe_2}$}},\ }\href {https://doi.org/10.1021/acs.jpcc.9b08445} {\bibfield  {journal} {\bibinfo  {journal} {J. Phys. Chem. C}\ }\textbf {\bibinfo {volume} {123}},\ \bibinfo {pages} {24988} (\bibinfo {year} {2019})}\BibitemShut {NoStop}%
\bibitem [{\citenamefont {Edstr\"om}\ \emph {et~al.}(2022)\citenamefont {Edstr\"om}, \citenamefont {Amoroso}, \citenamefont {Picozzi}, \citenamefont {Barone},\ and\ \citenamefont {Stengel}}]{edstrom2022}%
  \BibitemOpen
  \bibfield  {author} {\bibinfo {author} {\bibfnamefont {A.}~\bibnamefont {Edstr\"om}}, \bibinfo {author} {\bibfnamefont {D.}~\bibnamefont {Amoroso}}, \bibinfo {author} {\bibfnamefont {S.}~\bibnamefont {Picozzi}}, \bibinfo {author} {\bibfnamefont {P.}~\bibnamefont {Barone}},\ and\ \bibinfo {author} {\bibfnamefont {M.}~\bibnamefont {Stengel}},\ }\bibfield  {title} {\bibinfo {title} {{Curved Magnetism in ${\mathrm{CrI}}_{3}$}},\ }\href {https://doi.org/10.1103/PhysRevLett.128.177202} {\bibfield  {journal} {\bibinfo  {journal} {Phys. Rev. Lett.}\ }\textbf {\bibinfo {volume} {128}},\ \bibinfo {pages} {177202} (\bibinfo {year} {2022})}\BibitemShut {NoStop}%
\bibitem [{\citenamefont {Qiu}\ \emph {et~al.}(2023)\citenamefont {Qiu}, \citenamefont {Li}, \citenamefont {Zhou},\ and\ \citenamefont {Cai}}]{qiu2023}%
  \BibitemOpen
  \bibfield  {author} {\bibinfo {author} {\bibfnamefont {G.}~\bibnamefont {Qiu}}, \bibinfo {author} {\bibfnamefont {Z.}~\bibnamefont {Li}}, \bibinfo {author} {\bibfnamefont {K.}~\bibnamefont {Zhou}},\ and\ \bibinfo {author} {\bibfnamefont {Y.}~\bibnamefont {Cai}},\ }\bibfield  {title} {\bibinfo {title} {{Flexomagnetic noncollinear state with a plumb line shape spin configuration in edged two-dimensional magnetic ${\rm CrI_3}$}},\ }\href {https://doi.org/10.1038/s41535-023-00547-w} {\bibfield  {journal} {\bibinfo  {journal} {npj Quantum Mater.}\ }\textbf {\bibinfo {volume} {8}},\ \bibinfo {pages} {15} (\bibinfo {year} {2023})}\BibitemShut {NoStop}%
\bibitem [{\citenamefont {Qiao}\ \emph {et~al.}(2024)\citenamefont {Qiao}, \citenamefont {Sladek}, \citenamefont {Sladek}, \citenamefont {Kaminskiy}, \citenamefont {Pyatakov},\ and\ \citenamefont {Ren}}]{qiao2024}%
  \BibitemOpen
  \bibfield  {author} {\bibinfo {author} {\bibfnamefont {L.}~\bibnamefont {Qiao}}, \bibinfo {author} {\bibfnamefont {J.}~\bibnamefont {Sladek}}, \bibinfo {author} {\bibfnamefont {V.}~\bibnamefont {Sladek}}, \bibinfo {author} {\bibfnamefont {A.~S.}\ \bibnamefont {Kaminskiy}}, \bibinfo {author} {\bibfnamefont {A.~P.}\ \bibnamefont {Pyatakov}},\ and\ \bibinfo {author} {\bibfnamefont {W.}~\bibnamefont {Ren}},\ }\bibfield  {title} {\bibinfo {title} {{Curvature-induced magnetization in a ${\mathrm{CrI}}_{3}$ bilayer: Flexomagnetic effect enhancement in van der Waals antiferromagnets}},\ }\href {https://doi.org/10.1103/PhysRevB.109.014410} {\bibfield  {journal} {\bibinfo  {journal} {Phys. Rev. B}\ }\textbf {\bibinfo {volume} {109}},\ \bibinfo {pages} {014410} (\bibinfo {year} {2024})}\BibitemShut {NoStop}%
\bibitem [{\citenamefont {Tang}\ \emph {et~al.}(2025{\natexlab{b}})\citenamefont {Tang}, \citenamefont {Gong},\ and\ \citenamefont {Yi}}]{tang2025_prb}%
  \BibitemOpen
  \bibfield  {author} {\bibinfo {author} {\bibfnamefont {Z.}~\bibnamefont {Tang}}, \bibinfo {author} {\bibfnamefont {Q.}~\bibnamefont {Gong}},\ and\ \bibinfo {author} {\bibfnamefont {M.}~\bibnamefont {Yi}},\ }\bibfield  {title} {\bibinfo {title} {{Intrinsic flexomagnetism of antiferromagnetic monolayer ${\rm FeSe}$}},\ }\href {https://doi.org/10.1103/swjl-y8z1} {\bibfield  {journal} {\bibinfo  {journal} {Phys. Rev. B}\ }\textbf {\bibinfo {volume} {112}},\ \bibinfo {pages} {014432} (\bibinfo {year} {2025}{\natexlab{b}})}\BibitemShut {NoStop}%
\bibitem [{\citenamefont {Liu}\ \emph {et~al.}(2025)\citenamefont {Liu}, \citenamefont {Chen}, \citenamefont {Yin}, \citenamefont {Tang}, \citenamefont {Gong}, \citenamefont {Yi},\ and\ \citenamefont {Liu}}]{liu2025}%
  \BibitemOpen
  \bibfield  {author} {\bibinfo {author} {\bibfnamefont {Y.}~\bibnamefont {Liu}}, \bibinfo {author} {\bibfnamefont {W.}~\bibnamefont {Chen}}, \bibinfo {author} {\bibfnamefont {Y.}~\bibnamefont {Yin}}, \bibinfo {author} {\bibfnamefont {Z.}~\bibnamefont {Tang}}, \bibinfo {author} {\bibfnamefont {Q.}~\bibnamefont {Gong}}, \bibinfo {author} {\bibfnamefont {M.}~\bibnamefont {Yi}},\ and\ \bibinfo {author} {\bibfnamefont {Y.}~\bibnamefont {Liu}},\ }\bibfield  {title} {\bibinfo {title} {{Large piezo-/flexo-electric and flexomagnetic effects in a semiconducting cobalt telluride monolayer}},\ }\href {https://doi.org/10.1039/D5NH00287G} {\bibfield  {journal} {\bibinfo  {journal} {Nanoscale Horiz.}\ }\textbf {\bibinfo {volume} {10}},\ \bibinfo {pages} {2995} (\bibinfo {year} {2025})}\BibitemShut {NoStop}%
\bibitem [{\citenamefont {Shen}\ \emph {et~al.}(2018)\citenamefont {Shen}, \citenamefont {Song}, \citenamefont {Tong}, \citenamefont {Shen}, \citenamefont {Gong},\ and\ \citenamefont {Duan}}]{shen2018}%
  \BibitemOpen
  \bibfield  {author} {\bibinfo {author} {\bibfnamefont {Y.-H.}\ \bibnamefont {Shen}}, \bibinfo {author} {\bibfnamefont {Y.-X.}\ \bibnamefont {Song}}, \bibinfo {author} {\bibfnamefont {W.-Y.}\ \bibnamefont {Tong}}, \bibinfo {author} {\bibfnamefont {X.-W.}\ \bibnamefont {Shen}}, \bibinfo {author} {\bibfnamefont {S.-j.}\ \bibnamefont {Gong}},\ and\ \bibinfo {author} {\bibfnamefont {C.-G.}\ \bibnamefont {Duan}},\ }\bibfield  {title} {\bibinfo {title} {{Giant Flexomagnetoelectric Effect in Dilute Magnetic Monolayer}},\ }\href {https://doi.org/https://doi.org/10.1002/adts.201800048} {\bibfield  {journal} {\bibinfo  {journal} {Adv. Theor. Simul.}\ }\textbf {\bibinfo {volume} {1}},\ \bibinfo {pages} {1800048} (\bibinfo {year} {2018})}\BibitemShut {NoStop}%
\bibitem [{\citenamefont {Belyaev}\ \emph {et~al.}(2020)\citenamefont {Belyaev}, \citenamefont {Izotov}, \citenamefont {Solovev},\ and\ \citenamefont {Boev}}]{belyaev2020}%
  \BibitemOpen
  \bibfield  {author} {\bibinfo {author} {\bibfnamefont {B.~A.}\ \bibnamefont {Belyaev}}, \bibinfo {author} {\bibfnamefont {A.~V.}\ \bibnamefont {Izotov}}, \bibinfo {author} {\bibfnamefont {P.~N.}\ \bibnamefont {Solovev}},\ and\ \bibinfo {author} {\bibfnamefont {N.~M.}\ \bibnamefont {Boev}},\ }\bibfield  {title} {\bibinfo {title} {{Strain-Gradient-Induced Unidirectional Magnetic Anisotropy in Nanocrystalline Thin Permalloy Films}},\ }\href {https://doi.org/https://doi.org/10.1002/pssr.201900467} {\bibfield  {journal} {\bibinfo  {journal} {Phys. Status Solidi RRL}\ }\textbf {\bibinfo {volume} {14}},\ \bibinfo {pages} {1900467} (\bibinfo {year} {2020})}\BibitemShut {NoStop}%
\bibitem [{\citenamefont {Ling}\ \emph {et~al.}(2023)\citenamefont {Ling}, \citenamefont {Yu}, \citenamefont {Yuan}, \citenamefont {He}, \citenamefont {Han}, \citenamefont {Du}, \citenamefont {Fan}, \citenamefont {Yan},\ and\ \citenamefont {Xu}}]{ling2023}%
  \BibitemOpen
  \bibfield  {author} {\bibinfo {author} {\bibfnamefont {Y.}~\bibnamefont {Ling}}, \bibinfo {author} {\bibfnamefont {X.}~\bibnamefont {Yu}}, \bibinfo {author} {\bibfnamefont {S.}~\bibnamefont {Yuan}}, \bibinfo {author} {\bibfnamefont {A.}~\bibnamefont {He}}, \bibinfo {author} {\bibfnamefont {Z.}~\bibnamefont {Han}}, \bibinfo {author} {\bibfnamefont {J.}~\bibnamefont {Du}}, \bibinfo {author} {\bibfnamefont {Q.}~\bibnamefont {Fan}}, \bibinfo {author} {\bibfnamefont {S.}~\bibnamefont {Yan}},\ and\ \bibinfo {author} {\bibfnamefont {Q.}~\bibnamefont {Xu}},\ }\bibfield  {title} {\bibinfo {title} {{Flexomagnetic Effect Enhanced Ferromagnetism and Magnetoelectrochemistry in Freestanding High-Entropy Alloy Films}},\ }\href {https://doi.org/10.1021/acsnano.3c05255} {\bibfield  {journal} {\bibinfo  {journal} {ACS Nano}\ }\textbf {\bibinfo {volume} {17}},\ \bibinfo {pages} {17299} (\bibinfo {year} {2023})},\ \bibinfo {note} {doi: 10.1021/acsnano.3c05255}\BibitemShut {NoStop}%
\bibitem [{\citenamefont {Makushko}\ \emph {et~al.}(2022)\citenamefont {Makushko}, \citenamefont {Kosub}, \citenamefont {Pylypovskyi}, \citenamefont {Hedrich}, \citenamefont {Li}, \citenamefont {Pashkin}, \citenamefont {Avdoshenko}, \citenamefont {H^^c3^^bcbner}, \citenamefont {Ganss}, \citenamefont {Wolf}, \citenamefont {Lubk}, \citenamefont {Liedke}, \citenamefont {Butterling}, \citenamefont {Wagner}, \citenamefont {Wagner}, \citenamefont {Shields}, \citenamefont {Lehmann}, \citenamefont {Veremchuk}, \citenamefont {Fassbender}, \citenamefont {Maletinsky},\ and\ \citenamefont {Makarov}}]{makushko2022}%
  \BibitemOpen
  \bibfield  {author} {\bibinfo {author} {\bibfnamefont {P.}~\bibnamefont {Makushko}}, \bibinfo {author} {\bibfnamefont {T.}~\bibnamefont {Kosub}}, \bibinfo {author} {\bibfnamefont {O.}~\bibnamefont {Pylypovskyi}}, \bibinfo {author} {\bibfnamefont {N.}~\bibnamefont {Hedrich}}, \bibinfo {author} {\bibfnamefont {J.}~\bibnamefont {Li}}, \bibinfo {author} {\bibfnamefont {A.}~\bibnamefont {Pashkin}}, \bibinfo {author} {\bibfnamefont {S.}~\bibnamefont {Avdoshenko}}, \bibinfo {author} {\bibfnamefont {R.}~\bibnamefont {H^^c3^^bcbner}}, \bibinfo {author} {\bibfnamefont {F.}~\bibnamefont {Ganss}}, \bibinfo {author} {\bibfnamefont {D.}~\bibnamefont {Wolf}}, \bibinfo {author} {\bibfnamefont {A.}~\bibnamefont {Lubk}}, \bibinfo {author} {\bibfnamefont {M.}~\bibnamefont {Liedke}}, \bibinfo {author} {\bibfnamefont {M.}~\bibnamefont {Butterling}}, \bibinfo {author} {\bibfnamefont {A.}~\bibnamefont {Wagner}}, \bibinfo {author} {\bibfnamefont {K.}~\bibnamefont {Wagner}}, \bibinfo {author} {\bibfnamefont {B.}~\bibnamefont {Shields}}, \bibinfo {author} {\bibfnamefont {P.}~\bibnamefont {Lehmann}}, \bibinfo {author} {\bibfnamefont {I.}~\bibnamefont {Veremchuk}}, \bibinfo {author} {\bibfnamefont {J.}~\bibnamefont {Fassbender}}, \bibinfo {author} {\bibfnamefont {P.}~\bibnamefont {Maletinsky}},\ and\ \bibinfo {author} {\bibfnamefont {D.}~\bibnamefont {Makarov}},\ }\bibfield  {title} {\bibinfo {title} {{Flexomagnetism and vertically graded N^^c3^^a9el temperature of antiferromagnetic ${\rm Cr_2 O_3}$ thin films}},\ }\href {https://doi.org/10.1038/s41467-022-34233-5} {\bibfield  {journal} {\bibinfo  {journal} {Nat. Commun.}\ }\textbf {\bibinfo {volume} {13}},\ \bibinfo {pages} {6745} (\bibinfo {year} {2022})}\BibitemShut {NoStop}%
\bibitem [{\citenamefont {Callen}\ and\ \citenamefont {Callen}(1963)}]{callen1963}%
  \BibitemOpen
  \bibfield  {author} {\bibinfo {author} {\bibfnamefont {E.~R.}\ \bibnamefont {Callen}}\ and\ \bibinfo {author} {\bibfnamefont {H.~B.}\ \bibnamefont {Callen}},\ }\bibfield  {title} {\bibinfo {title} {{Static Magnetoelastic Coupling in Cubic Crystals}},\ }\href {https://doi.org/10.1103/PhysRev.129.578} {\bibfield  {journal} {\bibinfo  {journal} {Phys. Rev.}\ }\textbf {\bibinfo {volume} {129}},\ \bibinfo {pages} {578} (\bibinfo {year} {1963})}\BibitemShut {NoStop}%
\bibitem [{\citenamefont {Callen}\ and\ \citenamefont {Callen}(1965)}]{callen1965}%
  \BibitemOpen
  \bibfield  {author} {\bibinfo {author} {\bibfnamefont {E.}~\bibnamefont {Callen}}\ and\ \bibinfo {author} {\bibfnamefont {H.~B.}\ \bibnamefont {Callen}},\ }\bibfield  {title} {\bibinfo {title} {{Magnetostriction, Forced Magnetostriction, and Anomalous Thermal Expansion in Ferromagnets}},\ }\href {https://doi.org/10.1103/PhysRev.139.A455} {\bibfield  {journal} {\bibinfo  {journal} {Phys. Rev.}\ }\textbf {\bibinfo {volume} {139}},\ \bibinfo {pages} {A455} (\bibinfo {year} {1965})}\BibitemShut {NoStop}%
\bibitem [{\citenamefont {Ji}\ \emph {et~al.}(2007)\citenamefont {Ji}, \citenamefont {Song}, \citenamefont {Koo}, \citenamefont {Park}, \citenamefont {Park}, \citenamefont {Lee}, \citenamefont {Lee}, \citenamefont {Park}, \citenamefont {Kim}, \citenamefont {Cho}, \citenamefont {Hong}, \citenamefont {Lee},\ and\ \citenamefont {Iga}}]{ji2007}%
  \BibitemOpen
  \bibfield  {author} {\bibinfo {author} {\bibfnamefont {S.}~\bibnamefont {Ji}}, \bibinfo {author} {\bibfnamefont {C.}~\bibnamefont {Song}}, \bibinfo {author} {\bibfnamefont {J.}~\bibnamefont {Koo}}, \bibinfo {author} {\bibfnamefont {J.}~\bibnamefont {Park}}, \bibinfo {author} {\bibfnamefont {Y.~J.}\ \bibnamefont {Park}}, \bibinfo {author} {\bibfnamefont {K.-B.}\ \bibnamefont {Lee}}, \bibinfo {author} {\bibfnamefont {S.}~\bibnamefont {Lee}}, \bibinfo {author} {\bibfnamefont {J.-G.}\ \bibnamefont {Park}}, \bibinfo {author} {\bibfnamefont {J.~Y.}\ \bibnamefont {Kim}}, \bibinfo {author} {\bibfnamefont {B.~K.}\ \bibnamefont {Cho}}, \bibinfo {author} {\bibfnamefont {K.-P.}\ \bibnamefont {Hong}}, \bibinfo {author} {\bibfnamefont {C.-H.}\ \bibnamefont {Lee}},\ and\ \bibinfo {author} {\bibfnamefont {F.}~\bibnamefont {Iga}},\ }\bibfield  {title} {\bibinfo {title} {{Resonant X-Ray Scattering Study of Quadrupole-Strain Coupling in ${\mathrm{DyB}}_{4}$}},\ }\href {https://doi.org/10.1103/PhysRevLett.99.076401} {\bibfield  {journal} {\bibinfo  {journal} {Phys. Rev. Lett.}\ }\textbf {\bibinfo {volume} {99}},\ \bibinfo {pages} {076401} (\bibinfo {year} {2007})}\BibitemShut {NoStop}%
\bibitem [{\citenamefont {Rosenberg}\ \emph {et~al.}(2019)\citenamefont {Rosenberg}, \citenamefont {Chu}, \citenamefont {Ruff}, \citenamefont {Hristov},\ and\ \citenamefont {Fisher}}]{rosenberg2019}%
  \BibitemOpen
  \bibfield  {author} {\bibinfo {author} {\bibfnamefont {E.~W.}\ \bibnamefont {Rosenberg}}, \bibinfo {author} {\bibfnamefont {J.-H.}\ \bibnamefont {Chu}}, \bibinfo {author} {\bibfnamefont {J.~P.~C.}\ \bibnamefont {Ruff}}, \bibinfo {author} {\bibfnamefont {A.~T.}\ \bibnamefont {Hristov}},\ and\ \bibinfo {author} {\bibfnamefont {I.~R.}\ \bibnamefont {Fisher}},\ }\bibfield  {title} {\bibinfo {title} {{Divergence of the quadrupole-strain susceptibility of the electronic nematic system ${\rm YbRu_2Ge_2}$}},\ }\href {https://doi.org/10.1073/pnas.1818910116} {\bibfield  {journal} {\bibinfo  {journal} {Proc. Natl. Acad. Sci. U.S.A.}\ }\textbf {\bibinfo {volume} {116}},\ \bibinfo {pages} {7232} (\bibinfo {year} {2019})}\BibitemShut {NoStop}%
\bibitem [{\citenamefont {Luttinger}(1964)}]{luttinger1964}%
  \BibitemOpen
  \bibfield  {author} {\bibinfo {author} {\bibfnamefont {J.~M.}\ \bibnamefont {Luttinger}},\ }\bibfield  {title} {\bibinfo {title} {{Theory of Thermal Transport Coefficients}},\ }\href {https://doi.org/10.1103/PhysRev.135.A1505} {\bibfield  {journal} {\bibinfo  {journal} {Phys. Rev.}\ }\textbf {\bibinfo {volume} {135}},\ \bibinfo {pages} {A1505} (\bibinfo {year} {1964})}\BibitemShut {NoStop}%
\bibitem [{\citenamefont {Kubo}\ \emph {et~al.}(1957)\citenamefont {Kubo}, \citenamefont {Yokota},\ and\ \citenamefont {Nakajima}}]{kubo1957}%
  \BibitemOpen
  \bibfield  {author} {\bibinfo {author} {\bibfnamefont {R.}~\bibnamefont {Kubo}}, \bibinfo {author} {\bibfnamefont {M.}~\bibnamefont {Yokota}},\ and\ \bibinfo {author} {\bibfnamefont {S.}~\bibnamefont {Nakajima}},\ }\bibfield  {title} {\bibinfo {title} {{Statistical-Mechanical Theory of Irreversible Processes. II. Response to Thermal Disturbance}},\ }\href {https://doi.org/10.1143/JPSJ.12.1203} {\bibfield  {journal} {\bibinfo  {journal} {J. Phys. Soc. Jpn.}\ }\textbf {\bibinfo {volume} {12}},\ \bibinfo {pages} {1203} (\bibinfo {year} {1957})}\BibitemShut {NoStop}%
\bibitem [{Note1()}]{Note1}%
  \BibitemOpen
  \bibinfo {note} {The definition of $Q_{\lambda }(\protect \bm {r})$ is fixed by the requirement that the Hamiltonian be Hermitian. When modulation of hopping amplitudes is taken into account, $\protect \hat {Q}_\lambda $ can generally become a nonlocal operator. Nevertheless, as long as $Q_\lambda (\protect \bm {r}) = Q^\dagger _\lambda (\protect \bm {r})$~[Eq.~\protect \eqref {eq:Q(r)}] is satisfied, $\protect \mathcal {V}$ is guaranteed to be Hermitian, and the formulation in the present study remains valid even if $\protect \hat {Q}_\lambda $ is a nonlocal operator. A similar definition of a local density can also be found in Eq.~(21) of Ref.~\cite {qin2011}.}\BibitemShut {Stop}%
\bibitem [{\citenamefont {Ogawa}\ \emph {et~al.}(2023)\citenamefont {Ogawa}, \citenamefont {Funato},\ and\ \citenamefont {Kohno}}]{ogawa2023}%
  \BibitemOpen
  \bibfield  {author} {\bibinfo {author} {\bibfnamefont {Y.}~\bibnamefont {Ogawa}}, \bibinfo {author} {\bibfnamefont {T.}~\bibnamefont {Funato}},\ and\ \bibinfo {author} {\bibfnamefont {H.}~\bibnamefont {Kohno}},\ }\bibfield  {title} {\bibinfo {title} {{Microscopic Analysis of Lattice Distortion Effects in Rashba Systems}},\ }\href {https://doi.org/10.7566/JPSJ.92.113702} {\bibfield  {journal} {\bibinfo  {journal} {J. Phys. Soc. Jpn.}\ }\textbf {\bibinfo {volume} {92}},\ \bibinfo {pages} {113702} (\bibinfo {year} {2023})}\BibitemShut {NoStop}%
\bibitem [{\citenamefont {Uchino}\ \emph {et~al.}()\citenamefont {Uchino}, \citenamefont {Ogawa},\ and\ \citenamefont {Hayami}}]{uchino2025_arxiv}%
  \BibitemOpen
  \bibfield  {author} {\bibinfo {author} {\bibfnamefont {K.}~\bibnamefont {Uchino}}, \bibinfo {author} {\bibfnamefont {Y.}~\bibnamefont {Ogawa}},\ and\ \bibinfo {author} {\bibfnamefont {S.}~\bibnamefont {Hayami}},\ }\bibfield  {title} {\bibinfo {title} {{Analysis of Spin Current Generation by Elastic Waves in $f$-wave Altermagnets}},\ }\href {https://arxiv.org/abs/2508.06027} {\bibinfo  {journal} {arXiv:2508.06027}\ }\BibitemShut {NoStop}%
\bibitem [{\citenamefont {Chang}\ and\ \citenamefont {Niu}(1996)}]{chang1996}%
  \BibitemOpen
\bibfield  {journal} {  }\bibfield  {author} {\bibinfo {author} {\bibfnamefont {M.-C.}\ \bibnamefont {Chang}}\ and\ \bibinfo {author} {\bibfnamefont {Q.}~\bibnamefont {Niu}},\ }\bibfield  {title} {\bibinfo {title} {{Berry phase, hyperorbits, and the Hofstadter spectrum: Semiclassical dynamics in magnetic Bloch bands}},\ }\href {https://doi.org/10.1103/PhysRevB.53.7010} {\bibfield  {journal} {\bibinfo  {journal} {Phys. Rev. B}\ }\textbf {\bibinfo {volume} {53}},\ \bibinfo {pages} {7010} (\bibinfo {year} {1996})}\BibitemShut {NoStop}%
\bibitem [{\citenamefont {Sundaram}\ and\ \citenamefont {Niu}(1999)}]{sundaram1999}%
  \BibitemOpen
  \bibfield  {author} {\bibinfo {author} {\bibfnamefont {G.}~\bibnamefont {Sundaram}}\ and\ \bibinfo {author} {\bibfnamefont {Q.}~\bibnamefont {Niu}},\ }\bibfield  {title} {\bibinfo {title} {{Wave-packet dynamics in slowly perturbed crystals: Gradient corrections and Berry-phase effects}},\ }\href {https://doi.org/10.1103/PhysRevB.59.14915} {\bibfield  {journal} {\bibinfo  {journal} {Phys. Rev. B}\ }\textbf {\bibinfo {volume} {59}},\ \bibinfo {pages} {14915} (\bibinfo {year} {1999})}\BibitemShut {NoStop}%
\bibitem [{\citenamefont {Dong}\ \emph {et~al.}(2020)\citenamefont {Dong}, \citenamefont {Xiao}, \citenamefont {Xiong},\ and\ \citenamefont {Niu}}]{dong2020}%
  \BibitemOpen
  \bibfield  {author} {\bibinfo {author} {\bibfnamefont {L.}~\bibnamefont {Dong}}, \bibinfo {author} {\bibfnamefont {C.}~\bibnamefont {Xiao}}, \bibinfo {author} {\bibfnamefont {B.}~\bibnamefont {Xiong}},\ and\ \bibinfo {author} {\bibfnamefont {Q.}~\bibnamefont {Niu}},\ }\bibfield  {title} {\bibinfo {title} {{Berry Phase Effects in Dipole Density and the Mott Relation}},\ }\href {https://doi.org/10.1103/PhysRevLett.124.066601} {\bibfield  {journal} {\bibinfo  {journal} {Phys. Rev. Lett.}\ }\textbf {\bibinfo {volume} {124}},\ \bibinfo {pages} {066601} (\bibinfo {year} {2020})}\BibitemShut {NoStop}%
\bibitem [{\citenamefont {Freimuth}\ \emph {et~al.}(2014)\citenamefont {Freimuth}, \citenamefont {Bl\"ugel},\ and\ \citenamefont {Mokrousov}}]{freimuth2014CIM}%
  \BibitemOpen
  \bibfield  {author} {\bibinfo {author} {\bibfnamefont {F.}~\bibnamefont {Freimuth}}, \bibinfo {author} {\bibfnamefont {S.}~\bibnamefont {Bl\"ugel}},\ and\ \bibinfo {author} {\bibfnamefont {Y.}~\bibnamefont {Mokrousov}},\ }\bibfield  {title} {\bibinfo {title} {{Spin-orbit torques in Co/Pt(111) and Mn/W(001) magnetic bilayers from first principles}},\ }\href {https://doi.org/10.1103/PhysRevB.90.174423} {\bibfield  {journal} {\bibinfo  {journal} {Phys. Rev. B}\ }\textbf {\bibinfo {volume} {90}},\ \bibinfo {pages} {174423} (\bibinfo {year} {2014})}\BibitemShut {NoStop}%
\bibitem [{\citenamefont {\ifmmode~\check{Z}\else \v{Z}\fi{}elezn\'y}\ \emph {et~al.}(2017)\citenamefont {\ifmmode~\check{Z}\else \v{Z}\fi{}elezn\'y}, \citenamefont {Zhang}, \citenamefont {Felser},\ and\ \citenamefont {Yan}}]{zelezny2017CIM}%
  \BibitemOpen
  \bibfield  {author} {\bibinfo {author} {\bibfnamefont {J.}~\bibnamefont {\ifmmode~\check{Z}\else \v{Z}\fi{}elezn\'y}}, \bibinfo {author} {\bibfnamefont {Y.}~\bibnamefont {Zhang}}, \bibinfo {author} {\bibfnamefont {C.}~\bibnamefont {Felser}},\ and\ \bibinfo {author} {\bibfnamefont {B.}~\bibnamefont {Yan}},\ }\bibfield  {title} {\bibinfo {title} {{Spin-Polarized Current in Noncollinear Antiferromagnets}},\ }\href {https://doi.org/10.1103/PhysRevLett.119.187204} {\bibfield  {journal} {\bibinfo  {journal} {Phys. Rev. Lett.}\ }\textbf {\bibinfo {volume} {119}},\ \bibinfo {pages} {187204} (\bibinfo {year} {2017})}\BibitemShut {NoStop}%
\bibitem [{\citenamefont {Johansson}(2024)}]{johansson2024}%
  \BibitemOpen
  \bibfield  {author} {\bibinfo {author} {\bibfnamefont {A.}~\bibnamefont {Johansson}},\ }\bibfield  {title} {\bibinfo {title} {{Theory of spin and orbital Edelstein effects}},\ }\href {https://doi.org/10.1088/1361-648X/ad5e2b} {\bibfield  {journal} {\bibinfo  {journal} {J. Phys.: Condens. Matter}\ }\textbf {\bibinfo {volume} {36}},\ \bibinfo {pages} {423002} (\bibinfo {year} {2024})}\BibitemShut {NoStop}%
\bibitem [{\citenamefont {Juraschek}\ \emph {et~al.}(2017)\citenamefont {Juraschek}, \citenamefont {Fechner}, \citenamefont {Balatsky},\ and\ \citenamefont {Spaldin}}]{juraschek2017_prm}%
  \BibitemOpen
  \bibfield  {author} {\bibinfo {author} {\bibfnamefont {D.~M.}\ \bibnamefont {Juraschek}}, \bibinfo {author} {\bibfnamefont {M.}~\bibnamefont {Fechner}}, \bibinfo {author} {\bibfnamefont {A.~V.}\ \bibnamefont {Balatsky}},\ and\ \bibinfo {author} {\bibfnamefont {N.~A.}\ \bibnamefont {Spaldin}},\ }\bibfield  {title} {\bibinfo {title} {{Dynamical multiferroicity}},\ }\href {https://doi.org/10.1103/PhysRevMaterials.1.014401} {\bibfield  {journal} {\bibinfo  {journal} {Phys. Rev. Mater.}\ }\textbf {\bibinfo {volume} {1}},\ \bibinfo {pages} {014401} (\bibinfo {year} {2017})}\BibitemShut {NoStop}%
\bibitem [{\citenamefont {Juraschek}\ and\ \citenamefont {Spaldin}(2019)}]{juraschek2019_prm}%
  \BibitemOpen
  \bibfield  {author} {\bibinfo {author} {\bibfnamefont {D.~M.}\ \bibnamefont {Juraschek}}\ and\ \bibinfo {author} {\bibfnamefont {N.~A.}\ \bibnamefont {Spaldin}},\ }\bibfield  {title} {\bibinfo {title} {{Orbital magnetic moments of phonons}},\ }\href {https://doi.org/10.1103/PhysRevMaterials.3.064405} {\bibfield  {journal} {\bibinfo  {journal} {Phys. Rev. Mater.}\ }\textbf {\bibinfo {volume} {3}},\ \bibinfo {pages} {064405} (\bibinfo {year} {2019})}\BibitemShut {NoStop}%
\bibitem [{\citenamefont {Lou}\ \emph {et~al.}(2021)\citenamefont {Lou}, \citenamefont {Katailiha}, \citenamefont {Bhardwaj}, \citenamefont {Beyermann}, \citenamefont {Juraschek},\ and\ \citenamefont {Kumar}}]{lou2021}%
  \BibitemOpen
  \bibfield  {author} {\bibinfo {author} {\bibfnamefont {P.~C.}\ \bibnamefont {Lou}}, \bibinfo {author} {\bibfnamefont {A.}~\bibnamefont {Katailiha}}, \bibinfo {author} {\bibfnamefont {R.~G.}\ \bibnamefont {Bhardwaj}}, \bibinfo {author} {\bibfnamefont {W.~P.}\ \bibnamefont {Beyermann}}, \bibinfo {author} {\bibfnamefont {D.~M.}\ \bibnamefont {Juraschek}},\ and\ \bibinfo {author} {\bibfnamefont {S.}~\bibnamefont {Kumar}},\ }\bibfield  {title} {\bibinfo {title} {{Large Magnetic Moment in Flexoelectronic Silicon at Room Temperature}},\ }\href {https://doi.org/10.1021/acs.nanolett.1c00052} {\bibfield  {journal} {\bibinfo  {journal} {Nano Lett.}\ }\textbf {\bibinfo {volume} {21}},\ \bibinfo {pages} {2939} (\bibinfo {year} {2021})}\BibitemShut {NoStop}%
\bibitem [{\citenamefont {Fang}\ \emph {et~al.}(2015)\citenamefont {Fang}, \citenamefont {Kuate~Defo}, \citenamefont {Shirodkar}, \citenamefont {Lieu}, \citenamefont {Tritsaris},\ and\ \citenamefont {Kaxiras}}]{fang2015}%
  \BibitemOpen
  \bibfield  {author} {\bibinfo {author} {\bibfnamefont {S.}~\bibnamefont {Fang}}, \bibinfo {author} {\bibfnamefont {R.}~\bibnamefont {Kuate~Defo}}, \bibinfo {author} {\bibfnamefont {S.~N.}\ \bibnamefont {Shirodkar}}, \bibinfo {author} {\bibfnamefont {S.}~\bibnamefont {Lieu}}, \bibinfo {author} {\bibfnamefont {G.~A.}\ \bibnamefont {Tritsaris}},\ and\ \bibinfo {author} {\bibfnamefont {E.}~\bibnamefont {Kaxiras}},\ }\bibfield  {title} {\bibinfo {title} {{Ab initio tight-binding Hamiltonian for transition metal dichalcogenides}},\ }\href {https://doi.org/10.1103/PhysRevB.92.205108} {\bibfield  {journal} {\bibinfo  {journal} {Phys. Rev. B}\ }\textbf {\bibinfo {volume} {92}},\ \bibinfo {pages} {205108} (\bibinfo {year} {2015})}\BibitemShut {NoStop}%
\bibitem [{\citenamefont {Feng}\ \emph {et~al.}(2020)\citenamefont {Feng}, \citenamefont {Liu}, \citenamefont {Zhou}, \citenamefont {Gao}, \citenamefont {Gao}, \citenamefont {Zhuang}, \citenamefont {Xu}, \citenamefont {Hu}, \citenamefont {Wang}, \citenamefont {Chen}, \citenamefont {Zhao}, \citenamefont {Dou},\ and\ \citenamefont {Du}}]{feng2020}%
  \BibitemOpen
  \bibfield  {author} {\bibinfo {author} {\bibfnamefont {H.}~\bibnamefont {Feng}}, \bibinfo {author} {\bibfnamefont {C.}~\bibnamefont {Liu}}, \bibinfo {author} {\bibfnamefont {S.}~\bibnamefont {Zhou}}, \bibinfo {author} {\bibfnamefont {N.}~\bibnamefont {Gao}}, \bibinfo {author} {\bibfnamefont {Q.}~\bibnamefont {Gao}}, \bibinfo {author} {\bibfnamefont {J.}~\bibnamefont {Zhuang}}, \bibinfo {author} {\bibfnamefont {X.}~\bibnamefont {Xu}}, \bibinfo {author} {\bibfnamefont {Z.}~\bibnamefont {Hu}}, \bibinfo {author} {\bibfnamefont {J.}~\bibnamefont {Wang}}, \bibinfo {author} {\bibfnamefont {L.}~\bibnamefont {Chen}}, \bibinfo {author} {\bibfnamefont {J.}~\bibnamefont {Zhao}}, \bibinfo {author} {\bibfnamefont {S.}~\bibnamefont {Dou}},\ and\ \bibinfo {author} {\bibfnamefont {Y.}~\bibnamefont {Du}},\ }\bibfield  {title} {\bibinfo {title} {{Experimental Realization of Two-Dimensional Buckled Lieb Lattice}},\ }\href {https://doi.org/10.1021/acs.nanolett.9b05316} {\bibfield  {journal} {\bibinfo  {journal} {Nano Lett.}\ }\textbf {\bibinfo {volume} {20}},\ \bibinfo {pages} {2537} (\bibinfo {year} {2020})}\BibitemShut {NoStop}%
\bibitem [{\citenamefont {Samy}\ \emph {et~al.}(2021)\citenamefont {Samy}, \citenamefont {Zeng}, \citenamefont {Birowosuto},\ and\ \citenamefont {El~Moutaouakil}}]{samy2021MoS2}%
  \BibitemOpen
  \bibfield  {author} {\bibinfo {author} {\bibfnamefont {O.}~\bibnamefont {Samy}}, \bibinfo {author} {\bibfnamefont {S.}~\bibnamefont {Zeng}}, \bibinfo {author} {\bibfnamefont {M.~D.}\ \bibnamefont {Birowosuto}},\ and\ \bibinfo {author} {\bibfnamefont {A.}~\bibnamefont {El~Moutaouakil}},\ }\bibfield  {title} {\bibinfo {title} {{A Review on MoS2 Properties, Synthesis, Sensing Applications and Challenges}},\ }\href {https://doi.org/10.3390/cryst11040355} {\bibfield  {journal} {\bibinfo  {journal} {Crystals}\ }\textbf {\bibinfo {volume} {11}},\ \bibinfo {pages} {355} (\bibinfo {year} {2021})}\BibitemShut {NoStop}%
\bibitem [{\citenamefont {Zhang}\ \emph {et~al.}(2024)\citenamefont {Zhang}, \citenamefont {Zhang}, \citenamefont {Guo}, \citenamefont {Li},\ and\ \citenamefont {Li}}]{zhang2024MoS2}%
  \BibitemOpen
  \bibfield  {author} {\bibinfo {author} {\bibfnamefont {Y.}~\bibnamefont {Zhang}}, \bibinfo {author} {\bibfnamefont {R.}~\bibnamefont {Zhang}}, \bibinfo {author} {\bibfnamefont {Y.}~\bibnamefont {Guo}}, \bibinfo {author} {\bibfnamefont {Y.}~\bibnamefont {Li}},\ and\ \bibinfo {author} {\bibfnamefont {K.}~\bibnamefont {Li}},\ }\bibfield  {title} {\bibinfo {title} {{A review on ${\rm MoS_2}$ structure, preparation, energy storage applications and challenges}},\ }\href {https://doi.org/https://doi.org/10.1016/j.jallcom.2024.174916} {\bibfield  {journal} {\bibinfo  {journal} {J. Alloy. Compd.}\ }\textbf {\bibinfo {volume} {998}},\ \bibinfo {pages} {174916} (\bibinfo {year} {2024})}\BibitemShut {NoStop}%
\bibitem [{\citenamefont {Liu}\ \emph {et~al.}(2013)\citenamefont {Liu}, \citenamefont {Shan}, \citenamefont {Yao}, \citenamefont {Yao},\ and\ \citenamefont {Xiao}}]{liu2013}%
  \BibitemOpen
  \bibfield  {author} {\bibinfo {author} {\bibfnamefont {G.-B.}\ \bibnamefont {Liu}}, \bibinfo {author} {\bibfnamefont {W.-Y.}\ \bibnamefont {Shan}}, \bibinfo {author} {\bibfnamefont {Y.}~\bibnamefont {Yao}}, \bibinfo {author} {\bibfnamefont {W.}~\bibnamefont {Yao}},\ and\ \bibinfo {author} {\bibfnamefont {D.}~\bibnamefont {Xiao}},\ }\bibfield  {title} {\bibinfo {title} {{Three-band tight-binding model for monolayers of group-VIB transition metal dichalcogenides}},\ }\href {https://doi.org/10.1103/PhysRevB.88.085433} {\bibfield  {journal} {\bibinfo  {journal} {Phys. Rev. B}\ }\textbf {\bibinfo {volume} {88}},\ \bibinfo {pages} {085433} (\bibinfo {year} {2013})}\BibitemShut {NoStop}%
\bibitem [{\citenamefont {Yan}\ \emph {et~al.}(2019)\citenamefont {Yan}, \citenamefont {Khalsa}, \citenamefont {Schaefer}, \citenamefont {Jarjour}, \citenamefont {Rouvimov}, \citenamefont {Nowack}, \citenamefont {Xing},\ and\ \citenamefont {Jena}}]{yan2019}%
  \BibitemOpen
  \bibfield  {author} {\bibinfo {author} {\bibfnamefont {R.}~\bibnamefont {Yan}}, \bibinfo {author} {\bibfnamefont {G.}~\bibnamefont {Khalsa}}, \bibinfo {author} {\bibfnamefont {B.~T.}\ \bibnamefont {Schaefer}}, \bibinfo {author} {\bibfnamefont {A.}~\bibnamefont {Jarjour}}, \bibinfo {author} {\bibfnamefont {S.}~\bibnamefont {Rouvimov}}, \bibinfo {author} {\bibfnamefont {K.~C.}\ \bibnamefont {Nowack}}, \bibinfo {author} {\bibfnamefont {H.~G.}\ \bibnamefont {Xing}},\ and\ \bibinfo {author} {\bibfnamefont {D.}~\bibnamefont {Jena}},\ }\bibfield  {title} {\bibinfo {title} {{Thickness dependence of superconductivity in ultrathin ${\rm NbS_2}$}},\ }\href {https://doi.org/10.7567/1882-0786/aaff89} {\bibfield  {journal} {\bibinfo  {journal} {Appl. Phys. Express}\ }\textbf {\bibinfo {volume} {12}},\ \bibinfo {pages} {023008} (\bibinfo {year} {2019})}\BibitemShut {NoStop}%
\bibitem [{\citenamefont {Chen}\ \emph {et~al.}(2025)\citenamefont {Chen}, \citenamefont {Ma}, \citenamefont {Wang}, \citenamefont {Li}, \citenamefont {Li},\ and\ \citenamefont {Bian}}]{chen2025}%
  \BibitemOpen
  \bibfield  {author} {\bibinfo {author} {\bibfnamefont {J.}~\bibnamefont {Chen}}, \bibinfo {author} {\bibfnamefont {Z.}~\bibnamefont {Ma}}, \bibinfo {author} {\bibfnamefont {D.}~\bibnamefont {Wang}}, \bibinfo {author} {\bibfnamefont {X.}~\bibnamefont {Li}}, \bibinfo {author} {\bibfnamefont {S.}~\bibnamefont {Li}},\ and\ \bibinfo {author} {\bibfnamefont {B.}~\bibnamefont {Bian}},\ }\bibfield  {title} {\bibinfo {title} {{Electrical contact between 2D material ${\rm NbS_2}$ and WSSe}},\ }\href {https://doi.org/https://doi.org/10.1016/j.physe.2025.116179} {\bibfield  {journal} {\bibinfo  {journal} {Physica E}\ }\textbf {\bibinfo {volume} {168}},\ \bibinfo {pages} {116179} (\bibinfo {year} {2025})}\BibitemShut {NoStop}%
\bibitem [{\citenamefont {Yin}\ \emph {et~al.}(2021)\citenamefont {Yin}, \citenamefont {Tan}, \citenamefont {Ding}, \citenamefont {Wen}, \citenamefont {Li}, \citenamefont {Teobaldi},\ and\ \citenamefont {Liu}}]{yin2021janus}%
  \BibitemOpen
  \bibfield  {author} {\bibinfo {author} {\bibfnamefont {W.-J.}\ \bibnamefont {Yin}}, \bibinfo {author} {\bibfnamefont {H.-J.}\ \bibnamefont {Tan}}, \bibinfo {author} {\bibfnamefont {P.-J.}\ \bibnamefont {Ding}}, \bibinfo {author} {\bibfnamefont {B.}~\bibnamefont {Wen}}, \bibinfo {author} {\bibfnamefont {X.-B.}\ \bibnamefont {Li}}, \bibinfo {author} {\bibfnamefont {G.}~\bibnamefont {Teobaldi}},\ and\ \bibinfo {author} {\bibfnamefont {L.-M.}\ \bibnamefont {Liu}},\ }\bibfield  {title} {\bibinfo {title} {{Recent advances in low-dimensional Janus materials: theoretical and simulation perspectives}},\ }\href {https://doi.org/10.1039/D1MA00660F} {\bibfield  {journal} {\bibinfo  {journal} {Mater. Adv.}\ }\textbf {\bibinfo {volume} {2}},\ \bibinfo {pages} {7543} (\bibinfo {year} {2021})}\BibitemShut {NoStop}%
\bibitem [{\citenamefont {Li}\ \emph {et~al.}(2018)\citenamefont {Li}, \citenamefont {Cheng},\ and\ \citenamefont {Huang}}]{li2018janus}%
  \BibitemOpen
  \bibfield  {author} {\bibinfo {author} {\bibfnamefont {R.}~\bibnamefont {Li}}, \bibinfo {author} {\bibfnamefont {Y.}~\bibnamefont {Cheng}},\ and\ \bibinfo {author} {\bibfnamefont {W.}~\bibnamefont {Huang}},\ }\bibfield  {title} {\bibinfo {title} {{Recent Progress of Janus 2D Transition Metal Chalcogenides: From Theory to Experiments}},\ }\href {https://doi.org/https://doi.org/10.1002/smll.201802091} {\bibfield  {journal} {\bibinfo  {journal} {Small}\ }\textbf {\bibinfo {volume} {14}},\ \bibinfo {pages} {1802091} (\bibinfo {year} {2018})}\BibitemShut {NoStop}%
\bibitem [{\citenamefont {Lu}\ \emph {et~al.}(2017)\citenamefont {Lu}, \citenamefont {Zhu}, \citenamefont {Xiao}, \citenamefont {Chuu}, \citenamefont {Han}, \citenamefont {Chiu}, \citenamefont {Cheng}, \citenamefont {Yang}, \citenamefont {Wei}, \citenamefont {Yang}, \citenamefont {Wang}, \citenamefont {Sokaras}, \citenamefont {Nordlund}, \citenamefont {Yang}, \citenamefont {Muller}, \citenamefont {Chou}, \citenamefont {Zhang},\ and\ \citenamefont {Li}}]{lu2017}%
  \BibitemOpen
  \bibfield  {author} {\bibinfo {author} {\bibfnamefont {A.-Y.}\ \bibnamefont {Lu}}, \bibinfo {author} {\bibfnamefont {H.}~\bibnamefont {Zhu}}, \bibinfo {author} {\bibfnamefont {J.}~\bibnamefont {Xiao}}, \bibinfo {author} {\bibfnamefont {C.-P.}\ \bibnamefont {Chuu}}, \bibinfo {author} {\bibfnamefont {Y.}~\bibnamefont {Han}}, \bibinfo {author} {\bibfnamefont {M.-H.}\ \bibnamefont {Chiu}}, \bibinfo {author} {\bibfnamefont {C.-C.}\ \bibnamefont {Cheng}}, \bibinfo {author} {\bibfnamefont {C.-W.}\ \bibnamefont {Yang}}, \bibinfo {author} {\bibfnamefont {K.-H.}\ \bibnamefont {Wei}}, \bibinfo {author} {\bibfnamefont {Y.}~\bibnamefont {Yang}}, \bibinfo {author} {\bibfnamefont {Y.}~\bibnamefont {Wang}}, \bibinfo {author} {\bibfnamefont {D.}~\bibnamefont {Sokaras}}, \bibinfo {author} {\bibfnamefont {D.}~\bibnamefont {Nordlund}}, \bibinfo {author} {\bibfnamefont {P.}~\bibnamefont {Yang}}, \bibinfo {author} {\bibfnamefont {D.}~\bibnamefont {Muller}}, \bibinfo {author} {\bibfnamefont {M.-Y.}\ \bibnamefont {Chou}}, \bibinfo {author} {\bibfnamefont {X.}~\bibnamefont {Zhang}},\ and\ \bibinfo {author} {\bibfnamefont {L.-J.}\ \bibnamefont {Li}},\ }\bibfield  {title} {\bibinfo {title} {{Janus monolayers of transition metal dichalcogenides}},\ }\href {https://doi.org/10.1038/nnano.2017.100} {\bibfield  {journal} {\bibinfo  {journal} {Nat. Nanotechnol.}\ }\textbf {\bibinfo {volume} {12}},\ \bibinfo {pages} {744} (\bibinfo {year} {2017})}\BibitemShut {NoStop}%
\bibitem [{\citenamefont {Giannozzi}\ \emph {et~al.}(2017)\citenamefont {Giannozzi}, \citenamefont {Andreussi}, \citenamefont {Brumme}, \citenamefont {Bunau}, \citenamefont {Buongiorno~Nardelli}, \citenamefont {Calandra}, \citenamefont {Car}, \citenamefont {Cavazzoni}, \citenamefont {Ceresoli}, \citenamefont {Cococcioni}, \citenamefont {Colonna}, \citenamefont {Carnimeo}, \citenamefont {Dal~Corso}, \citenamefont {de~Gironcoli}, \citenamefont {Delugas}, \citenamefont {DiStasio}, \citenamefont {Ferretti}, \citenamefont {Floris}, \citenamefont {Fratesi}, \citenamefont {Fugallo}, \citenamefont {Gebauer}, \citenamefont {Gerstmann}, \citenamefont {Giustino}, \citenamefont {Gorni}, \citenamefont {Jia}, \citenamefont {Kawamura}, \citenamefont {Ko}, \citenamefont {Kokalj}, \citenamefont {K^^c3^^bc^^c3^^a7^^c3^^bckbenli}, \citenamefont {Lazzeri}, \citenamefont {Marsili}, \citenamefont {Marzari}, \citenamefont {Mauri}, \citenamefont {Nguyen}, \citenamefont {Nguyen}, \citenamefont {Otero-de-la Roza}, \citenamefont {Paulatto}, \citenamefont {Ponc^^c3^^a9}, \citenamefont {Rocca}, \citenamefont {Sabatini}, \citenamefont {Santra}, \citenamefont {Schlipf}, \citenamefont {Seitsonen}, \citenamefont {Smogunov}, \citenamefont {Timrov}, \citenamefont {Thonhauser}, \citenamefont {Umari}, \citenamefont {Vast}, \citenamefont {Wu},\ and\ \citenamefont {Baroni}}]{giannozzi_2017}%
  \BibitemOpen
  \bibfield  {author} {\bibinfo {author} {\bibfnamefont {P.}~\bibnamefont {Giannozzi}}, \bibinfo {author} {\bibfnamefont {O.}~\bibnamefont {Andreussi}}, \bibinfo {author} {\bibfnamefont {T.}~\bibnamefont {Brumme}}, \bibinfo {author} {\bibfnamefont {O.}~\bibnamefont {Bunau}}, \bibinfo {author} {\bibfnamefont {M.}~\bibnamefont {Buongiorno~Nardelli}}, \bibinfo {author} {\bibfnamefont {M.}~\bibnamefont {Calandra}}, \bibinfo {author} {\bibfnamefont {R.}~\bibnamefont {Car}}, \bibinfo {author} {\bibfnamefont {C.}~\bibnamefont {Cavazzoni}}, \bibinfo {author} {\bibfnamefont {D.}~\bibnamefont {Ceresoli}}, \bibinfo {author} {\bibfnamefont {M.}~\bibnamefont {Cococcioni}}, \bibinfo {author} {\bibfnamefont {N.}~\bibnamefont {Colonna}}, \bibinfo {author} {\bibfnamefont {I.}~\bibnamefont {Carnimeo}}, \bibinfo {author} {\bibfnamefont {A.}~\bibnamefont {Dal~Corso}}, \bibinfo {author} {\bibfnamefont {S.}~\bibnamefont {de~Gironcoli}}, \bibinfo {author} {\bibfnamefont {P.}~\bibnamefont {Delugas}}, \bibinfo {author} {\bibfnamefont {R.~A.}\ \bibnamefont {DiStasio}}, \bibinfo {author} {\bibfnamefont {A.}~\bibnamefont {Ferretti}}, \bibinfo {author} {\bibfnamefont {A.}~\bibnamefont {Floris}}, \bibinfo {author} {\bibfnamefont {G.}~\bibnamefont {Fratesi}}, \bibinfo {author} {\bibfnamefont {G.}~\bibnamefont {Fugallo}}, \bibinfo {author} {\bibfnamefont {R.}~\bibnamefont {Gebauer}}, \bibinfo {author} {\bibfnamefont {U.}~\bibnamefont {Gerstmann}}, \bibinfo {author} {\bibfnamefont {F.}~\bibnamefont {Giustino}}, \bibinfo {author} {\bibfnamefont {T.}~\bibnamefont {Gorni}}, \bibinfo {author} {\bibfnamefont {J.}~\bibnamefont {Jia}}, \bibinfo {author} {\bibfnamefont {M.}~\bibnamefont {Kawamura}}, \bibinfo {author} {\bibfnamefont {H.-Y.}\ \bibnamefont {Ko}}, \bibinfo {author} {\bibfnamefont {A.}~\bibnamefont {Kokalj}}, \bibinfo {author} {\bibfnamefont {E.}~\bibnamefont {K^^c3^^bc^^c3^^a7^^c3^^bckbenli}}, \bibinfo {author} {\bibfnamefont {M.}~\bibnamefont {Lazzeri}}, \bibinfo {author} {\bibfnamefont {M.}~\bibnamefont {Marsili}}, \bibinfo {author} {\bibfnamefont {N.}~\bibnamefont {Marzari}}, \bibinfo {author} {\bibfnamefont {F.}~\bibnamefont {Mauri}}, \bibinfo {author} {\bibfnamefont {N.~L.}\ \bibnamefont {Nguyen}}, \bibinfo {author} {\bibfnamefont {H.-V.}\ \bibnamefont {Nguyen}}, \bibinfo {author} {\bibfnamefont {A.}~\bibnamefont {Otero-de-la Roza}}, \bibinfo {author} {\bibfnamefont {L.}~\bibnamefont {Paulatto}}, \bibinfo {author} {\bibfnamefont {S.}~\bibnamefont {Ponc^^c3^^a9}}, \bibinfo {author} {\bibfnamefont {D.}~\bibnamefont {Rocca}}, \bibinfo {author} {\bibfnamefont {R.}~\bibnamefont {Sabatini}}, \bibinfo {author} {\bibfnamefont {B.}~\bibnamefont {Santra}}, \bibinfo {author} {\bibfnamefont {M.}~\bibnamefont {Schlipf}}, \bibinfo {author} {\bibfnamefont {A.~P.}\ \bibnamefont {Seitsonen}}, \bibinfo {author} {\bibfnamefont {A.}~\bibnamefont {Smogunov}}, \bibinfo {author} {\bibfnamefont {I.}~\bibnamefont {Timrov}}, \bibinfo {author} {\bibfnamefont {T.}~\bibnamefont {Thonhauser}}, \bibinfo {author} {\bibfnamefont {P.}~\bibnamefont {Umari}}, \bibinfo {author} {\bibfnamefont {N.}~\bibnamefont {Vast}}, \bibinfo {author} {\bibfnamefont {X.}~\bibnamefont {Wu}},\ and\ \bibinfo {author} {\bibfnamefont {S.}~\bibnamefont {Baroni}},\ }\bibfield  {title} {\bibinfo {title} {{Advanced capabilities for materials modelling with Quantum ESPRESSO}},\ }\href {https://doi.org/10.1088/1361-648X/aa8f79} {\bibfield  {journal} {\bibinfo  {journal} {J. Phys.: Condens. Matter}\ }\textbf {\bibinfo {volume} {29}},\ \bibinfo {pages} {465901} (\bibinfo {year} {2017})}\BibitemShut {NoStop}%
\bibitem [{\citenamefont {Cheng}\ \emph {et~al.}(2013)\citenamefont {Cheng}, \citenamefont {Zhu}, \citenamefont {Tahir},\ and\ \citenamefont {Schwingenschl^^c3^^b6gl}}]{cheng2013}%
  \BibitemOpen
  \bibfield  {author} {\bibinfo {author} {\bibfnamefont {Y.~C.}\ \bibnamefont {Cheng}}, \bibinfo {author} {\bibfnamefont {Z.~Y.}\ \bibnamefont {Zhu}}, \bibinfo {author} {\bibfnamefont {M.}~\bibnamefont {Tahir}},\ and\ \bibinfo {author} {\bibfnamefont {U.}~\bibnamefont {Schwingenschl^^c3^^b6gl}},\ }\bibfield  {title} {\bibinfo {title} {{Spin-orbit^^e2^^80^^93induced spin splittings in polar transition metal dichalcogenide monolayers}},\ }\href {https://doi.org/10.1209/0295-5075/102/57001} {\bibfield  {journal} {\bibinfo  {journal} {Europhys. Lett.}\ }\textbf {\bibinfo {volume} {102}},\ \bibinfo {pages} {57001} (\bibinfo {year} {2013})}\BibitemShut {NoStop}%
\bibitem [{\citenamefont {Mostofi}\ \emph {et~al.}(2008)\citenamefont {Mostofi}, \citenamefont {Yates}, \citenamefont {Lee}, \citenamefont {Souza}, \citenamefont {Vanderbilt},\ and\ \citenamefont {Marzari}}]{wannier90}%
  \BibitemOpen
  \bibfield  {author} {\bibinfo {author} {\bibfnamefont {A.~A.}\ \bibnamefont {Mostofi}}, \bibinfo {author} {\bibfnamefont {J.~R.}\ \bibnamefont {Yates}}, \bibinfo {author} {\bibfnamefont {Y.-S.}\ \bibnamefont {Lee}}, \bibinfo {author} {\bibfnamefont {I.}~\bibnamefont {Souza}}, \bibinfo {author} {\bibfnamefont {D.}~\bibnamefont {Vanderbilt}},\ and\ \bibinfo {author} {\bibfnamefont {N.}~\bibnamefont {Marzari}},\ }\bibfield  {title} {\bibinfo {title} {{wannier90: A tool for obtaining maximally-localised Wannier functions}},\ }\href {https://doi.org/https://doi.org/10.1016/j.cpc.2007.11.016} {\bibfield  {journal} {\bibinfo  {journal} {Comput. Phys. Commun.}\ }\textbf {\bibinfo {volume} {178}},\ \bibinfo {pages} {685} (\bibinfo {year} {2008})}\BibitemShut {NoStop}%
\bibitem [{\citenamefont {Pizzi}\ \emph {et~al.}(2020)\citenamefont {Pizzi}, \citenamefont {Vitale}, \citenamefont {Arita}, \citenamefont {Bl^^c3^^bcgel}, \citenamefont {Freimuth}, \citenamefont {G^^c3^^a9ranton}, \citenamefont {Gibertini}, \citenamefont {Gresch}, \citenamefont {Johnson}, \citenamefont {Koretsune}, \citenamefont {Iba^^c3^^b1ez-Azpiroz}, \citenamefont {Lee}, \citenamefont {Lihm}, \citenamefont {Marchand}, \citenamefont {Marrazzo}, \citenamefont {Mokrousov}, \citenamefont {Mustafa}, \citenamefont {Nohara}, \citenamefont {Nomura}, \citenamefont {Paulatto}, \citenamefont {Ponc^^c3^^a9}, \citenamefont {Ponweiser}, \citenamefont {Qiao}, \citenamefont {Th^^c3^^b6le}, \citenamefont {Tsirkin}, \citenamefont {Wierzbowska}, \citenamefont {Marzari}, \citenamefont {Vanderbilt}, \citenamefont {Souza}, \citenamefont {Mostofi},\ and\ \citenamefont {Yates}}]{pizzi_2020}%
  \BibitemOpen
  \bibfield  {author} {\bibinfo {author} {\bibfnamefont {G.}~\bibnamefont {Pizzi}}, \bibinfo {author} {\bibfnamefont {V.}~\bibnamefont {Vitale}}, \bibinfo {author} {\bibfnamefont {R.}~\bibnamefont {Arita}}, \bibinfo {author} {\bibfnamefont {S.}~\bibnamefont {Bl^^c3^^bcgel}}, \bibinfo {author} {\bibfnamefont {F.}~\bibnamefont {Freimuth}}, \bibinfo {author} {\bibfnamefont {G.}~\bibnamefont {G^^c3^^a9ranton}}, \bibinfo {author} {\bibfnamefont {M.}~\bibnamefont {Gibertini}}, \bibinfo {author} {\bibfnamefont {D.}~\bibnamefont {Gresch}}, \bibinfo {author} {\bibfnamefont {C.}~\bibnamefont {Johnson}}, \bibinfo {author} {\bibfnamefont {T.}~\bibnamefont {Koretsune}}, \bibinfo {author} {\bibfnamefont {J.}~\bibnamefont {Iba^^c3^^b1ez-Azpiroz}}, \bibinfo {author} {\bibfnamefont {H.}~\bibnamefont {Lee}}, \bibinfo {author} {\bibfnamefont {J.-M.}\ \bibnamefont {Lihm}}, \bibinfo {author} {\bibfnamefont {D.}~\bibnamefont {Marchand}}, \bibinfo {author} {\bibfnamefont {A.}~\bibnamefont {Marrazzo}}, \bibinfo {author} {\bibfnamefont {Y.}~\bibnamefont {Mokrousov}}, \bibinfo {author} {\bibfnamefont {J.~I.}\ \bibnamefont {Mustafa}}, \bibinfo {author} {\bibfnamefont {Y.}~\bibnamefont {Nohara}}, \bibinfo {author} {\bibfnamefont {Y.}~\bibnamefont {Nomura}}, \bibinfo {author} {\bibfnamefont {L.}~\bibnamefont {Paulatto}}, \bibinfo {author} {\bibfnamefont {S.}~\bibnamefont {Ponc^^c3^^a9}}, \bibinfo {author} {\bibfnamefont {T.}~\bibnamefont {Ponweiser}}, \bibinfo {author} {\bibfnamefont {J.}~\bibnamefont {Qiao}}, \bibinfo {author} {\bibfnamefont {F.}~\bibnamefont {Th^^c3^^b6le}}, \bibinfo {author} {\bibfnamefont {S.~S.}\ \bibnamefont {Tsirkin}}, \bibinfo {author} {\bibfnamefont {M.}~\bibnamefont {Wierzbowska}}, \bibinfo {author} {\bibfnamefont {N.}~\bibnamefont {Marzari}}, \bibinfo {author} {\bibfnamefont {D.}~\bibnamefont {Vanderbilt}}, \bibinfo {author} {\bibfnamefont {I.}~\bibnamefont {Souza}}, \bibinfo {author} {\bibfnamefont {A.~A.}\ \bibnamefont {Mostofi}},\ and\ \bibinfo {author} {\bibfnamefont {J.~R.}\ \bibnamefont {Yates}},\ }\bibfield  {title} {\bibinfo {title} {{Wannier90 as a community code: new features and applications}},\ }\href {https://doi.org/10.1088/1361-648X/ab51ff} {\bibfield  {journal} {\bibinfo  {journal} {J. Phys.: Condens. Matter}\ }\textbf {\bibinfo {volume} {32}},\ \bibinfo {pages} {165902} (\bibinfo {year} {2020})}\BibitemShut {NoStop}%
\bibitem [{\citenamefont {Wilson}\ and\ \citenamefont {Feher}(1961)}]{wilson1961}%
  \BibitemOpen
  \bibfield  {author} {\bibinfo {author} {\bibfnamefont {D.~K.}\ \bibnamefont {Wilson}}\ and\ \bibinfo {author} {\bibfnamefont {G.}~\bibnamefont {Feher}},\ }\bibfield  {title} {\bibinfo {title} {{Electron Spin Resonance Experiments on Donors in Silicon. III. Investigation of Excited States by the Application of Uniaxial Stress and Their Importance in Relaxation Processes}},\ }\href {https://doi.org/10.1103/PhysRev.124.1068} {\bibfield  {journal} {\bibinfo  {journal} {Phys. Rev.}\ }\textbf {\bibinfo {volume} {124}},\ \bibinfo {pages} {1068} (\bibinfo {year} {1961})}\BibitemShut {NoStop}%
\bibitem [{\citenamefont {Thonhauser}\ \emph {et~al.}(2005)\citenamefont {Thonhauser}, \citenamefont {Ceresoli}, \citenamefont {Vanderbilt},\ and\ \citenamefont {Resta}}]{thonhauser2005}%
  \BibitemOpen
  \bibfield  {author} {\bibinfo {author} {\bibfnamefont {T.}~\bibnamefont {Thonhauser}}, \bibinfo {author} {\bibfnamefont {D.}~\bibnamefont {Ceresoli}}, \bibinfo {author} {\bibfnamefont {D.}~\bibnamefont {Vanderbilt}},\ and\ \bibinfo {author} {\bibfnamefont {R.}~\bibnamefont {Resta}},\ }\bibfield  {title} {\bibinfo {title} {{Orbital Magnetization in Periodic Insulators}},\ }\href {https://doi.org/10.1103/PhysRevLett.95.137205} {\bibfield  {journal} {\bibinfo  {journal} {Phys. Rev. Lett.}\ }\textbf {\bibinfo {volume} {95}},\ \bibinfo {pages} {137205} (\bibinfo {year} {2005})}\BibitemShut {NoStop}%
\bibitem [{\citenamefont {Xiao}\ \emph {et~al.}(2005)\citenamefont {Xiao}, \citenamefont {Shi},\ and\ \citenamefont {Niu}}]{xiao2005}%
  \BibitemOpen
  \bibfield  {author} {\bibinfo {author} {\bibfnamefont {D.}~\bibnamefont {Xiao}}, \bibinfo {author} {\bibfnamefont {J.}~\bibnamefont {Shi}},\ and\ \bibinfo {author} {\bibfnamefont {Q.}~\bibnamefont {Niu}},\ }\bibfield  {title} {\bibinfo {title} {{Berry Phase Correction to Electron Density of States in Solids}},\ }\href {https://doi.org/10.1103/PhysRevLett.95.137204} {\bibfield  {journal} {\bibinfo  {journal} {Phys. Rev. Lett.}\ }\textbf {\bibinfo {volume} {95}},\ \bibinfo {pages} {137204} (\bibinfo {year} {2005})}\BibitemShut {NoStop}%
\bibitem [{\citenamefont {Shi}\ \emph {et~al.}(2007)\citenamefont {Shi}, \citenamefont {Vignale}, \citenamefont {Xiao},\ and\ \citenamefont {Niu}}]{shi2007}%
  \BibitemOpen
  \bibfield  {author} {\bibinfo {author} {\bibfnamefont {J.}~\bibnamefont {Shi}}, \bibinfo {author} {\bibfnamefont {G.}~\bibnamefont {Vignale}}, \bibinfo {author} {\bibfnamefont {D.}~\bibnamefont {Xiao}},\ and\ \bibinfo {author} {\bibfnamefont {Q.}~\bibnamefont {Niu}},\ }\bibfield  {title} {\bibinfo {title} {{Quantum Theory of Orbital Magnetization and Its Generalization to Interacting Systems}},\ }\href {https://doi.org/10.1103/PhysRevLett.99.197202} {\bibfield  {journal} {\bibinfo  {journal} {Phys. Rev. Lett.}\ }\textbf {\bibinfo {volume} {99}},\ \bibinfo {pages} {197202} (\bibinfo {year} {2007})}\BibitemShut {NoStop}%
\bibitem [{\citenamefont {Xiao}\ \emph {et~al.}(2010)\citenamefont {Xiao}, \citenamefont {Chang},\ and\ \citenamefont {Niu}}]{xiao2010}%
  \BibitemOpen
  \bibfield  {author} {\bibinfo {author} {\bibfnamefont {D.}~\bibnamefont {Xiao}}, \bibinfo {author} {\bibfnamefont {M.-C.}\ \bibnamefont {Chang}},\ and\ \bibinfo {author} {\bibfnamefont {Q.}~\bibnamefont {Niu}},\ }\bibfield  {title} {\bibinfo {title} {{Berry phase effects on electronic properties}},\ }\href {https://doi.org/10.1103/RevModPhys.82.1959} {\bibfield  {journal} {\bibinfo  {journal} {Rev. Mod. Phys.}\ }\textbf {\bibinfo {volume} {82}},\ \bibinfo {pages} {1959} (\bibinfo {year} {2010})}\BibitemShut {NoStop}%
\bibitem [{\citenamefont {Fuchs}\ \emph {et~al.}(2010)\citenamefont {Fuchs}, \citenamefont {Pi^^c3^^a9chon}, \citenamefont {Goerbig},\ and\ \citenamefont {Montambaux}}]{fuchs2010}%
  \BibitemOpen
  \bibfield  {author} {\bibinfo {author} {\bibfnamefont {J.}~\bibnamefont {Fuchs}}, \bibinfo {author} {\bibfnamefont {F.}~\bibnamefont {Pi^^c3^^a9chon}}, \bibinfo {author} {\bibfnamefont {M.}~\bibnamefont {Goerbig}},\ and\ \bibinfo {author} {\bibfnamefont {G.}~\bibnamefont {Montambaux}},\ }\bibfield  {title} {\bibinfo {title} {{Topological Berry phase and semiclassical quantization of cyclotron orbits for two dimensional electrons in coupled band models}},\ }\href {https://doi.org/10.1140/epjb/e2010-00259-2} {\bibfield  {journal} {\bibinfo  {journal} {Eur. Phys. J. B}\ }\textbf {\bibinfo {volume} {77}},\ \bibinfo {pages} {351} (\bibinfo {year} {2010})}\BibitemShut {NoStop}%
\bibitem [{\citenamefont {Osumi}\ \emph {et~al.}(2021)\citenamefont {Osumi}, \citenamefont {Zhang},\ and\ \citenamefont {Murakami}}]{osumi2021}%
  \BibitemOpen
  \bibfield  {author} {\bibinfo {author} {\bibfnamefont {K.}~\bibnamefont {Osumi}}, \bibinfo {author} {\bibfnamefont {T.}~\bibnamefont {Zhang}},\ and\ \bibinfo {author} {\bibfnamefont {S.}~\bibnamefont {Murakami}},\ }\bibfield  {title} {\bibinfo {title} {{Kinetic magnetoelectric effect in topological insulators}},\ }\href {https://doi.org/10.1038/s42005-021-00702-4} {\bibfield  {journal} {\bibinfo  {journal} {Commun. Phys.}\ }\textbf {\bibinfo {volume} {4}},\ \bibinfo {pages} {211} (\bibinfo {year} {2021})}\BibitemShut {NoStop}%
\bibitem [{\citenamefont {Pezo}\ \emph {et~al.}(2022)\citenamefont {Pezo}, \citenamefont {Garc\'{\i}a~Ovalle},\ and\ \citenamefont {Manchon}}]{pezo2022}%
  \BibitemOpen
  \bibfield  {author} {\bibinfo {author} {\bibfnamefont {A.}~\bibnamefont {Pezo}}, \bibinfo {author} {\bibfnamefont {D.}~\bibnamefont {Garc\'{\i}a~Ovalle}},\ and\ \bibinfo {author} {\bibfnamefont {A.}~\bibnamefont {Manchon}},\ }\bibfield  {title} {\bibinfo {title} {{Orbital Hall effect in crystals: Interatomic versus intra-atomic contributions}},\ }\href {https://doi.org/10.1103/PhysRevB.106.104414} {\bibfield  {journal} {\bibinfo  {journal} {Phys. Rev. B}\ }\textbf {\bibinfo {volume} {106}},\ \bibinfo {pages} {104414} (\bibinfo {year} {2022})}\BibitemShut {NoStop}%
\bibitem [{\citenamefont {Bhowal}\ and\ \citenamefont {Vignale}(2021)}]{bhowal2021}%
  \BibitemOpen
  \bibfield  {author} {\bibinfo {author} {\bibfnamefont {S.}~\bibnamefont {Bhowal}}\ and\ \bibinfo {author} {\bibfnamefont {G.}~\bibnamefont {Vignale}},\ }\bibfield  {title} {\bibinfo {title} {{Orbital Hall effect as an alternative to valley Hall effect in gapped graphene}},\ }\href {https://doi.org/10.1103/PhysRevB.103.195309} {\bibfield  {journal} {\bibinfo  {journal} {Phys. Rev. B}\ }\textbf {\bibinfo {volume} {103}},\ \bibinfo {pages} {195309} (\bibinfo {year} {2021})}\BibitemShut {NoStop}%
\bibitem [{\citenamefont {L\"auchli}\ \emph {et~al.}(2006)\citenamefont {L\"auchli}, \citenamefont {Mila},\ and\ \citenamefont {Penc}}]{Lauchli2006}%
  \BibitemOpen
  \bibfield  {author} {\bibinfo {author} {\bibfnamefont {A.}~\bibnamefont {L\"auchli}}, \bibinfo {author} {\bibfnamefont {F.}~\bibnamefont {Mila}},\ and\ \bibinfo {author} {\bibfnamefont {K.}~\bibnamefont {Penc}},\ }\bibfield  {title} {\bibinfo {title} {{Quadrupolar {P}hases of the ${S=1}$ {B}ilinear-{B}iquadratic {H}eisenberg {M}odel on the {T}riangular {L}attice}},\ }\href {https://doi.org/10.1103/PhysRevLett.97.087205} {\bibfield  {journal} {\bibinfo  {journal} {Phys. Rev. Lett.}\ }\textbf {\bibinfo {volume} {97}},\ \bibinfo {pages} {087205} (\bibinfo {year} {2006})}\BibitemShut {NoStop}%
\bibitem [{\citenamefont {Tsunetsugu}\ and\ \citenamefont {Arikawa}(2006)}]{tsunetsugu2006}%
  \BibitemOpen
  \bibfield  {author} {\bibinfo {author} {\bibfnamefont {H.}~\bibnamefont {Tsunetsugu}}\ and\ \bibinfo {author} {\bibfnamefont {M.}~\bibnamefont {Arikawa}},\ }\bibfield  {title} {\bibinfo {title} {{Spin Nematic Phase in S=1 Triangular Antiferromagnets}},\ }\href {https://doi.org/10.1143/JPSJ.75.083701} {\bibfield  {journal} {\bibinfo  {journal} {J. Phys. Soc. Jpn.}\ }\textbf {\bibinfo {volume} {75}},\ \bibinfo {pages} {083701} (\bibinfo {year} {2006})}\BibitemShut {NoStop}%
\bibitem [{\citenamefont {Martin}(1972)}]{martin1972}%
  \BibitemOpen
  \bibfield  {author} {\bibinfo {author} {\bibfnamefont {R.~M.}\ \bibnamefont {Martin}},\ }\bibfield  {title} {\bibinfo {title} {Piezoelectricity},\ }\href {https://doi.org/10.1103/PhysRevB.5.1607} {\bibfield  {journal} {\bibinfo  {journal} {Phys. Rev. B}\ }\textbf {\bibinfo {volume} {5}},\ \bibinfo {pages} {1607} (\bibinfo {year} {1972})}\BibitemShut {NoStop}%
\bibitem [{\citenamefont {Baroni}\ \emph {et~al.}(2001)\citenamefont {Baroni}, \citenamefont {de~Gironcoli}, \citenamefont {Dal~Corso},\ and\ \citenamefont {Giannozzi}}]{baroni2001}%
  \BibitemOpen
  \bibfield  {author} {\bibinfo {author} {\bibfnamefont {S.}~\bibnamefont {Baroni}}, \bibinfo {author} {\bibfnamefont {S.}~\bibnamefont {de~Gironcoli}}, \bibinfo {author} {\bibfnamefont {A.}~\bibnamefont {Dal~Corso}},\ and\ \bibinfo {author} {\bibfnamefont {P.}~\bibnamefont {Giannozzi}},\ }\bibfield  {title} {\bibinfo {title} {{Phonons and related crystal properties from density-functional perturbation theory}},\ }\href {https://doi.org/10.1103/RevModPhys.73.515} {\bibfield  {journal} {\bibinfo  {journal} {Rev. Mod. Phys.}\ }\textbf {\bibinfo {volume} {73}},\ \bibinfo {pages} {515} (\bibinfo {year} {2001})}\BibitemShut {NoStop}%
\bibitem [{\citenamefont {Ueda}\ \emph {et~al.}(2023)\citenamefont {Ueda}, \citenamefont {Garc^^c3^^ada-Fern^^c3^^a1ndez}, \citenamefont {Agrestini}, \citenamefont {Romao}, \citenamefont {van~den Brink}, \citenamefont {Spaldin}, \citenamefont {Zhou},\ and\ \citenamefont {Staub}}]{ueda2023}%
  \BibitemOpen
  \bibfield  {author} {\bibinfo {author} {\bibfnamefont {H.}~\bibnamefont {Ueda}}, \bibinfo {author} {\bibfnamefont {M.}~\bibnamefont {Garc^^c3^^ada-Fern^^c3^^a1ndez}}, \bibinfo {author} {\bibfnamefont {S.}~\bibnamefont {Agrestini}}, \bibinfo {author} {\bibfnamefont {C.}~\bibnamefont {Romao}}, \bibinfo {author} {\bibfnamefont {J.}~\bibnamefont {van~den Brink}}, \bibinfo {author} {\bibfnamefont {N.}~\bibnamefont {Spaldin}}, \bibinfo {author} {\bibfnamefont {K.-J.}\ \bibnamefont {Zhou}},\ and\ \bibinfo {author} {\bibfnamefont {U.}~\bibnamefont {Staub}},\ }\bibfield  {title} {\bibinfo {title} {{Chiral phonons in quartz probed by X-rays}},\ }\href {https://doi.org/10.1038/s41586-023-06016-5} {\bibfield  {journal} {\bibinfo  {journal} {Nature}\ }\textbf {\bibinfo {volume} {618}},\ \bibinfo {pages} {946} (\bibinfo {year} {2023})}\BibitemShut {NoStop}%
\bibitem [{\citenamefont {Ohe}\ \emph {et~al.}(2024)\citenamefont {Ohe}, \citenamefont {Shishido}, \citenamefont {Kato}, \citenamefont {Utsumi}, \citenamefont {Matsuura},\ and\ \citenamefont {Togawa}}]{ohe2024}%
  \BibitemOpen
  \bibfield  {author} {\bibinfo {author} {\bibfnamefont {K.}~\bibnamefont {Ohe}}, \bibinfo {author} {\bibfnamefont {H.}~\bibnamefont {Shishido}}, \bibinfo {author} {\bibfnamefont {M.}~\bibnamefont {Kato}}, \bibinfo {author} {\bibfnamefont {S.}~\bibnamefont {Utsumi}}, \bibinfo {author} {\bibfnamefont {H.}~\bibnamefont {Matsuura}},\ and\ \bibinfo {author} {\bibfnamefont {Y.}~\bibnamefont {Togawa}},\ }\bibfield  {title} {\bibinfo {title} {{Chirality-Induced Selectivity of Phonon Angular Momenta in Chiral Quartz Crystals}},\ }\href {https://doi.org/10.1103/PhysRevLett.132.056302} {\bibfield  {journal} {\bibinfo  {journal} {Phys. Rev. Lett.}\ }\textbf {\bibinfo {volume} {132}},\ \bibinfo {pages} {056302} (\bibinfo {year} {2024})}\BibitemShut {NoStop}%
\bibitem [{\citenamefont {Kageyama}\ \emph {et~al.}(1999)\citenamefont {Kageyama}, \citenamefont {Yoshimura}, \citenamefont {Stern}, \citenamefont {Mushnikov}, \citenamefont {Onizuka}, \citenamefont {Kato}, \citenamefont {Kosuge}, \citenamefont {Slichter}, \citenamefont {Goto},\ and\ \citenamefont {Ueda}}]{kageyama1999}%
  \BibitemOpen
  \bibfield  {author} {\bibinfo {author} {\bibfnamefont {H.}~\bibnamefont {Kageyama}}, \bibinfo {author} {\bibfnamefont {K.}~\bibnamefont {Yoshimura}}, \bibinfo {author} {\bibfnamefont {R.}~\bibnamefont {Stern}}, \bibinfo {author} {\bibfnamefont {N.~V.}\ \bibnamefont {Mushnikov}}, \bibinfo {author} {\bibfnamefont {K.}~\bibnamefont {Onizuka}}, \bibinfo {author} {\bibfnamefont {M.}~\bibnamefont {Kato}}, \bibinfo {author} {\bibfnamefont {K.}~\bibnamefont {Kosuge}}, \bibinfo {author} {\bibfnamefont {C.~P.}\ \bibnamefont {Slichter}}, \bibinfo {author} {\bibfnamefont {T.}~\bibnamefont {Goto}},\ and\ \bibinfo {author} {\bibfnamefont {Y.}~\bibnamefont {Ueda}},\ }\bibfield  {title} {\bibinfo {title} {{Exact Dimer Ground State and Quantized Magnetization Plateaus in the Two-Dimensional Spin System ${\mathrm{SrCu}}_{2}({\mathrm{BO}}_{3}){}_{2}$}},\ }\href {https://doi.org/10.1103/PhysRevLett.82.3168} {\bibfield  {journal} {\bibinfo  {journal} {Phys. Rev. Lett.}\ }\textbf {\bibinfo {volume} {82}},\ \bibinfo {pages} {3168} (\bibinfo {year} {1999})}\BibitemShut {NoStop}%
\bibitem [{\citenamefont {Sipe}\ and\ \citenamefont {Shkrebtii}(2000)}]{sipe2000}%
  \BibitemOpen
  \bibfield  {author} {\bibinfo {author} {\bibfnamefont {J.~E.}\ \bibnamefont {Sipe}}\ and\ \bibinfo {author} {\bibfnamefont {A.~I.}\ \bibnamefont {Shkrebtii}},\ }\bibfield  {title} {\bibinfo {title} {{Second-order optical response in semiconductors}},\ }\href {https://doi.org/10.1103/PhysRevB.61.5337} {\bibfield  {journal} {\bibinfo  {journal} {Phys. Rev. B}\ }\textbf {\bibinfo {volume} {61}},\ \bibinfo {pages} {5337} (\bibinfo {year} {2000})}\BibitemShut {NoStop}%
\bibitem [{\citenamefont {{Parker, Daniel E. and Morimoto, Takahiro and Orenstein, Joseph and Moore, Joel E.}}(2019)}]{parker2019}%
  \BibitemOpen
  \bibfield  {author} {\bibinfo {author} {\bibnamefont {{Parker, Daniel E. and Morimoto, Takahiro and Orenstein, Joseph and Moore, Joel E.}}},\ }\bibfield  {title} {\bibinfo {title} {{Diagrammatic approach to nonlinear optical response with application to Weyl semimetals}},\ }\href {https://doi.org/10.1103/PhysRevB.99.045121} {\bibfield  {journal} {\bibinfo  {journal} {Phys. Rev. B}\ }\textbf {\bibinfo {volume} {99}},\ \bibinfo {pages} {045121} (\bibinfo {year} {2019})}\BibitemShut {NoStop}%
\bibitem [{\citenamefont {Qin}\ \emph {et~al.}(2011)\citenamefont {Qin}, \citenamefont {Niu},\ and\ \citenamefont {Shi}}]{qin2011}%
  \BibitemOpen
  \bibfield  {author} {\bibinfo {author} {\bibfnamefont {T.}~\bibnamefont {Qin}}, \bibinfo {author} {\bibfnamefont {Q.}~\bibnamefont {Niu}},\ and\ \bibinfo {author} {\bibfnamefont {J.}~\bibnamefont {Shi}},\ }\bibfield  {title} {\bibinfo {title} {{Energy Magnetization and the Thermal Hall Effect}},\ }\href {https://doi.org/10.1103/PhysRevLett.107.236601} {\bibfield  {journal} {\bibinfo  {journal} {Phys. Rev. Lett.}\ }\textbf {\bibinfo {volume} {107}},\ \bibinfo {pages} {236601} (\bibinfo {year} {2011})}\BibitemShut {NoStop}%
\end{thebibliography}%

\end{document}